\documentclass[%
twocolumn,
reprint,
 amsmath,amssymb,
 aps, physrev,
]{revtex4-2}

\usepackage{graphicx}
\usepackage{dcolumn}
\usepackage{bm}
\usepackage{xcolor}
\usepackage{soul}
\usepackage[labelformat=simple]{subcaption}

\newcommand{\Ren}{\mbox{Re}}

\newcommand{\rme}{\mathrm{e}}
\newcommand{\rmi}{\mathrm{i}}
\newcommand{\rmd}{\mathrm{d}}

\newcommand{\bz}{\mathbf{0}}
\newcommand{\bu}{\mathbf{u}}
\newcommand{\nab}{\mathbf{\nabla}}
\newcommand{\bUb}{\boldsymbol{U}_{\rme}}

\newcommand{\pb}{P_{\rme}}
\newcommand{\bq}{\mathbf{q}}

\newcommand{\hbf}{\hat{\mathbf{f}}}
\newcommand{\bx}{\mathbf{x}}
\newcommand{\hbU}{\hat{\boldsymbol{U}}}
\newcommand{\pbU}{\boldsymbol{U}^{\prime}}
\newcommand{\bU}{\boldsymbol{U}}

\newcommand{\bff}{\mathbf{f}}
\newcommand{\ez}{\epsilon_o}
\newcommand{\eps}{\epsilon}
\newcommand{\wz}{\omega_o}
\newcommand{\cc}{\mbox{c.c.}}
\newcommand{\pa}{\partial}
\newcommand{\sig}{\sigma}
\newcommand{\de}{\delta}
\newcommand{\ga}{\gamma}
\newcommand{\bga}{\Gamma}
\newcommand{\om}{\omega}
\newcommand{\tet}{\theta}
\newcommand{\Tet}{\Theta}
\newcommand{\al}{\alpha}
\newcommand{\bet}{\beta}
\newcommand{\hbu}{\hat{\mathbf{u}}}
\newcommand{\ebu}{\tilde{\mathbf{u}}}
\newcommand{\ebq}{\tilde{\mathbf{q}}}
\newcommand{\bbu}{\breve{\mathbf{u}}}

\newcommand{\hxi}{\hat{\xi}}
\newcommand{\obu}{\overline{\mathbf{u}}}
\newcommand{\obup}{\overline{\mathbf{u}}^{\perp}}
\newcommand{\obfp}{\hat{\overline{\mathbf{f}}}^{\perp}}
\newcommand{\op}{\overline{p}}
\newcommand{\opp}{\overline{p}^{\perp}}
\newcommand{\hobu}{\hat{\overline{\mathbf{u}}}}

\newcommand{\fnl}{\hat{\mathbf{f}}_{nl}}
\newcommand{\fnlm}{\hat{\mathbf{f}}_{nl,0}}
\newcommand{\fnlh}{\hat{\mathbf{f}}_{nl,2}}

\newcommand{\pfnlm}{\mathbf{f}^{\prime}_{nl,0}}
\newcommand{\pfnlh}{\mathbf{f}^{\prime}_{nl,2}}
\newcommand{\punlm}{\mathbf{u}^{\prime}_{nl,0}}
\newcommand{\punlh}{\mathbf{u}^{\prime}_{nl,2}}

\newcommand{\pnlm}{p^{\prime}_{nl,0}}
\newcommand{\pnlh}{p^{\prime}_{nl,2}}

\newcommand{\cbu}{\check{\mathbf{u}}}

\newcommand{\bbup}{\breve{\mathbf{u}}^{\perp}}

\newcommand{\ccp}{\check{p}}

\newcommand{\bpp}{\breve{p}^{\perp}}
\newcommand{\bp}{\breve{p}}
\newcommand{\cbf}{\check{\mathbf{f}}}
\newcommand{\bg}{\mathbf{g}}

\newcommand{\pbu}{\mathbf{u}^{\prime}}
\newcommand{\pbq}{\mathbf{q}^{\prime}}

\newcommand{\pbf}{\mathbf{f}^{\prime}}
\newcommand{\pxi}{\xi^{\prime}}

\newcommand{\Gt}{G_{(\al,\bet)}(t_o)}

\newcommand{\bC}{\mathbf{C}}
\newcommand{\bL}{\mathbf{L}}
\newcommand{\bP}{\mathbf{P}}
\newcommand{\bB}{\mathbf{B}}
\newcommand{\bBi}{\mathbf{B}^{\infty}}
\newcommand{\bI}{\mathbf{I}}
\newcommand{\bR}{\mathbf{R}}
\newcommand{\bPhi}{\mathbf{\Phi}}
\newcommand{\eixz}{\rme^{\rmi(\alpha x + \beta z)}}

\newcommand\ssp[2]{\left \langle #1 \middle| #2 \right \rangle}

\newcommand\spt[2]{\left \langle #1 \middle| #2 \right \rangle_{v,\mathcal{T}}}

\newcommand\npt[1]{|| #1 ||_{v,\mathcal{T}}}

\newcommand\spf[2]{\left \langle #1 \middle| #2 \right \rangle_{v,\mathcal{F}}}

\newcommand\npf[1]{|| #1 ||_{v,\mathcal{F}}}

\newcommand\set[1]{\left\{ #1 \right\}}
\newcommand\spn[1]{\mbox{Span}\left\{ #1 \right\}}
\newcommand\pae[1]{\left( #1 \right) }
\newcommand\sae[1]{\left[ #1 \right] }
\newcommand\ei[1]{\rme^{\rmi #1 t}}
\newcommand\ve[2]{\begin{pmatrix} #1 \\ #2 \end{pmatrix}}
\newcommand\veh[2]{\begin{pmatrix} #1 & #2 \end{pmatrix}}
\newcommand\maa[4]{\begin{pmatrix} #1 & #2 \\ #3 & #4 \end{pmatrix}}
\newcommand\ea[1]{\mathbb{E}\left[ #1 \right]}
\newcommand \ta[1]{ \overline{#1} }

\newcommand\eixzm[1]{\rme^{\rmi #1 (\alpha x + \beta z)}}

\newcommand \Fo[1]{ \mathcal{F}[ #1 ] }
\newcommand \Foi[1]{ \mathcal{F}^{-1}[ #1 ] }

\newcommand{\iinf}{\int_{-\infty}^{\infty}}

\begin{document}

\preprint{APS/123-QED}

\title{\textbf{Nonmodal amplitude equations} 
}%

\author{Yves-Marie Ducimeti\`ere}
 \email{Contact author: yd3213@nyu.edu}
\affiliation{%
Courant Institute of Mathematical Sciences, New York University, New York, NY 10012, USA
}%

\author{Fran\c{c}ois Gallaire}
 \email{Contact author: francois.gallaire@epfl.ch}
\affiliation{
Laboratory of Fluid Mechanics and Instabilities, \'Ecole Polytechnique F\'ed\'erale de Lausanne, Lausanne CH-1015, Switzerland}

\date{\today}

\begin{abstract}
We consider fluid flows for which the linearized Navier-Stokes operator is strongly non-normal. The responses of such flows to external perturbations are spanned by a generically very large number of non-orthogonal eigenmodes. They are therefore qualified as ``nonmodal" responses, to insist on the inefficiency of the eigenbasis to describe them. In the aim of the article to reduce the system to a lower-dimensional one free of spatial degrees of freedom, (eigen)modal reduction techniques, such as the center manifold, are thus inappropriate precisely because the leading-order dynamics cannot be restricted to a low-dimensional eigensubspace. On the other hand, it is often true that only a small number (we assume only one) of singular modes is sufficient to reconstruct the nonmodal responses at the leading order. By adopting the latter paradigm, we propose a general method to analytically derive a weakly nonlinear amplitude equation for the nonmodal response of a fluid flow to a small harmonic forcing, stochastic forcing, and initial perturbation, respectively. In these last two problems, we assumed a parallel base flow with a spatially monochromatic external excitation. The present approach for deriving nonmodal amplitude equations is simpler than the one we previously proposed, for neither the operator perturbation nor the ensuing compatibility condition is formally necessary. Additionally, it provides an explicit, rigorous treatment of the sub-optimal responses. When applied to the stochastic response, the present method makes it possible to derive an amplitude equation that is substantially easier to solve and interpret than the one we proposed previously. 
Eventually, the three derived amplitude equations are tested in three distinct, two and three-dimensional flows. For sufficiently small excitation amplitudes, yet up to values large enough for the flow to depart from the linear regime, they can systematically predict the weakly nonlinear modification of the gains as the amplitude of the external excitation is increased. This, at an extremely low numerical cost as compared to fully nonlinear techniques. However, the proposed weakly nonlinear approach, precisely because it condemns the flow spatial structure to be mostly along the leading singular mode, is often too simplistic to predict the occurrence of subcritical transitions as the excitation amplitude is increased to too large values. 
\end{abstract}

\maketitle


\section{Introduction: responses to external disturbances}

Incompressible fluid flows, governed by the incompressible Navier-Stokes equations, are of considerable phenomenological richness. This includes forming complex spatio-temporal patterns, chaos, turbulence, and many others. Some of these observed phenomena could find elements of explanations by characterizing the linear response, i.e., the response to infinitesimally small disturbances, of the Navier-Stokes equations, which we briefly recall in the following lines. In doing so, we introduce some specific notations that will be used throughout this article. Some fully nonlinear or semi-linear extensions proposed in the literature are also succinctly presented.

\subsection{Response to an initial perturbation}

Let $\bq(\bx,t)=(\bu(\bx,t),p(\bx,t))^T$, with $\bu$ the velocity perturbation field and $p$ the pressure field, designates an infinitesimal perturbation of the flow state around one of its (possibly many) fixed points, whose velocity field we name $\bUb(\bx)$ in what follows (the knowledge of the corresponding pressure field $\pb(\bx)$ is not found necessary in the sequel). The velocity $\bUb$ is sometimes also called ``base flow". The dynamics of $\bq$ is linear and governed by
\begin{align}
\bB\pa_t\bq = \bL \bq,
\label{eq:lin}
\end{align}
subject to
\begin{align}
\bq(\bx,0)=\bq_0(\bx) = \ve{\bu_0(\bx)}{p_0(\bx)},
\label{eq:linBIS}
\end{align}
with $\bq_0$ the initial condition of the perturbation and where no sustained external forcing is considered for the moment. The linear operator $\bL$ results from linearizing the Navier-Stokes equations around $\bUb$. The explicit expression for $\bL$ shall be given later in this article. The operator $\bB$ is defined as 
\begin{align}
\bB = \maa{\bI}{\bz}{\bz}{0}, 
\label{eq:project}
\end{align}
with $\bI$ the identity operator of the dimension of the velocity field. The presence of the singular operator $\bB$ in Eq.~(\ref{eq:lin}) ensues from the fact that the pressure field is not differentiated with respect to time. Indeed, for incompressible flows considered here, the role of the pressure is solely to ensure the instantaneous satisfaction of the divergence-free (incompressibility) condition on the velocity. 
The system in Eq.~(\ref{eq:lin}) is also equipped with appropriate boundary conditions of $\bq$ on $\pa \Omega$, the latter symbol designating the boundary of the considered spatial domain $\Omega$.

Assuming the system in Eq.~(\ref{eq:lin}) to be diagonalizable and omitting the possible presence of continuous spectra, it is associated with a family of generically complex eigenmodes. Some of them belong to the non-trivial kernel of $\bB$, corresponding to eigenmodes with null velocity but non-null pressure, and thus are associated with infinite eigenvalues. The latter eigenmodes are unimportant for our purposes and will not be considered in what follows. Instead, we denote by $\set{\ebq_j}_{j\geq 1} =\set{\ebq_1,\ebq_2,...}$ the eigenmodes which do not belong to the kernel of $\bB$, and thus are associated with finite eigenvalues. By hypothesis, they provide a complete basis for the velocity field, but not for the full state variable consisting of both velocity and pressure, precisely because the ignored eigenmodes have a null velocity field.  
Each of these eigenmodes solves by definition 
\begin{align}
\sig_j \bB \ebq_j =  \bL \ebq_j,\quad j=1,2,...,
\label{eq:eig}
\end{align}
where the scalar $\sig_j$, finite and generically complex-valued, is the eigenvalue associated with the eigenmode $\ebq_j$.

It is instructive to expand the velocity field solution of Eq.~(\ref{eq:lin}) in the basis formed by the family $\set{\ebq_j}_{j\geq 1}$, which by abuse of language we call the eigenbasis. For this purpose, it is necessary to construct another basis, bi-orthogonal to the eigenbasis under some inner product, such as to determine the component of the solution on each of the eigenmodes. The $L^2(\Omega)$ inner product for generally complex functions on $\Omega$, is commonly considered and is expressed as 
\begin{align}
\ssp{\bq_a}{\bq_b} = \int_{\Omega}\bq_a^{H}\bq_b \rmd \Omega,
\label{eq:L2ip}
\end{align}
with the superscript $H$ designating the Hermitian transpose. Following this choice of inner product, it is possible to construct an operator that is said to be ``adjoint" to $\bL$, denoted by $\bL^{\dag}$, and ensuing from the definition
\begin{align}
\ssp{\bL\bq}{\bq^{\dag}} = \ssp{\bq}{\bL^{\dag}\bq^{\dag}}, \quad \forall \bq\in\mathcal{D}(\bL), \bq^{\dag}\in\mathcal{D}(\bL^{\dag}).
\label{eq:adj}
\end{align}
The function space $\mathcal{D}(\bL)$ (resp. $\mathcal{D}(\bL^{\dag})$) is the domain of the operator $\bL$ (resp. $\bL^{\dag}$) and contains appropriate
boundary and regularity conditions on the direct field $\bq$ (resp. adjoint field $\bq^{\dag}$). Furthermore, it is easy to show that $\bB^{\dag}=\bB$. The eigenmodes of the system adjoint to that in Eq.~(\ref{eq:lin}), denoted $\set{\ebq^{\dag}_j}_{j\geq1}$ solve
\begin{align}
\sig^*_j \bB \ebq^{\dag}_j =  \bL^{\dag} \ebq^{\dag}_j, \quad j=1,2,...
\label{eq:eigad}
\end{align}
with the superscript ``$*$" standing for the complex conjugation. Again, we have ignored the eigenmodes belonging to the kernel of $\bB$. The eigenmodes of the adjoint system, $\set{\ebq^{\dag}_j}_{j\geq1}$, will be referred to as the ``adjoint eigenmodes", as opposed to $\set{\ebq_j}_{j\geq 1}$, referred to as the ``direct eigenmodes". It is easy to demonstrate that the direct and adjoint eigenmodes are such that 
\begin{align}
\ssp{\bB\ebq_k^{\dag}}{\ebq_j} = \ssp{\ebu_k^{\dag}}{\ebu_j} = 0, \quad \mbox{if} \quad j\neq k.
\label{eq:biorth}
\end{align}
In other terms, as desired, the adjoint and direct eigenmodes form bi-orthogonal bases under the inner product in Eq.~(\ref{eq:L2ip}) weighted by $\bB$. Note that, as a consequence of this inclusion of $\bB$, the bi-orthogonality property of the adjoint and direct eigenmodes applies directly to their respective velocity fields.

From here, by using the bi-orthogonality property, the velocity field solution of Eq.~(\ref{eq:lin}) can be expanded on the eigenbasis as  
\begin{align}
\bu(t) = \sum_{j\geq 1} \ebu_j \rme^{\sig_j t} \frac{\ssp{\bB\ebq_j^{\dag}}{\bq_0}}{\ssp{\bB\ebq_j^{\dag}}{\ebq_j}} = \sum_{j\geq 1} \ebu_j \rme^{\sig_j t} \frac{\ssp{\ebu_j^{\dag}}{\bu_0}}{\ssp{\ebu_j^{\dag}}{\ebu_j}},
\label{eq:dyad}
\end{align}
where only the temporal dependencies are highlighted. Note that, as expected, the initial condition on the pressure perturbation field, $p_0$, does not influence $\bu(t)$, while the initial velocity $\bu_0$ does. By measuring the amplitude of $\bu(t)$ according to the norm induced by the $L^2$ inner product, which is directly proportional to the kinetic energy of the flow, we can derive from Eq.~(\ref{eq:dyad}) the following expression 
\begin{widetext}
\begin{align}
||\bu(t)||^2 = \ssp{\bu(t)}{\bu(t)}  
=\sum_{j\geq 1} \rme^{2\sigma_{j,r}t} \frac{\left | \ssp{\ebu^{\dag}_j}{\bu_0}\right |^2}{\left | \ssp{\ebu^{\dag}_j}{\ebu_j}\right |^2} ||\ebu_j||^2
        + \underbrace{\sum_{j\geq 1} \sum_{k \neq j} \rme^{(\sigma_j^*+\sigma_k)t}\frac{\ssp{\ebu^{\dag}_j}{\bu_0}^*\ssp{\ebu^{\dag}_k}{\bu_0}}{\ssp{\ebu^{\dag}_j}{\ebu_j}^*\ssp{\ebu^{\dag}_k}{\ebu_k}}\ssp{\ebu_j}{\ebu_k}}_{\parbox{5cm}{$= 0$ if $\bL$ is normal or if $\bu_0$ is along an eigenmode.}} 
\label{eq:kine}
\end{align}
\end{widetext}
where the subscript ``$r$" in $\sigma_{j,r}$ designates the real part. Equation~(\ref{eq:kine}), which can also be found for instance in Ref.~\cite{MeligHDR}, is enlightening in many ways, some of which we develop now.

\subsubsection{Linear time-asymptotic response}

Let us first consider the time asymptotic response for $t \rightarrow \infty$, and distinguish three cases.

\begin{itemize}
\item In this limit $t\rightarrow \infty$, the energy expressed in Eq.~(\ref{eq:kine}) is dominated by the term associated with the eigenvalue (or the pair of eigenvalues if complex conjugates) with the largest real part, as a consequence of the exponential dependencies. Thereby, if the largest real part is strictly negative, then the perturbation eventually vanishes, i.e. $||\bu(t)||^2 \rightarrow 0$ for $t \rightarrow \infty$. Accordingly, the fixed point $\bUb$ around which the equations were linearized is said to be ``linearly stable". 
\item If, on the contrary, the largest real part is strictly positive, then the perturbation eventually grows exponentially as $t\rightarrow \infty$ and $\bUb$ is then said to be ``linearly unstable". 
\item Eventually, if the largest real part is null, then the present linear theory does not conclude as to the stability of $\bUb$, and the latter is said to be ``linearly neutral" (or also ``linearly marginal").
\end{itemize}

In the rest of this article, whenever the adjectives ``stable", ``unstable" or ``neutral" are used, they will systematically refer to linear stability analysis.

\subsubsection{Linear finite-time response \label{sec:iftr}}

We emphasize that the discussion above, where conclusions can be drawn from the sole knowledge of the eigenvalues, concerned only the time asymptotic limit $t \rightarrow \infty$. However, we argue now that the finite time behavior might be just as relevant if $\bL$ is non-normal. The operator $\bL$ is said to be non-normal if it does not commute with its adjoint, i.e., if
\begin{align}
\bL\bL^{\dag} \neq \bL^{\dag}\bL,
\label{eq:nndef}
\end{align}
(which can also be caused by the fact that the domains $\mathcal{D}(\bL)$ and $\mathcal{D}(\bL^{\dag})$ differ from each other).
Accordingly, it is said to be normal if $\bL\bL^{\dag} = \bL^{\dag}\bL$ (where the equality also implies that the domains $\mathcal{D}(\bL)$ and $\mathcal{D}(\bL^{\dag})$ are the same), and to be self-adjoint if $\bL=\bL^{\dag}$. Note that a self-adjoint operator is necessarily normal, but the reciprocal is not true. Important properties concerning the eigenmodes of $\bL$ arise from this classification. In particular:

\begin{itemize}
\item if $\bL$ is normal, then each eigenmode is proportional to its adjoint, i.e. $\ebq_j \propto \ebq_j^{\dag}$ (equal up to the phase if normalized the same way) for $j=1,2,...$. Consequently, the relation in Eq.~(\ref{eq:biorth}) implies that the eigenmodes $\set{\ebq_j}_{j\geq 1}$ constitute an orthogonal family under the inner product in Eq.~(\ref{eq:L2ip}), with the inclusion of the operator $\bB$. An important consequence is that in Eq.~(\ref{eq:kine}), the double-sum term is necessarily null if $\bL$ is normal, even when $\bu_0$ is not along the velocity field of one of the eigenmodes. That is because $\ssp{\ebu_j}{\ebu_k} = 0$ if $k\neq j$. Therefore, the energy of the perturbation, $||\bu(t)||^2$, decays monotonously if the largest real part of the eigenvalue is negative, for it reduces to the sum of decaying exponentials pondered by positive coefficients. Moreover, $||\bu(t)||^2$ converges exponentially fast to a single exponential decay (resp. growth) with the rate given by the least stable (resp. most unstable) eigenvalue.  
\item If $\bL$ is non-normal, however, then the eigenmodes do not form an orthogonal set. Accordingly, the double sum term in Eq.~(\ref{eq:kine}), which involves eigenmode-eigenmode interactions through the inner product $\ssp{\ebu_j}{\ebu_k}$, has no reason to vanish if $\bu_0$ projects over more than one eigenmode. Even if all the eigenvalues have a negative real part and thus all the exponential terms $ \propto \rme^{(\sigma_j^*+\sigma_k)t}$ decrease, their multiplying coefficients can be negative. This implies that the energy $||\bu(t)||^2$ can grow at finite times, as these terms, which contribute negatively to the sum, can decay faster than those that contribute positively. Furthermore, such growth can be shown to be algebraic (see §$4.1.1$ in Ref.~\cite{SH01}). This phenomenon, of a stable system showing an energy increase at finite times, is referred to as ``transient growth". For larger times, the exponential behavior ultimately takes over if the linearization is still valid. Generically, a very large number of eigenmodes are needed to characterize the transient growth, which suggests that these latter form an inefficient basis for the flow dynamics. In other terms, at finite time, the dynamics cannot be reduced to the least stable or most unstable eigenmode. As an example, Ref.~\cite{Akervik07} considered the transient growth in the energy of a fluid flow over a separated boundary-layer flow. Using eigenmodes as a projection basis for analyzing the flow dynamics, they have shown that about one hundred modes are required for converged results of optimal growth (please see their Fig.~4a).
\end{itemize}

Indeed, at least under the inner product in Eq.~(\ref{eq:L2ip}), the linearized Navier-Stokes operator generically is non-normal whenever the base flow $\bUb$ is non-zero \cite{Chomaz05, Reddy93, Trefethen93, SH01, KerswellAnnuRev2018}. That is because the sign of the term representing the advection of the perturbation by the base flow has been reversed in $\bL^{\dag}$ with respect to that in $\bL$. The direct and adjoint eigenmodes thus present different spatial supports and cannot be equal, for they have been advected in opposite directions. Nonetheless, it is important to realize that, by definition, the degree of non-normality of an operator depends on the choice of the inner product under which the adjoint operator is constructed. For instance, in the context of plane parallel shear flows, it was shown in Ref.~\cite{Heifetz05} that a certain measure of the flow non-normality differs if the adjoint is constructed under the inner product in Eq.~(\ref{eq:L2ip}), or if it is under another inner product inducing the enstrophy norm, i.e., the $L^2$ norm of the vorticity field. 

\subsubsection{Linear optimal transient growth \label{sec:ifoptg}}

While eigenmodes are sufficient to characterize the time asymptotic dynamics, they typically form an inefficient basis for describing the finite-time dynamics for non-normal systems. Thereby, fluid flow studies are often extended to computing the initial condition that maximizes the gain in energy at a certain time $t_o$, called a ``temporal horizon". One of the many benefits of this approach is to directly reveal the full potential of the system for transient growth. This amounts to solving the maximization problem
\begin{align}
      G(t_o) &= \max_{\bu_0}\frac{||\bu(t_o)||}{||\bu_0||} = \max_{\bu_0}\frac{||\bPhi(t_o,0)\bu_0||}{||\bu_0||}=||\bPhi(t_o,0)|| \nonumber \\
      &=\max_{\bu_0}\sqrt{\frac{\ssp{\bPhi(t_o,0)^{\dag}\bPhi(t_o,0)\bu_0}{\bu_0}}{\ssp{\bu_0}{\bu_0}}}.  
    \label{eq:Gto}
\end{align}
In Eq.~(\ref{eq:Gto}), we have introduced the ``propagator" operator $\bPhi(t_a,t_b)$, whose application on the velocity field at time $t=t_b$, i.e. $\bu(t_b)$, maps it onto its evolution at time $t=t_a$, i.e. $\bu(t_a)=\bPhi(t_a,t_b)\bu(t_b)$. Specifically, the propagator $\bPhi(t_a,t_b)$ is a synthetic manner to represent the temporal integration of Eq.~(\ref{eq:lin}) from $t=t_b$, knowing $\bu(t_b)$, to $t=t_a$, leading to $\bu(t_a)$. Note that writing $\bu(t_a) = \bPhi(t_a,t_b)\bu(t_b)$ implies that the pressure field at time $t_b$ has no influence on the velocity field at time $t_a$. Indeed, as we mentioned previously and as it appears from the dyadic expansion in Eq.~(\ref{eq:dyad}), this is a consequence of the specific expression of the singular operator $\bB$.    

The solution to the maximization problem in Eq.~(\ref{eq:Gto}) can be found by computing the operator norm of $\bPhi(t_o,0)$, or, equivalently, the largest eigenvalue of the operator $\bPhi(t_o,0)^{\dag}\bPhi(t_o,0)$. We denote by $\set{\bbu_{0,j}}_{j\geq 1}$ the eigenmodes of $\bPhi(t_o,0)^{\dag}\bPhi(t_o,0)$, each solving by definition
\begin{align}
\bPhi(t_o,0)^{\dag}\bPhi(t_o,0)\bbu_{0,j}=G^2_j\bbu_{0,j}, \quad j=1,2,...
    \label{eq:satg}
\end{align}
The eigenvalue problem in Eq.~(\ref{eq:satg}) is self-adjoint, even if $\bL$ is not, which implies that the eigenvalues $G_1 > G_2 > ... $, sorted by decreasing magnitude, are real and positive. Each eigenmode $\bbu_{0,j}$ is normalized such that $||\bbu_{0,j}||=1$, for all $j \geq 1$. The maximum transient gain, at time $t=t_o$, is directly given by $G(t_o)=G_1$. The associated optimal initial condition corresponds to $\bbu_{0,1}$. The family of structures $\set{\bbu_{0,j}}_{j \geq 2}$ correspond to sub-optimal initial conditions. The eigenmode $\bbu_{0,j}$ also coincides with the $j$th right ``singular" mode of $\bPhi(t_o,0)$, associated with the singular value $G_j$. To be specific, it is possible to write
\begin{align}
G_j \bbu_j(t_o) = \bPhi(t_o,0)\bbu_{0,j}, \quad G_j \bbu_{0,j} = \bPhi(t_o,0)^{\dag}\bbu_j(t_o),
\end{align}
where $\bbu_j(t_o)$, also normalized such that $||\bbu_j(t_o)||=1$ for $j\geq 1$, is the $j$th left singular mode of $\bPhi(t_o,0)$. Thereby, $\bbu_j(t_o)$ is the normalized response at $t=t_o$ of the system seeded by $\bbu_{0,j}$. Of central importance, and inherited from the fact that the operator in Eq.~(\ref{eq:satg}) is self-adjoint, both the $\set{\bbu_{0,j}}_{j \geq 1}$ and the $\set{\bbu_j(t_o)}_{j \geq 1}$ families are respectively orthonormal. Thereby, some arbitrary initial condition $\bu_0$ can be decomposed as  
\begin{align}
\bu_0 = \sum_{j \geq 1}\ssp{\bbu_{0,j}}{\bu_0}\bbu_{0,j},
\end{align}
leading to a response at $t=t_o$ described by
\begin{align}
\bu(t_o) &= \bPhi(t_o,0)\bu_0 \nonumber \\
&= \sum_{j \geq 1}\ssp{\bbu_{0,j}}{\bu_0} \bPhi(t_o,0)\bbu_{0,j} \nonumber \\
&= \sum_{j \geq 1}\ssp{\bbu_{0,j}}{\bu_0} G_j \bbu_j(t_o).
    \label{eq:respde}
\end{align}
Equation (\ref{eq:respde}) helps understand why amending the paradigm from computing the eigenmodes of $\bL$ to solving Eq.~(\ref{eq:satg}), is fruitful. In doing so, we replaced an inefficient, non-orthogonal eigenbasis, with two orthonormal ones into which both the initial condition and the response at $t=t_o$ can be projected, respectively. Above all, the component of $\bu(t_o)$ onto $\bbu_j(t_o)$ is directly given by the component of $\bu_0$ onto $\bbu_{0,j}$, multiplied by the associated gain $G_j$. This makes it possible to prioritize the respective contribution of each structure $\bbu_j(t_o)$ in the response $\bu(t_o)$. For instance, in a configuration where $G(t_o)=G_1 \gg G_2 >G_3>...$, i.e. where $\bPhi(t_o,0)$ is close to being of unitary rank, and where an arbitrary initial condition $\bu_0$ supposedly does not project particularly well on one of the sub-optimal $\set{\bbu_{0,j}}_{j\geq 2}$, the response in Eq.~(\ref{eq:respde}) is well approximated by   
\begin{align}
\bu(t_o) \approx \ssp{\bbu_{0,1}}{\bu_0} G_1 \bbu_1(t_o).
    \label{eq:respdeapp}
\end{align}
This simple result has profound consequences: the structure (not the amplitude) of the response does not depend on the structure of the initial condition but becomes inherent to the propagator itself. Indeed, $\bbu_1(t_o)$ is computed solely from the knowledge of $\bPhi(t_o,0)$. As soon as a few singular modes $\bbu_j(t_o)$ are associated with gains much larger than all the others, these leading structures are expected to dominate that of the response for some times around $t_o$, regardless of how the latter was initiated (and ignoring the unlikely initial conditions that are orthogonal to the set of corresponding leading $\bbu_{0,j}$). This implies that it is possible to predict the flow structure even in an uncontrolled environment, where $\bu_0$ can't be known or prepared. In this perspective, note that the configuration where solely $\bbu_1(t_o)$ has a much larger gain than all the others is particularly favorable. 

The maximum transient gain and associated optimal structures are parameterized by the temporal horizon $t_o$, and thus need to be computed for each value of the latter. They can be very different depending on $t_o$, and the desired property $G_1 \gg G_2 > G_3 > ...$ generically only holds over a restricted range of intermediate $t_o$ values.

Solving Eq.~(\ref{eq:Gto}) and computing the associated singular modes is referred to in the literature as a ``nonmodal" analysis, where ``non-eigenmodal" should be understood. Indeed, the adjective nonmodal emphasizes the difference with a stability analysis, concerned with the computation of eigenmodes.

Due to the non-normality of the linearized Navier-Stokes operator, fluid flows furnish numerous examples of transient growth phenomena (please see Ref.~\cite{SH01} for a review). To the authors' knowledge, nonmodal studies have been first carried out in the context of parallel shear flows by Refs.~\cite{Butler92, Schmid94, Corbett00} and revealed two non-normal growth mechanisms, Orr and ``lift-up". The adjective ``parallel" means that the base flow is invariant in the direction of the main advection. 


Non-parallelism of the base flow was shown in Ref.~\cite{Cossu97} and later in Refs.~\cite{Blackburn08, Marquet08, Marquet09} to be an additional source of non-normality. This was also formalized in the more general framework of variable-coefficients operators in Ref.~\cite{Tref05} Chapter~$11$. 
The authors of Ref.~\cite{Cossu97} argued that this additional source of non-normality is related to the presence of a region of space within which the base flow profile is (locally) convectively unstable. 
Accordingly, a considerable body of work has been devoted to computing transient growth properties of non-parallel flows. Among them, Refs.~\cite{Ehrenstein05, Akervik08, Monkrousos10} concerned with a spatially evolving Blasius boundary layer flow, could be mentioned. As for them, Refs.~\cite{Akervik07, Ehrenstein08, Alizard09} were concerned about a separated boundary layer flow. 

\subsubsection{Nonlinear effects and possible bypass transition}

Until now, the developments and conclusions were within the framework of linear dynamics, exact only in the limit of infinitesimal perturbations. However, even if a perturbation is small enough for the linearization to be valid at initial times, precisely because its response can be substantially amplified through linear non-normal mechanisms, the nonlinear interactions of the latter may not remain negligible. This calls for the study of nonlinear effects on transient growth, raising the question of the relevance of linear optimal structures in characterizing nonlinear flow regimes. As we argue below, although nonlinear behaviors might be categorized, there is no generic answer to this question.

The authors of Ref.~\cite{Trefethen93} argue that nonlinear terms, the amplitude of which is made significant by linear transient growth, can make the flow escape from its original attractor. This can, for instance, happen as the perturbation nonlinearly feeds back onto the base flow, thus modifying it to render it unstable. Such phenomenology, where linear transient growth and nonlinearities act collaboratively and successively to bring about a flow transition, was called the ``bypass scenario" in Ref.~\cite{Trefethen93}. It is nowadays a well-accepted scenario for the nonlinear transition of some linearly stable flows to another state, including a turbulent attractor. 

The bypass transition scenario is exemplified in Ref.~\cite{Pringle12} and Ref.~\cite{Farano15}, where the linearly stable pipe and plane Poiseuille flows, respectively, were shown to transition to turbulence following a low-amplitude initial perturbation (provided the Reynolds number is sufficiently large). The bypass scenario was also found to be relevant in Refs.~\cite{Rossi97, Ducimetiere23} for the transition of the Lamb-Oseen vortex flow to another state, whose vorticity field shows three poles instead of one. In boundary layer flow, the non-normal ``lift-up" mechanism appears to be an essential ingredient for the transition to turbulence \cite{Matsubara01, Andersson99, Brandt04, Song24}, in addition to the role it plays in the so-called self-sustained cycle of turbulence \cite{Hamilton95, Waleffe97}.


In the bypass scenario presented above, the perturbation is even more amplified in a nonlinear regime than in a purely linear one, but this need not be the case. Nonlinearities can have a ``saturating" effect on the energy of the perturbation, inducing a decrease in the transient gain with the amplitude of the initial perturbation. This is, for instance, occurring in the backward-facing step flow at $\Ren=500$ considered in Ref.~\cite{Blackburn08}, as visible in Fig.~$12$ therein. 

To systematically assess the effect of nonlinearities on the optimal transient gain and its associated structure, a comprehensive analysis framework has recently been condensed in Ref.~\cite{KerswellAnnuRev2018}. There, the maximization problem in Eq.~(\ref{eq:Gto}) is generalized to include the nonlinearities of the Navier-Stokes equations, and the solution is found through a direct-adjoint looping algorithm. Preceding this review article, this method was applied to compute the nonlinear optimal initial perturbations leading to the flow transition to turbulence, in the Blasius boundary layer flow \cite{Cherubini10, Cherubini11}, in the plane Couette flow \cite{Duguet10, Monokrousos11, Rabin12, Cherubini13}, as well as in the pipe \cite{Pringle12} and plane \cite{Farano15, Farano16} Poiseuille flows.

\subsection{Response to a harmonic forcing}

Until now, we have focused on the response to an initial perturbation. Importantly, some of the reported observations can be easily adapted to the responses to other types of excitations. Among them, the response of a stable system to a harmonic forcing and a stochastic forcing in the momentum equations, respectively, are also of interest in this article. We briefly introduce these two problems in the following lines. 

\subsubsection{Linear optimal harmonic response}

Let a harmonic forcing in momentum have a frequency $\wz$ and a structure $\hbf_h(\bx)$. Its linear response over a stable system, implying $\rmi \wz$ not to be an eigenvalue of $\bL$, also oscillates at $\wz$ in the time asymptotic regime $t\rightarrow\infty$ and reads $\bu(\bx,t) \rightarrow \hbu_h(\bx)\ei{\wz} + \cc$, with
\begin{align}
\hbu_h = \bR(\wz)\hbf_h = \sum_{j\geq 1}\frac{\ebu_j}{\rmi \wz - \sigma_j}\frac{\ssp{\ebu_j^{\dag}}{\hbf_h}}{\ssp{\ebu_j^{\dag}}{\ebu_j}}.
\label{eq:resh}
\end{align}
In Eq.~(\ref{eq:resh}) we have introduced the resolvent operator $\bR(\wz)$, mapping the forcing structure $\hbf_h$, onto the response structure $\hbu_h$. The latter shall be referred to as the ``harmonic response". The energy $||\bu(t)||^2$ of the response averaged over one oscillation period is proportional to $||\hbu_h||^2$. Using the dyadic represention of the resolvent operator in Eq.~(\ref{eq:resh}), the full expression for $||\hbu_h||^2$ is easily shown to also involve a double sum term $\sum_{j\geq 1}\sum_{j\neq k} ... \ssp{\ebu_j}{\ebu_k}$ as in Eq.~(\ref{eq:kine}). Thereby it is also true here that, if $\bL$ is non-normal and thus the eigenmodes are not orthogonal, then the energy of the harmonic response is determined by a generically very large number of eigenmode-eigenmode interactions. Again, the eigenmodes thus form a very inefficient basis, in the sense that the harmonic response is inefficiently described by a single or even a few of them. This is in stark contrast with vibration analysis in structural mechanics, precisely called modal analysis. 

For this reason, nonmodal tools are also typically employed to characterize the harmonic response. They typically consist of computing the singular values and associated left and right singular modes of the resolvent operator $\bR(\wz)$. Again, these two singular modes families both constitute orthonormal bases, into which some unknown forcing structure and its response can be efficiently projected, respectively. Crucially, the respective contributions of the pair of singular modes to the overall harmonic response can also be prioritized according to their associated singular value, or ``harmonic gain".

For these reasons, the computation of the largest singular value, i.e., maximum harmonic gain, of the resolvent operator and associated optimal structures was widely performed in fluid mechanics literature. This is true for both parallel (please see Ref.~\cite{Schmid07} for a review) and non-parallel shear flows \cite{Akervik07, Alizard09, Sipp12, Boujo15,Garnaud13B}. Large values are often found for the maximum harmonic gain, which, as argued,  
is not necessarily related to a specific eigenmodal mechanism but can result from an interaction between a vast amount of eigenmodes. This is, for instance, the case in the mean turbulent jet flow studied in Ref.~\cite{Garnaud13B}. Indeed, their Fig.~$4$ shows a large harmonic gain of $O(10^2)$ over a range of frequencies for which, in their Fig.~$3$, no eigenvalues stand out.

As for the transient gain, large values of harmonic gain in non-parallel flows can sometimes be linked to the presence of a convectively unstable region \cite{Huerre98}. In such cases, there is typically a strong gap between the leading singular value of the resolvent operator and the others~\cite{Beneddine16}. This was illustrated in Ref.~\cite{Boujo15} for the flow past a backward-facing step, as well as in Ref.~\cite{Garnaud13B} and Refs.~\cite{Jeun16, Semeraro16, Schmidt18} for incompressible and compressible turbulent jets, respectively. In consequence, the response of the flow to a structurally arbitrary forcing, at a frequency associated with convective mechanisms, will systematically resemble the most amplified response. This is again an appreciable progress: by changing the paradigm from computing eigenmodes to computing singular modes, it is possible to reduce the dynamics of the flow to a single (singular) mode.

\subsubsection{Nonlinear extensions}

A large harmonic gain over some frequency interval implies the flow to largely amplify any sustained small external disturbance whose Fourier decomposition contains these frequencies. Therefore, as for the transient gain, it also indicates the propensity for the flow to trigger nonlinearities, which could make it transit to another state or regime, for instance, turbulence. This motivated the development of methods to extend the resolvent analysis in nonlinear regimes.

Among them, the ``self-consistent" method outlined in Ref.~\cite{Lugo16} considered the effect of nonlinearities on the harmonic response to external harmonic forcing, under some simplifying hypothesis. 
The velocity field is first split into a mean (in time) component and a fluctuation. The problem is then closed by assuming the fluctuation to be monochromatic and neglecting the effects of all harmonics. Under this hypothesis, the nonlinear fluctuation-fluctuation interaction term does not contribute to the fluctuation itself, as it produces different frequencies, and thus the fluctuation obeys a linear equation. The closed system was then solved iteratively for the flow past a backward-facing step at $\Ren=500$. It successfully compared with fully nonlinear simulations, and revealed the harmonic gain to decrease with nonlinearities. 

The technique advanced in Ref.~\cite{Lugo16}, which is equivalent to the idea of harmonic balance, can be generalized by including an arbitrary number of harmonics, as performed in Ref.~\cite{Rigas21}. This comes at the cost of augmenting the dimension of the system and making the equation for the fluctuations nonlinear. The number of harmonics to be included for an accurate description of the nonlinear flow is case-dependent, and the truncation is difficult to justify \textit{a priori}. Indeed, contrary to asymptotic expansion methods, the method does not formalize a hierarchical ordering between the amplitude of the harmonics.
Nevertheless, since the method in Ref.~\cite{Rigas21} describes a nonlinear harmonic response, it also provides a systematic framework to generalize the problem of maximizing the harmonic gain in nonlinear regimes, in the same way as Ref.~\cite{KerswellAnnuRev2018} did for the transient growth problem.

As another approach, Refs.~\cite{Pier01b,Pier03} considered the response of a spatially slowly-varying flow or system to a harmonic forcing localized in space. In the absence of an absolutely unstable region, they have shown that the nonlinear modulation of the linear response, in terms of both amplitude and wavenumber, can be simply deduced by solving a one-dimensional nonlinear spatial dispersion relation, parameterized by the streamwise coordinate. The theoretical framework proposed in Refs.~\cite{Pier01b, Pier03} comes at a very cheap computation cost with modern techniques and makes possible considerable conceptual progress.

\subsection{Response to a stochastic forcing}

The question of the flow response to some stochastic forcing is also appraised in this article. Such stochastic forcing can, for instance, be chosen as a sum of uncorrelated white noise processes, each multiplying a frozen spatial structure, as in Ref.~\cite{Farrell93}. Again, if $\bL$ is non-normal, then the response of the system in Eq.~(\ref{eq:lin}) to such stochastic forcing is generically inefficiently described by eigenmodes. They may all be stable with large damping rates, and yet a very large number of them can combine to bring about a substantial response-to-forcing variance amplification. Thus, here too nonmodal tools are employed to propose an efficient description of the linear flow dynamics. By ``efficient", we mean a description in which only one or a few basis modes are needed. 

\subsubsection{Linear optimal stochastic response}

The theoretical framework for a nonmodal stochastic forcing analysis of the linearized Navier-Stokes equations was introduced in Refs.~\cite{Farrell93, Farrell94, Farrell96}. Taking the stochastic forcing as white noise processes, the stochastic gains and the associated optimal and sub-optimal forcing and response structures were computed for the plane Poiseuille and Couette flow \cite{Farrell93}. The analysis was also carried out for the Lamb-Oseen vortex flow \cite{Fontane08} and the non-parallel flow past a backward-facing step \cite{Boujo15, Dergham2013}. In all these studies, the stochastic gains have revealed considerable.

\subsubsection{Nonlinear extensions \label{sec:stocnlex}}

Once again, linear nonmodal analysis of the receptivity to stochastic forcing can be extended to nonlinear flow regimes. 

A possible \textit{a priori} approach, named ``SSST" for stochastic structural stability theory, was introduced in Ref.~\cite{Farrell03} in the context of two-dimensional turbulent jets in a streamwise-periodic domain. The idea of the SSST is also to decompose the flow into a mean flow, averaged in the streamwise direction, and the sum of fluctuations, each with a different streamwise wavenumber. 
The fluctuation equations, one for each wavenumber, are forced by a nonlinear convolution term, accounting for the fluctuations-fluctuations interactions generating the considered wavenumber. To close the problem, this nonlinear fluctuations-fluctuations forcing term is modeled by a series of uncorrelated random noise processes which are white in time, with the intensity and the spatial auto-correlation function as free parameters. The equations for the fluctuations thus become linear. 
Among many other examples, the SSST system successfully described the interaction between rolls and streaks structure appearing in the transition to turbulence in the three-dimensional Couette flow \cite{Farrell12}.  

The fluctuations-fluctuations nonlinear interaction terms in the equation for the fluctuations, instead of being replaced by white noise, are sometimes simply ignored, leading to a class of models called ``semi-linear": nonlinear for the mean, and linear for the fluctuations. The semi-linear approach has been generalized in Ref.~\cite{Marston16} to include large-scale fluctuations-fluctuations interactions, and tested on externally stochastically driven jets over the spherical surface and $\beta$-plane (see Ref.~\cite{Marston23} for a recent review).

The authors of Ref.~\cite{Lugo16s} considered the effect of nonlinearities on the response of the backward-facing step to an external white noise forcing. Neglecting the fluctuations-fluctuations interaction terms, they arrived a to a coupled system that they solved iteratively, predicting nonlinear stochastic gain in good agreement with fully nonlinear results. Specifically, the stochastic gain significantly decreased while increasing the forcing amplitude.

\subsection{In this article: deriving weakly nonlinear amplitude equations for nonmodal responses}

This introductory part was relatively exhaustive, for it was concerned with the flow response to three different types of external excitations: an initial perturbation, a harmonic forcing, and a stochastic forcing. Nevertheless, the statements made can be summarized under the same general comments as follows:

\begin{itemize}
    \item An operator is non-normal if its application does not commute with that of its adjoint. A central consequence of non-normality is that the eigenmodes are not orthogonal under the inner product with which the adjoint was constructed. 
    \item Non-orthogonality of the eigenmodes implies that the corresponding induced norm of the linear flow response can take substantial values, determined by interactions between a generically large number of eigenmodes. Eigenmodes thus form an inefficient basis for describing the flow response, at least when the response is observed under this specific induced norm. 
    \item Instead, nonmodal tools make it possible to construct an orthonormal basis for the structure of the flow excitation. This is true whether the latter is an initial condition, a harmonic forcing, or a stochastic forcing. Furthermore, the respective contribution, to the induced norm of the response, of each element of this basis, can be prioritized. 
    \item In fluid mechanics, it is often the case that the responses to only a few of these forcing structures dominate the linear response, in terms of this induced norm, a property referred to as the low-rank approximation. In other words, projecting the linear response in the subspace spanned by the few dominant nonmodal responses extracts the leading-order response. This means that, at least in the linear regime, the Navier-Stokes equations can be rigorously reduced to a low-dimensional system of equations for the coordinates within this subspace. 
\end{itemize}

In the present article, we argue that projecting the flow response in the subspace spanned by the few dominant linear nonmodal responses, also extracts the leading-order response in a weakly nonlinear regime. Thereby, we derive a low-dimensional system of equations for the amplitudes of the dominant nonmodal responses, which incorporate the leading-order nonlinearities of the Navier-Stokes equations. They are valid in a regime where the error resulting from neglecting the higher-order nonlinearities is small according to the chosen induced norm. Owing to their simplicity, such nonmodal amplitude equations bring insight into the weakly nonlinear mechanisms that modify the gains as one increases the amplitude of the initial condition, the harmonic forcing, or a stochastic forcing, respectively. 

Nevertheless, precisely because of the perturbative nature of the method, where the leading-order is restricted to the low-dimensional dominant nonmodal subspace and/or to a low number of spatial or temporal harmonics, we shall see that the derived amplitude equations can be too simplistic to predict subcritical transitions of the flow.

As mentioned above, nonmodal responses to these three types of excitations are already provided with techniques capable of predicting fully nonlinear or semi-linear effects. Amplitude equations, concerned with the leading-order expression of nonlinear terms, are necessarily less predictive. However, they are easier to solve and interpret. In addition, fully nonlinear techniques highlighted in this introduction were each developed for their respective purposes. It is perhaps difficult to extract a general nonlinear theory linking these methods, whereas a clear analogy exists at the linear level. Instead, the amplitude equations that we derive in this article may each describe the response to a different nature of excitation, and yet, as we shall expose, they all ensue from the same backbone principle.     

\subsection{Existing model reduction methods}
There are already well-established techniques for reducing the Navier-Stokes equations to a low-dimensional system of weakly nonlinear ones for the amplitudes of (linear) eigenmodes. These modal amplitude equations are \textit{a priori} relevant for flows where only one or a few eigenmodes dominate the long-term linear dynamics (and thus, we expect, also the weakly nonlinear one). This is the case, for instance, if $\bL$ is neutral with a low-dimensional neutral eigenspace, or if $\bL$ is stable and its spectrum presents a large ``spectral gap". This means that a few eigenvalues have a damping rate much smaller (in amplitude) than all the others. In other terms, there are a few eigenmodes damped much more slowly than all the rest, such that the long-term linear dynamics can be well approximated by its projection into this low-dimensional, slow linear eigenspace. It is then \textit{a priori} relevant to project the weakly nonlinear dynamics into the slow manifold, i.e., to reduce the Navier-Stokes equations to a system of amplitude equations for the coordinates along the slowly decaying eigenmodes. Whether $\bL$ is neutral or stable, the overall principle is to slave the coordinates along the many quickly damped eigenmodes, to these along the few comparatively slow eigenmodes. From here, it is possible to reconstruct an asymptotic solution of the total nonlinear flow field. 

As reviewed in Ref.~\cite{Fauve98}, such reduction can be performed by relying on weakly nonlinear multiple-scale expansions. It has been performed for a variety of parallel \cite{Golubitsky85, Cross86, Crawford88, Chossat94, Chiffaudel87} and more recently non-parallel \cite{Sipp07, Meliga09, Meliga12} flows. The mathematical formalism of the multiple-scale method requires a nonempty neutral or close-to-neutral eigenspace in which the dynamics is projected. The center manifold theory, outlined in Refs.~\cite{Guckenheimer83, Haragus11, Cox91} and recently applied, for instance, in Refs.~\cite{Carini15, Negi24}, constitutes another technique. Among others, the center manifold theory was employed in Ref.~\cite{Knobloch83}, concerned with the convection in a horizontal layer rotating about a vertical axis. The authors also demonstrate that the amplitude equation they derived is identical to that obtained by using instead the multiple-scale technique. 
Recent developments from Refs.~\cite{Haller16, Li22}, restricted to stable systems, give the mathematical conditions for the existence and uniqueness of a ``spectral submanifold" (sometimes abbreviated in ``SSM"). A SSM is an invariant manifold that is the smoothest nonlinear continuation of some \textit{a priori} chosen (linear) eigensubspace. It is not a mathematical requirement that the latter must emanate from a slow eigenspace. However, as mentioned in Ref.~\cite{Haller16}, it is in this last configuration that a SSM yields the most physically relevant reduced-order model. The question of how to optimally select the slow or ``master" eigenspace onto which to project the dynamics, i.e. the eigenspace whose coordinates constitute the master coordinates to which all the remaining ones are slaved, is addressed in Refs.~\cite{Buza21, Touze21}. In the former article, the authors propose to compute the scalar curvature of the SSM in the direction of, say, the $j$th eigenmode and to include the latter in the master eigenspace if such curvature is large. Indeed, a large curvature implies that the $j$th eigenmode is substantially excited by weakly nonlinear interactions, thus, it must be associated with a master coordinate to capture its nonlinear retroaction on other master coordinates. As examples of application, the SSM theory was at the basis of the numerical tools deployed in Ref.~\cite{Kaszas22} for transitions among exact coherent states in the plane Couette flow. Eventually, the normal form theory \cite{Guckenheimer83, Haragus11} is another reduction technique amenable to nonlinear dynamical systems. The works in Refs.~\cite{Gallaire16, Pham18} consider extending the multiple scale technique above the instability threshold, where an eigenmode is unstable with a growth rate that does not enter at a perturbative level. As another approach, the Galerkin method is commonly employed to reduce the dimension of the flow dynamics \cite{Noack94, Noack03, Noack05}. The idea is to directly project the original equation into a subspace spanned by a certain number of eigenmodes, without prioritizing \textit{a priori} their contributions, e.g., in an asymptotic expansion. Consequently, the number of eigenmodes that should be included for describing accurately the flow dynamics is determined in an \textit{ad-hoc} manner, for the truncation error cannot be quantified according to some small parameter. As a famous example, the Lorentz system results from such a truncated Fourier-Galerkin expansion.

None of these modal reduction techniques is directly relevant to the current problem. 
That is because, as we already mentioned, by using the eigenmodes as a projection basis for the nonmodal response of a stable system, a generically considerable number of them is necessary for an accurate description. While for weakly non-parallel flows it is sometimes possible to link the linear nonmodal response to a single convectively unstable eigenmode per streamwise location \cite{Chomaz05, Boujo15} (i.e., to a single 'local' eigenmode, in opposition with 'global' eigenmodes, which are these considered throughout this article), the weakly nonlinear extension of such local analysis was shown in Ref.~\cite{LeDizes93} and Ref.~\cite{LeDizes94} (Section 2.2.3) to be mostly ill-posed. 

We refer again to Ref.~\cite{Akervik07}: no spectral gap is visible in their Fig.~$2$ and, indeed, about a hundred eigenmodes were necessary to reproduce the transient response in their Fig.~$4$. Consequently, even the SSM approach is inappropriate here, for the master eigenspace in which to project the dynamics cannot be low-dimensional. Indeed, a system of (at least) one hundred nonlinearly coupled amplitude equations could not be reasonably called a reduced model.
Another practical difficulty that may arise in using reduction methods based on modal quantities is that, for non-normal operators, eigenvalues and associated eigenmodes are excessively sensitive to small perturbations of the latter. Specifically, a $O(\eps)$ perturbation of a non-normal operator can modify the locations of its eigenvalues over distances substantially larger than $O(\eps)$ (please see Fig.~$15$ in Ref.~\cite{Schmid14} or Fig.~$3$ in Ref.~\cite{Ducimetiere22} for illustrations). As mentioned in Ref.~\cite{Chomaz05}, this implies in particular that a small perturbation of the operator can stabilize or destabilize the flow even far from the threshold. That is why a non-normal operator is typically characterized by its pseudospectrum (i.e., a set bounding the displacement of the eigenvalues under an $\eps$-perturbation of the operator) rather than by its spectrum, because the latter is generically, largely non-robust. We refer to Ref.~\cite{Tref05} Chapter~$2$ for an exhaustive survey on the topic.

Instead, we reorient the paradigm from assuming a spectral gap and projecting the weakly nonlinear flow dynamics onto the dominant eigenmodes, to assuming a large gap in the singular values of the response and projecting the weakly nonlinear flow dynamics onto the dominant nonmodal responses. It may be important to mention that a large spectral gap and a large gap in singular values are not mutually exclusive properties. It may be that the former is the only reason for the latter, even for non-normal operators (please see, for instance, Figs.~$4$ and $7$ in Ref.~\cite{Symon18}). In this situation, where the flow exhibits weakly damped eigenvalues responsible for the large gains, our method still applies. Simply, a classical modal amplitude equation is recovered in the limit of vanishing damping rate, as the optimal excitation structure and its response converge to the weakly damped adjoint and direct eigenmodes, respectively, and thus our method does not bring any originality compared to the modal methods introduced above. For that reason, in this article, the discussion and application cases will put the focus on nonmodal responses.

In increasing order of technicality, we will dedicate Sec.~\ref{sec:har} to the response to a harmonic forcing, Sec.~\ref{sec:sto} to the response to a stochastic forcing, and, eventually, Sec.~\ref{sec:tra} to the response to an initial perturbation. In these last two sections, to keep the calculations as light as possible, the analysis will be restricted to parallel base flows with spatially monochromatic external excitations (pure harmonic disturbance at a single wavelength).

\section{Weakly nonlinear response to a harmonic forcing \label{sec:har}}

\subsection{Linear regime \label{sec:subhar1}}

In this section, we propose a method to reduce the Navier-Stokes equations to an amplitude equation for the linearly optimal harmonic response. The operator $\bL$ is not assumed to possess any close-to-neutral eigenvalue(s), or any spectral gap, but to be stable and non-normal. 

The first step is to characterize the linear response by solving
\begin{align}
\bB\pa_t\bq &= \bL \bq + \ve{\hbf_h\ei{\wz}+\cc}{0} \nonumber  \\
&= \bL \bq + \bP^T\pae{\hbf_h\ei{\wz}+\cc},
\label{eq:linhar}
\end{align}
where we recall that $\bq(\bx,t) = (\bu(\bx,t) , p(\bx,t))^T$ contains both the velocity and the pressure field. Equation (\ref{eq:linhar}) is subject to a null initial condition, i.e. $\bq(\bx,0)=(\mathbf{0},0)^{T}$, although a precise expression of the latter is unimportant. We have also introduced $\hbf_h(\bx)$ as a complex-valued arbitrary forcing structure, and $\wz$ is the associated frequency. In Eq.~(\ref{eq:linhar}), we have defined the projection operator 
\begin{align}
\bP = \veh{\bI}{\bz},
\end{align}
such that
\begin{align}
\bP \ve{\bu}{p} = \bu,  \quad \bP^T\bu = \ve{\bu}{0}.
\end{align}
Specifically, applying $\bP$ on the state variable removes the pressure field, and the operator $\bB$ can be written as $\bB=\bP^T\bP$. Including $\bP$ in Eq.~(\ref{eq:linhar}) guarantees that the forcing acts as a source of momentum only, and not as a source of mass in the continuity equation. Note that it is also possible to force the pressure field. However, a forcing in pressure can be translated into a forcing in momentum via the incompressibility condition. 

Since $\bL$ is assumed stable $\rmi \wz$ cannot be one of its eigenvalues and the resolvent operator 
\begin{align}
\bR(\wz)=\bP(\rmi \wz \bB-\bL)^{-1}\bP^T
\end{align}
exists. The solution of Eq.~(\ref{eq:linhar}) for the velocity fields reads 
\begin{align}
\bu =  \bR(\wz)\hbf_h\ei{\wz} - \bPhi(t,0)\bR(\wz)\hbf_h + \cc.
\label{eq:linre}
\end{align}
The second term on the right-hand side of Eq.~(\ref{eq:linre}) must be present to enforce the velocity field to be initially null, but vanishes exponentially with time if $\bL$ is stable, which we assume in the rest of this section. Thereby, this term represents the transient part of the response. After it fades away, the response relaxes towards the harmonic response $\bR(\wz)\hbf_h\ei{\wz} + \cc$ such that, as we introduced already, in the limit $t\rightarrow \infty$, we have $\bu \rightarrow \hbu_h\ei{\wz} + \cc$ with 
\begin{align}
\hbu_h = \bR(\wz)\hbf_h.
\label{eq:resh2}
\end{align}

For the reasons advanced in the introductory part, if $\bL$ is non-normal, then it is particularly relevant to seek the forcing structure leading to the largest possible ratio between the norm induced by the $L^2$ inner product of the response, divided by that of the forcing. This $L^2$ inner product will be the only one considered in the present section, and all orthogonality properties are defined relatively to the latter. We aim to solve the maximization problem  
\begin{align}
G(\wz) &= \max_{\hbf_h}\frac{||\hbu_h||}{||\hbf_h||} = \max_{\hbf_h}\sqrt{\frac{\ssp{\hbf_h}{\bR(\wz)^{\dag}\bR(\wz)\hbf_h}}{\ssp{\hbf_h}{\hbf_h}}} \nonumber \\
&= \frac{1}{\ez},
\label{eq:hg1}
\end{align}
where $G(\wz)$ is the maximum harmonic gain. The eigenmode family of the self-adjoint operator $\bR(\wz)^{\dag}\bR(\wz)$ is denoted by $\set{\cbf_j}_{j\geq 1}$, where each eigenmode $\cbf_j$ is normalized according to $||\cbf_j||=1$, for all $j \geq 1$. It is associated with the real and strictly positive eigenvalue $G^2_j$, such that 
\begin{align}
\bR(\wz)^{\dag}\bR(\wz)\cbf_j = G_j^2 \cbf_j, \quad j =1,2, ...
\label{eq:hg2}
\end{align}
The eigenvalues $\set{G^2_j}_{j\geq 1}$ are sorted by decreasing values, and the largest eigenvalue, $G^2_1$, is equal to the optimal harmonic gain squared. Furthermore, as a consequence of the self-adjoint property of the operator $\bR(\wz)^{\dag}\bR(\wz)$, the family $\set{\cbf_j}_{j\geq 1}$ forms an orthonormal family of forcing structure. Among them, $\cbf_1$ is the most amplified, i.e., the optimal one, such that the following modes $\set{\cbf_j}_{j\geq 2}$ constitute sub-optimal forcing structures. The mode $\cbf_j$ is also the $j$th right singular mode of $\bR(\wz)$, associated with the singular value $G_j$ and with the left singular mode $\cbu_j$, also normalized as $||\cbu_j||=1$ for all $j \geq 1$. Specifically, it is possible to write 
\begin{align}
G_j \cbu_j =\bR(\wz)\cbf_j, \quad G_j \cbf_j =\bR(\wz)^{\dag}\cbu_j, \quad j\geq 1.
\label{eq:sm1}
\end{align}
This makes $\cbu_1$ the optimal harmonic response and $\set{\cbu_j}_{j\geq 2}$ the sub-optimal ones. 

Decomposing the harmonic forcing structure $\hbf_h$ in the $\set{\cbf_j}_{j \geq 1}$ orthonormal family, then applying the resolvent operator, leads to the following decomposition of the harmonic response
\begin{align}
\hbu_h = \sum_{j\geq 1} G_j \cbu_j \ssp{\cbf_j}{\hbf_h}.
\label{eq:harr}
\end{align}
Importantly, in Eq.~(\ref{eq:harr}), the contribution of each structure $\cbu_j$, to $\hbu_h$, is weighted by the corresponding $G_j$. 

Through the nonmodal mechanisms mentioned in the introduction, it is consistent to assume the optimal gain $G_1$ to be large. Furthermore, it is plausible to place ourselves in a configuration where $G_1$ is much larger than the sub-optimal gains, i.e., $G_1 \gg G_2 > G_3>...$ (as supported at least by Refs.~\cite{Boujo15, Beneddine16, Garnaud13B, Jeun16, Semeraro16, Schmidt18} for nonmodal mechanisms of a convective nature). Note that this supposedly large gap between the first and the subsequent singular values amounts to assuming that the resolvent operator is close to being of unit rank. We show in this section that, under these assumptions, it is possible to reduce the weakly nonlinear dynamics to solely a scalar equation for the amplitude of the optimal response $\cbu_1$. Due to the assumed large gap between the leading singular value and all others, the amplitude of the weakly nonlinear response along the sub-optimal structures $\set{\cbu_j}_{j \geq 2}$, can be slaved to that for $\cbu_1$. Quantitatively, the sub-optimal gains are assumed to be either of the order of the square root of the optimal gain, or of order one. The reasons behind these specific scalings will become clear in a moment. Mathematically, we assume that $G_j = \ga_j G_1^{1/2}$ for $N \geq j \geq 2$ and $G_j = \ga_j $ for $j\geq N+1$, with, $\ga_j = O(1)$ for $j\geq 2$. The integer $N$ is the index of the smallest sub-optimal gain which scales like the square root of the optimal one. It is introduced for formal reasons only, and its precise value has no practical importance. The optimal gain $G_1$ must be sufficiently large for its square root to be much smaller. Note that this square root scaling of the largest sub-optimals is, for instance, appropriate in the (laminar) flow past the backward-facing step at $\Ren = 500$ in Ref.~\cite{Boujo15}, as visible in the inset of their Fig.~$4$(b). There, the optimal gain was found to be $G_1 = 7453$ whereas the first-sub-optimal is $G_2 = 40.36 \approx 0.5\sqrt{G_1}$. As further examples, this specific scaling also applies to the incompressible jet turbulent mean flow at $\Ren = 10^3$ studied in Ref.~\cite{Garnaud13B} (please see their Fig.~$4$(a) for the gain curves). It is also relevant, at least in a certain region of the parameter space, for the compressible jet turbulent mean flows considered in Ref.~\cite{Schmidt18} (Figs.~$8$ and $9$ therein), in Ref.~\cite{Jeun16} (Fig.~$2$ therein) and in Ref.~\cite{Semeraro16} (Fig.~$4$ therein).

In Eq.~(\ref{eq:hg1}), the notation $\ez$ was introduced to designate the inverse of the optimal harmonic gain, i.e. $\ez = 1/G_1$, and thus $\ez$ is small by assumption. In terms of this $\ez$, we have assumed
\begin{align}
&\set{G_1,G_2,...,G_N,G_{N+1},...} \nonumber \\
& = \set{\frac{1}{\ez},\frac{\ga_2}{\ez^{1/2}},...,\frac{\ga_N}{\ez^{1/2}},\ga_{N+1},...}, 
\quad \ga_j = O(1), \quad j\geq 2,
\label{eq:scg}
\end{align}
(and $\ga_1=1$). Using Eq.~(\ref{eq:harr}), the scalings in Eq.~(\ref{eq:scg}) imply the linear harmonic response to some forcing structure $\hbf_h$ to read
\begin{align}
\hbu_h =& \frac{1}{\ez}\Biggl( \cbu_1 \ssp{\cbf_1}{\hbf_h} + \ez^{1/2}\sum_{j=2}^{N} \ga_j \cbu_j \ssp{\cbf_j}{\hbf_h} \nonumber \\
&+ \ez \sum_{j>N} \ga_j \cbu_j \ssp{\cbf_j}{\hbf_h}\Biggl).
\label{eq:harr2}
\end{align}
In Eq.~(\ref{eq:harr2}), there is, by assumption, an order separation between the contribution along the optimal response, and that along all of the sub-optimal ones which are at least $\ez^{-1/2} \gg 1$ smaller. Thereby, the linear harmonic response $\hbu_h$ is along the optimal one $\hbu_1$ at leading-order even though the arbitrary forcing structure $\hbf_h$ generically excites all those in the family $\set{\cbf_j}_{j\geq 1}$.

We shall assume the latter assertion to remain true also in a weakly nonlinear regime and then exploit this assumption to reduce there the Navier-Stokes equations to a scalar one for the amplitude of $\hbu_1$.

\subsection{Weakly nonlinear continuation \label{sec:subhar2}}

We now describe the harmonic response in a weakly nonlinear regime. For this, let us first come back to the (nonlinear) Navier-Stokes equations, subjected to a harmonic forcing according to
\begin{align}
\bB \pa_t \ve{\bU}{P} =& \ve{-(\bU \cdot \nab)\bU + \Ren^{-1}\Delta \bU - \nab P}{\nab \cdot \bU} \nonumber \\
&+ \ez^{3/2}\ve{\phi\hbf_h \ei{\wz}+\cc}{0}.
\label{eq:NSf}
\end{align}
From now on, the arbitrary forcing structure $\hbf_h$ is normalized such that $||\hbf_h||^2=1$. We have introduced in Eq.~(\ref{eq:NSf}) the free parameter $\phi=O(1)$, such that overall the harmonic forcing amplitude scales like $\ez^{3/2}$ (this specific scaling explained in a moment). In a weakly nonlinear regime, we approach the solution with an asymptotic expansion in terms of integer powers of $\ez^{1/2}\ll 1$, according to
\begin{widetext}
\begin{align}
\ve{\bU}{P}=& \ve{\bUb}{\pb} + \ez^{1/2}\sae{A_1\ve{\cbu_1}{\ccp_1}\ei{\wz} + \cc} + \ez\sae{\sum_{j=2}^{N}A_j\ve{\cbu_j}{\ccp_j}\ei{\wz} + \cc + \ve{\obu_2}{\op_2}} \nonumber \\
&+ \ez^{3/2}\sae{\sum_{j>N}A_j\ve{\cbu_j}{\ccp_j}\ei{\wz} + \cc + \ve{\obu_3}{\op_3}} + \underbrace{\sum_{m \geq 4}\ez^{m/2}\ve{\obu_m}{\op_m}}_{=O(\ez^2)},
\label{eq:asy1}
\end{align}
\end{widetext}
where the complex-valued, scalar amplitudes $\set{A_j}_{j\geq 1}$, each $A_j=O(1)$, are for the moment unknown. We recall that $(\bUb,\pb)^T$ denotes a steady solution of the unforced Navier-Stokes equation.

Coming back to Eq.~(\ref{eq:NSf}), the fact that the harmonic forcing was introduced at $O(\ez^{3/2})$ ensues from the following reasoning. Say that the asymptotic expansion in Eq.~(\ref{eq:asy1}) is made in terms of integer powers of $\ez^n$, with $n$ some rational number (chosen as $n=1/2$ in Eq.~(\ref{eq:asy1})). The first order of the expansion, containing the optimal response oscillating at the forcing frequency $\wz$, is thus of order $\ez^n$. Consequently, because of the quadratic nature of the nonlinearity in Eq.~(\ref{eq:NSf}), we anticipate that the first nonlinear interaction term oscillating at $\wz$ will arise at order $\ez^{3n}$. Indeed, at order $\ez^{2n}$, the quadratic nonlinear interaction of the first order with itself will only produce harmonics terms oscillating at $\wz-\wz=0$ and $\pm(\wz+\wz)=\pm 2\wz$; consequently, at order $\ez^{3n}$, the nonlinear interaction of the first order with the second will produce terms oscillating at $\pm(2\wz-\wz)=\pm\wz$ and $\pm(2\wz+\wz)=\pm 3\wz$, which this time contains the fundamental frequency $\wz$. Furthermore, if the optimal response appears at order $\ez^n$, then it must result from an external forcing appearing at order $\ez^{n+1}$ since the gain in amplitude is of order $\ez^{-1}$. We impose that the external forcing and the leading-order nonlinear interaction term appear in the same order, which leads to solving $3n = n+1$ and thus $n=1/2$. This explains both why the expansion in Eq.~(\ref{eq:asy1}) is made in terms of powers of $\ez^{1/2}$ and why the external forcing appears at order $\ez^{3/2}$. Choosing to scale the sub-optimal gains also in terms of powers of $\ez^{1/2}$ in Eq.~(\ref{eq:scg}) is the consequence of this fact, not the cause. More precisely, it was made so that the contributions of the sub-optimal responses could be ordered rigorously in the expansion in Eq.~(\ref{eq:asy1}), which, as just explained, had to be done in terms of powers of $\ez^{1/2}$.

We have also introduced in Eq.~(\ref{eq:asy1}) the nonlinearly-induced corrections terms $\obu_m$ for $m\geq 2$, the overbar denoting harmonics fields which oscillate at multiples of $\wz$, but not at $\wz$. These are generated by nonlinear interactions only and can be described by the Fourier series
\begin{align}
\obu_m = \obu_{m,0} + \pae{\sum_{ n \geq 2}\obu_{m,n}\ei{n\wz} + \cc}, \quad m\geq 2,
\label{eq:fs}
\end{align}
which, we insist, does not include the fundamental frequency $\wz$. In other terms, the equality $\obu_{m,1}=\bz$ holds for all $m\geq 2$. This is made on purpose so that the weakly nonlinear harmonic response, say $\hbU(\wz)$ for the velocity field, by definition oscillating at $\wz$ only, directly reads
\begin{align}
\hbU(\wz) =& \ez^{1/2}A_1\cbu_1 + \ez\sum_{j=2}^{N}A_j\cbu_j + \ez^{3/2}\sum_{j>N}A_j\cbu_j.
\label{eq:Uasy1}
\end{align}
without the need to incorporate contributions from the $\obu_m$. Such contributions would necessarily be redundant, since the $\set{\cbu_j}_{j\geq 1}$ family constitutes a complete basis for the velocity field. Thereby, $\hbU(\wz)$ can be fully reconstituted solely from the knowledge of the amplitudes $\set{A_j}_{j\geq 1}$ (there are no $O(\ez^2)$ terms in Eq.~(\ref{eq:Uasy1})). Owing to the orthonormality property of the family $\set{\cbu_j}_{j\geq 1}$, each $A_j$ immediately corresponds to the amplitude of the (orthogonal) projection of $\hbU(\wz)$ onto $\cbu_j$, rescaled by the appropriate power of $\ez^{1/2}$.

The weakly nonlinear response in Eq.~(\ref{eq:Uasy1}) is constructed to be very similar to its linear counterpart in Eq.~(\ref{eq:harr2}), rescaled by $\phi\ez^{3/2}$, but where at each order the prefactor multiplying $\cbu_j$ is replaced by the unknown amplitude $A_j$, for all $j\geq 1$. This similarity is meant to ensure the (assumed) continuity between the weakly nonlinear and linear regimes. It implies in particular that each amplitude $A_j$ must thus reduce to $\phi \ga_j \ssp{\cbf_j}{\hbf_h}$ in the linear regime, and for all $j\geq 1$ (with $\ga_1=1$). In addition, as the forcing amplitude is increased, the amplitudes typically are modified by weakly nonlinear effects and thus depart from their linear values.

We note that it would be more asymptotically rigorous to expand each amplitude according to the power series 
\begin{align}
A_j = \sum_{k\geq 0}\ez^{k/2}A_j^{(k)} = A_j^{(0)} + \ez^{1/2}A_j^{(1)} + \ez A_j^{(2)} + O(\ez^{3/2}),
\label{eq:apw}
\end{align}
for $j\geq 1$. This way, the contributions to each $A_j$ can also be ordered asymptotically in Eq.~(\ref{eq:asy1}). However, this would make the calculations heavier while we expect not to change the final equation for $A_j$. For the interested reader, evidence of the latter assertion is proposed in Appendix~\ref{app:Fuji} for an $O(\ez^{3/2})$ truncation (and we expect the results to generalize up to an arbitrarily large one). 
Therefore, we shall solve for the $\set{A_j}_{j\geq 1}$ directly, with each $A_j$ absorbing its contributions of all orders.

Since the amplitudes $\set{A_j}_{j\geq 1}$ are assumed to remain of order one, Eq.~(\ref{eq:Uasy1}) inherently restricts the weakly nonlinear harmonic response to remain along the (linearly) optimal one, $\cbu_1$, at leading-order, and with an amplitude $\ez^{1/2} A_1$. In addition, Eq.~(\ref{eq:Uasy1}) includes higher-order correction terms, smaller by a factor of at least $\ez^{-1/2} \gg 1$, along the sub-optimal structures. These correction terms can be already present at a linear level, as they appear in Eq.~(\ref{eq:harr2}), but also, as we shall see in Secs.~\ref{sec:sp} and \ref{sec:sp}, nonlinearly excited.

\begin{figure*}
\centering
\includegraphics[trim={0cm 0cm 0cm 0cm},clip,width=0.75\linewidth]{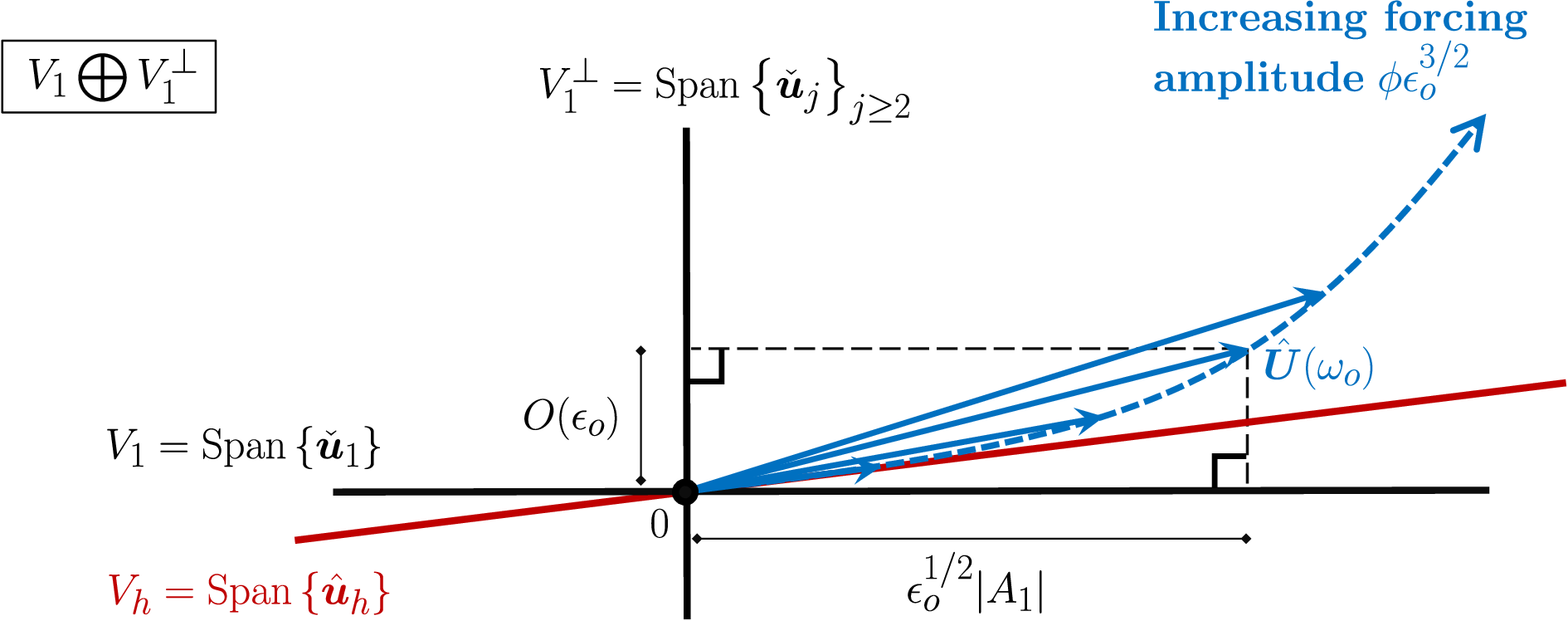}
\caption{Schematic representation of the weakly nonlinear harmonic response $\hat{\bU}(\wz)$ (blue arrows, one for each forcing amplitude) in the phase space, as described by the expansion in Eq.~(\ref{eq:Uasy1}). The phase space is shown here in the orthonormal basis formed by the $\set{\cbu_j}_{j\geq 1}$ family. The subspace $V_h =\spn{\hbu_h}$, where the linear response $\hbu_h$ to some arbitrary harmonic forcing $\hbf_h$ is expressed in Eq.~(\ref{eq:harr2}), is represented in red. By hypothesis, the contribution to $\hbu_h$ which is spanned by the sub-optimal responses, i.e., in $V_1^{\perp}=\spn{\cbu_j}_{j\geq2}$, is at least $\ez^{-1/2} = \sqrt{G_1}\gg 1$ times smaller than that along the optimal one $\cbu_1$. Thereby $V_h$ makes only a small angle with the subspace $V_1=\spn{\cbu_1}$. For vanishing harmonic forcing amplitude, the weakly nonlinear harmonic response must be tangent to $V_h$. However, as the forcing amplitude is increased, they become increasingly misaligned (please see the evolution of the blue arrows for increasing forcing amplitude). Consequently, as the forcing amplitude is increased, the locus of the weakly nonlinear responses (blue dashed line) is curved in the phase space. Yet, by continuity, the weakly nonlinear response still makes a small angle with the subspace $V_1$, such that its projection on $V_1$ still gives the leading-order response. }
\label{fig:sketch}
\end{figure*}

The velocity field corresponding to the expansion in Eq.~(\ref{eq:Uasy1}) is sketched in Fig.~\ref{fig:sketch}, in the phase space represented in the $\set{\cbu_j}_{j\geq 1}$ basis.    
Let us insist on the fact that, for Eq.~(\ref{eq:Uasy1}) to be a proper description of the harmonic response, the largest sub-optimal gain must be no larger than $O(\ez^{-1/2})$, as expressed in Eq.~(\ref{eq:scg}). This hypothesis will be systematically verified in all application cases. If there exist one or more sub-optimal gains of the same order as the optimal one, then Eq.~(\ref{eq:asy1}) must be amended so that the amplitude(s) of the corresponding sub-optimal response(s) also appear at leading order, alongside $A_1$. This yields no fundamental difficulties, but would make the subsequent calculations heavier and is beyond the scope of this article. 

Furthermore, the central assumption that the weakly nonlinear response is along the linear optimal one at leading order, i.e., that the angle between the responses locus (blue dashed line) and the subspace $V_1$ (black horizontal line) remains small in Fig.~\ref{fig:sketch}, is expected to be valid for sufficiently small forcing amplitude $\phi$ (yet up to large enough values for the flow to depart from is purely linear regime). That is in the continuity of the linear regime, where both curves must be tangent. Nevertheless, as we shall see in Sec.~\ref{sec:tra3} and Sec.~\ref{sec:substo3}, this assumption may be rapidly put into default as $\phi$ increases if the flow experiences a subcritical transition to another state (e.g., a turbulent attractor), for the latter typically exhibits a considerable enrichment both in terms of spatial structure and temporal harmonics. More generally, this assumption becomes quickly false as soon as the weakly nonlinear responses locus in Fig.~\ref{fig:sketch} exhibits a severe curvature in the phase space, for instance, because the nonlinear forcing term favors a specific sub-optimal response over the optimal despite the large gap in the gains. In the latter case, a solution in the spirit of Ref.~\cite{Buza21} could be proposed, and the proper set of optimal and sub-optimal responses to be included at leading-order could be chosen \textit{a priori}, still by gains values, but also by including the sub-optimal structure whose directions are associated with the largest local curvature of the weakly nonlinear responses locus.

Eventually, each pressure field $\ccp_j$ introduced in Eq.~(\ref{eq:asy1}) is the one associated with the velocity structure $\cbu_j$, such that  
\begin{align}
\ve{\cbu_j}{\ccp_j} =  G^{-1}_j\pae{\rmi \wz\bB-\bL}^{-1} \bP^{T}\cbf_j, \quad j\geq 1.
\label{eq:depp}
\end{align}

Before proceeding with the calculations, let us briefly emphasize that specifying a base flow, $\bUb$, and a Reynolds number, $\Ren$, completely determines $\ez$. Expansion in Eq.~(\ref{eq:asy1}) is then not truly asymptotic, since the small parameter $\ez$ is fixed and cannot be chosen arbitrarily small. Only in the limit $\Ren \rightarrow \infty$ does it hold that $\ez \rightarrow 0$, such that Eq.~(\ref{eq:asy1}) only is truly asymptotic in the limit of high Reynolds numbers, provided the scalings in Eq.~(\ref{eq:scg}) hold. Nevertheless, as we shall demonstrate, for a fixed value of $\ez$ the difference between the weakly nonlinear description in Eq.~(\ref{eq:asy1}), and the linear solution, can still be made arbitrarily small in the limit of vanishing forcing $\phi \rightarrow 0$.

Injecting both Eqs.~(\ref{eq:asy1}) and (\ref{eq:fs}) in the Navier-Stokes equations (\ref{eq:NSf}), leads to a new expansion for each of the Fourier components. The first, for the Fourier component oscillating at $0\times \wz=0$, writes
\begin{align}
&\bz = \ez \sae{\bL\ve{\obu_{2,0}}{\op_{2,0}} - 2|A_1|^2\bP^{T}\bC(\cbu_1,\cbu_1^*)} + O(\ez^{3/2}),
\label{eq:aF0}
\end{align}
where the superscript $``*"$ denotes the complex conjugation operation. In Eq.~(\ref{eq:aF0}), we have introduced the nonlinear, bilinear operator
\begin{align}
\bC(\bu_a,\bu_b) = \frac{1}{2} \sae{(\bu_a \cdot \nab)\bu_b +(\bu_b \cdot \nab)\bu_a}.
\label{eq:conv}
\end{align}
Additionally, the application of the linear operator $\bL$ over some $\pae{\bu_a,p_a}^T$ is computed as
\begin{align}
\bL\ve{\bu_a}{p_a} = \maa{-2\bC(\bUb,\bu_a)+\Ren^{-1}\Delta\bu_a}{-\nab p_a}{\nab \cdot \bu_a}{0}.
\label{eq:Le}
\end{align}

The expansion for the Fourier component oscillating at $2\wz$ reads
\begin{align}
\bz =& \ez \sae{-\pae{2\rmi \wz\bB-\bL}\ve{\obu_{2,2}}{\op_{2,2}} - A_1^2\bP^{T}\bC(\cbu_1,\cbu_1)} \nonumber \\
&+ O(\ez^{3/2}).
\label{eq:aF2}
\end{align}
The expansions for the even higher harmonics are not written down explicitly, as they will not be necessary for the following. In Eq.~(\ref{eq:aF0}) (resp., Eq.~(\ref{eq:aF2})), each $O(\ez^{m/2})$ is associated with its own unknown field $\obu_{m,0}$ (resp., $\obu_{m,2}$). Thereby, to have as many equations as unknown fields, the terms are collected at each specific order. In particular, collecting terms at $O(\ez)$ leads to
\begin{align}
\obu_{2,0} = |A_1|^2\obu_{2,0}^{|A_1|^2}, \quad \obu_{2,0}^{|A_1|^2} = -2\bR(0)\bC(\cbu_1,\cbu_1^*), \label{eq:so20}
\end{align}
together with
\begin{align}
\obu_{2,2} = A_1^2\obu_{2,2}^{A_1^2}, \quad \obu_{2,2}^{A_1^2} = -\bR(2\wz)\bC(\cbu_1,\cbu_1).
\label{eq:so22}
\end{align}
for the zero and second harmonics velocity fields, respectively. Let us insist that the superscripts $A_1^2$ and $|A_1|^2$ are not exponents but simply notations evoking the prefactors of the fields. We recall that the resolvent operator is such that $\bR(z) = \bP\pae{\rmi z \bB-\bL}^{-1}\bP^{T}$. Note that, to guarantee that the asymptotic expansion is well-posed, we implicitly assume both $\obu_{2,0}$ and $\obu_{2,2}$ to have an $O(1)$ norm. This way, these two fields indeed remain at $O(\ez)$ and do not need to the rescaled to appear at higher orders.
\begin{widetext}
The expansion for the Fourier component oscillating at $\wz$ is
\begin{align}
\bz =& \ez^{1/2}\sae{-A_1 \pae{\rmi \wz\bB-\bL} \ve{\cbu_1}{\ccp_1}} +\ez \sae{-\sum_{j=2}^{N}A_j \pae{\rmi \wz\bB-\bL} \ve{\cbu_j}{\ccp_j}}\nonumber  \\
& +\ez^{3/2}\sae{ -\sum_{j > N}A_j \pae{\rmi \wz\bB-\bL} \ve{\cbu_j}{\ccp_j} - 2\bP^{T}\pae{A_1\bC(\cbu_1,\obu_{2,0})  + A_1^*\bC(\cbu_1^*,\obu_{2,2})} + \phi \bP^{T} \hbf_h} +O(\ez^2).
\label{eq:aF1}
\end{align}
Eq.~(\ref{eq:aF1}) can be simplified by remembering the definition in Eq.~(\ref{eq:depp}), and becomes 
\begin{align}
& \bz = \ez^{1/2}\sae{-A_1 G^{-1}_1 \bP^T \cbf_1} \nonumber +\ez \sae{-\sum_{j=2}^{N}A_j G^{-1}_j \bP^T \cbf_j}\nonumber  \\
& +\ez^{3/2} \Biggl[ -\sum_{j > N}A_jG^{-1}_j \bP^T \cbf_j - 2\bP^{T}\pae{A_1\bC(\cbu_1,\obu_{2,0})  + A_1^*\bC(\cbu_1^*,\obu_{2,2})} + \phi \bP^{T} \hbf_h\Biggl] +O(\ez^2).
\label{eq:aF12}
\end{align}
By then using scalings of the gains in Eq.~(\ref{eq:scg}), as well as Eqs.~(\ref{eq:so20}) and (\ref{eq:so22}), and eventually applying the operator $\bP$ to project in the velocity space, we obtain
\begin{align}
& \bz = \ez^{3/2} \Biggl [ -A_1\cbf_1 -\sum_{j\geq 2} \ga^{-1}_j A_j \cbf_j - A_1|A_1|^2\pae{2\bC(\cbu_1,\obu_{2,0}^{|A_1|^2})  + 2\bC(\cbu_1^*,\obu_{2,2}^{A_1^2})} + \phi \hbf_h \Biggl ] + O(\ez^2),
\label{eq:aF13}
\end{align}
where the terms pre-multiplied by the gains at $O(\ez^{1/2})$ and $O(\ez)$ in Eq.~(\ref{eq:aF12}) are eventually appearing at $O(\ez^{3/2})$ in Eq.~(\ref{eq:aF13}). 
\end{widetext}

By construction, the $\set{A_j}_{j\geq 1}$ are the only unknowns in Eq.~(\ref{eq:aF13}) (all orders considered). This implies that the terms at each order should not be collected in Eq.~(\ref{eq:aF13}), for this would lead to as many equations (the term at $O(\ez^{3/2})$ would need to be equal to zero, but also the term at $O(\ez^{2})$, and also that at $O(\ez^{5/2})$, etc...) whose solutions each fix all the $\set{A_j}_{j\geq 1}$ once for all. In other terms, unlike the expansions for the harmonics fields, Eq.~(\ref{eq:aF13}) must here be solved all at once and not order-by-order, or the $\set{A_j}_{j\geq 1}$ are over-determined. The leading-order equation for each $A_j$ can be obtained by projecting Eq.~(\ref{eq:aF13}) over $\cbf_j$ and invoking on the orthonormality property of the $\set{\cbf_j}_{j\geq 1}$ family. In particular, taking the inner product with $\cbf_1$ leads to the following equation for $A_1$,
\begin{align}
\boxed{0 = -A_1 - A_1|A_1|^2(\mu + \nu) + \phi \ssp{\cbf_1}{\hbf_h}}+O(\ez^{1/2}), 
\label{eq:ampeq}
\end{align}
where we have defined the coefficients
\begin{subequations}
\begin{align}
\mu &= \ssp{\cbf_1}{2\bC(\cbu_1,\obu_{2,0}^{|A_1|^2})}, \label{eq:coeffshmu}\\
\nu &= \ssp{\cbf_1}{2\bC(\cbu_1^*,\obu_{2,2}^{A_1^2})}.
\label{eq:coeffshnu}
\end{align}
\end{subequations}
We obtain similarly for each $A_j$ with $j\geq 2$ the equation
\begin{align}
&-\ga^{-1}_j A_j - A_1|A_1|^2\ssp{\cbf_j}{2\bC(\cbu_1,\obu_{2,0}^{|A_1|^2})  + 2\bC(\cbu_1^*,\obu_{2,2}^{A_1^2})} \nonumber \\
&+ \phi \ssp{\cbf_j}{\hbf_h}+O(\ez^{1/2})=0.
\label{eq:ampeqj}
\end{align}
We note that projecting Eq.~(\ref{eq:aF13}) into the $\set{\cbf_j}_{j\geq 1}$ basis, specifically, has the advantage of, for each $j$, removing the contributions from the $\set{\cbf_k}_{k\geq 1,k\neq j}$. In Eq.~(\ref{eq:ampeq}) and Eq.~(\ref{eq:ampeqj}), the terms at $O(\ez^{1/2})$ embed higher-order nonlinearities.

Eq.~(\ref{eq:ampeq}) for $A_1$ is similar to that derived in Ref.~\cite{Ducimetiere22}, in the steady regime (please see Eq.~(2.12) therein, where the notations $A$, $\hbf_o$, $\hbu_o$, $\bu_{2,0}$ and $\hbu_{2,2}$ correspond in the present paper to $A_1$, $\cbf_1$, $\cbu_1$, $\obu_{2,0}^{|A_1|^2}$ and $\obu_{2,2}^{A_1^2}$, respectively). However, in Ref.~\cite{Ducimetiere22}, this equation was derived differently. It required perturbing the inverse resolvent operator to make it singular, then encompassing such operator perturbation in a multiple-scale asymptotic expansion closed by a standard compatibility condition. Although the development in Ref.~\cite{Ducimetiere22} has the advantage of linking with classical multiple-scale developments, in the present approach, neither the operator perturbation nor the compatibility condition was formally necessary. Furthermore, while it was stated without proof in Ref.~\cite{Ducimetiere22} that Eq.~(\ref{eq:ampeq}) is valid on condition that the sub-optimal gains are $O(\ez^{-1/2})$, the present approach gives the mathematical reason behind this, as it proposes a rigorous treatment of the sub-optimal responses. 

As expected, Eq.~(\ref{eq:ampeq}) reduces to the linear solution $A_1 =  \phi \ssp{\cbf_1}{\hbf_h}$ in the limit of vanishing forcing amplitude $\phi \rightarrow 0$. The first nonlinear term $A_1|A_1|^2\mu$ takes into account the fact that the nonlinear interaction of the optimal response $A_1\cbu_1$ with its complex conjugate yields a non-zero temporal average $|A_1|^2\obu_{2,0}^{|A_1|^2}$, which in turn retro-act on $A_1\cbu_1$ thus modifying its amplitude as compared to the linear regime. The second nonlinear term $A_1|A_1|^2\nu$ takes into consideration that the nonlinear interaction of the optimal response $A_1\cbu_1$ with itself produces a second harmonic field $A_1^2\obu_{2,0}^{|A_1|^2}$, whose retro-action on $A_1^*\cbu_1^*$ also modifies the amplitude of $\cbu_1$. 

It is in principle possible to introduce a dynamical term $\propto \rmd A_1 / \rmd t$ in Eq.~(\ref{eq:ampeq}), which could characterize some transient regime before reaching the steady equilibrium solution. For this, $A_1$ must be made dependent on a slow time scale according to $A_1=A_1(\ez t)$, and thereby the temporal derivative in the Navier-Stokes equations together with the chain rule derivative will naturally make appear a temporal derivative of $A_1$ in Eq.~(\ref{eq:ampeq}). This generalization was not performed in the calculation above because (i) we are interested in the harmonic gain which is defined on the steady amplitude, and (ii) the transient evolution of $A_1$ would still only give the transient evolution of the component oscillating at $\wz$ whereas we understand from the transient term in the right-hand side of Eq.~(\ref{eq:linre}), that the latter may involve many different instantaneous frequencies. 

The equations for the sub-optimal amplitude in Eq.~(\ref{eq:ampeqj}) are linear in $A_j$ at leading-order (i.e., by neglecting the $O(\ez^{1/2})$ term). They translate the fact that the sub-optimal response $\cbu_j$ is excited by the component on $\cbf_j$ onto (i) the externally applied forcing structure $\hbf_h$, and (ii) the third-order nonlinear interaction term $\propto A_1|A_1|^2$, involving the optimal structure only. The solution $A_j = \ga_j\phi\ssp{\cbf_j}{\hbf_h}$ is also consistently recovered in the limit of vanishing forcing amplitude. Interestingly, at leading-order, solving for the sub-optimal amplitudes $\set{A_j}_{j\geq 2}$ requires the knowledge of the optimal one $A_1$, whereas the reciprocal is not true and $A_1$ is determined by solving Eq.~(\ref{eq:ampeqj}) only. Such a slaving of the $\set{A_j}_{j\geq 2}$ to $A_1$ is a direct consequence of the assumed large singular value gap. The latter results in the asymptotic ordering in Eq.~(\ref{eq:asy1}), stipulating that the sub-optimal responses appear one order below the optimal one in terms of the expansion parameter. 



The harmonic gain associated with $\hbU(\wz)$, the component of the weakly nonlinear solution oscillating at $\wz$ and expressed in Eq.~(\ref{eq:Uasy1}), reads 
\begin{align}
G(\wz) &= \frac{||\hbU(\wz)||}{\ez^{3/2}\phi ||\hbf_h||} \nonumber \\
&= \frac{\pae{ \ez|A_1|^2 + \ez^2\sum_{N \geq j \geq 2}|A_j|^2 + O(\ez^3)}^{1/2} }{\ez^{3/2}\phi} \nonumber \\
&= \frac{\ez^{1/2}|A_1|(1 + O(\ez))}{\ez^{3/2}\phi} \nonumber \\
&= \frac{|A_1|}{\ez\phi}(1 + O(\ez)).
\label{eq:ghw}
\end{align}
Between the first and the second line in Eq.~(\ref{eq:ghw}), we have used the orthonormality property of the family $\set{\cbu_j}_{j\geq 1}$, together with the normalization $||\hbf_h||=1$. From Eq.~(\ref{eq:ghw}), the harmonic gain is given at leading order solely by a rescaled value of $A_1$, which incorporates leading-order nonlinear effects as $\phi$ is increased towards $O(1)$ values.

Eventually, we note that the nonlinear corrections fields $\obu^{A_1^2}_{2,2}$ and $\obu^{|A_1|^2}_{2,0}$ have prefactors quadratic in $A_1$. Thereby, since $A_1$ becomes proportional to $\phi$ in the linear limit $\phi \rightarrow 0$, these two fields can indeed be made arbitrarily smaller than the linear solution by decreasing the forcing amplitude. It can be shown that the latter fact is also true for nonlinear correction terms of higher orders.

\subsection{Application case: the flow past a BFS \label{sec:subhar3}}

For the sake of illustration, let us now appraise the validity of the amplitude equation (\ref{eq:ampeq}) on the two-dimensional flow past a backward-facing step (BFS). The latter flow is notoriously known for presenting a high degree of non-normality, which can be related to convective (Kelvin-Helmholtz) mechanisms \cite{Boujo15}. The results reported below have already been presented in Ref.~\cite{Ducimetiere22}, and more technical details can be found in the latter article.  

\begin{figure}
\centering
\includegraphics[trim={0.0cm 0.0cm 0.0cm 0.0cm},clip,width=0.975\linewidth]{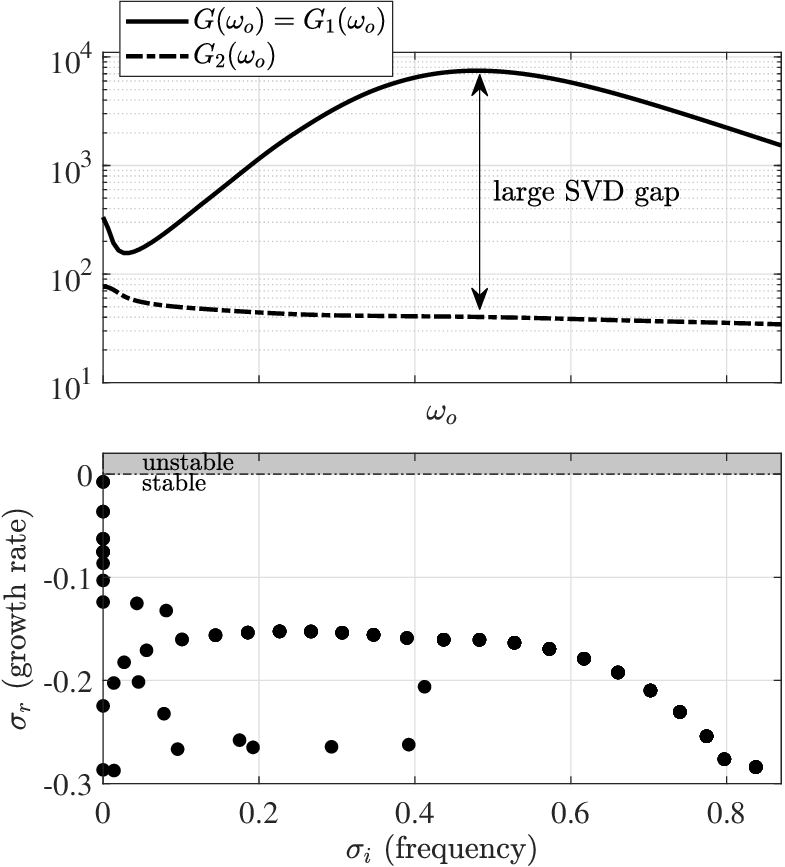}
\caption{Top: optimal linear harmonic gain (full line) for the BFS flow at $\Ren=500$ considered in Ref.~\cite{Ducimetiere22} (please see Fig.~$2$(a) therein for the geometry). The first sub-optimal gain is also reported as the dashed-dotted line. Gains are shown as a function of the forcing frequency. Bottom: spectrum of this same flow in the complex plane (each dot an eigenvalue). The frequency ranges on the $x$-axis are the same for both the top and the bottom frames. }
\label{fig:Ha}
\end{figure}

First, we show in Fig.~\ref{fig:Ha}, top frame, the (linear) optimal and first sub-optimal harmonic gain curves for the BFS flow at $\Ren=500$, as a function of the forcing frequency. 
Its spectrum is also revealed at the bottom frame, over the same frequency range. A comparison between both frames illustrates well the statements made in the introduction. Indeed, it appears that extremely large harmonic gains $O(10^4)$ are attained over a frequency range around $\wz \approx 0.5$, for which the eigenvalues are nevertheless all significantly damped and none of them stand out. In effect, these large gains are due to interactions between a substantial number of eigenmodes, and thus cannot be distinctly related to one or even a few of them. It also appears in Fig.~\ref{fig:Ha} that the optimal gain is substantially larger than the first sub-optimal one (mind the log scale of the $y$-axis of the top frame) and thus it is true that $G_1 \gg G_2 > G_3 > ...$, as hypothesized, for the whole considered range of frequencies but the ones around zero. This hypothesis made it possible to formally reduce both the linear and weakly nonlinear dynamics to solely the optimal response at leading order, ultimately leading to Eq.~(\ref{eq:ampeq}). Consequently, for the present application case, we \textit{a priori} expect the latter equation to be relevant.

\begin{figure*}
\centering
  \begin{subfigure}{1\columnwidth}
  \includegraphics[trim={0cm 0cm 0cm 0cm},clip,width=0.95\linewidth]{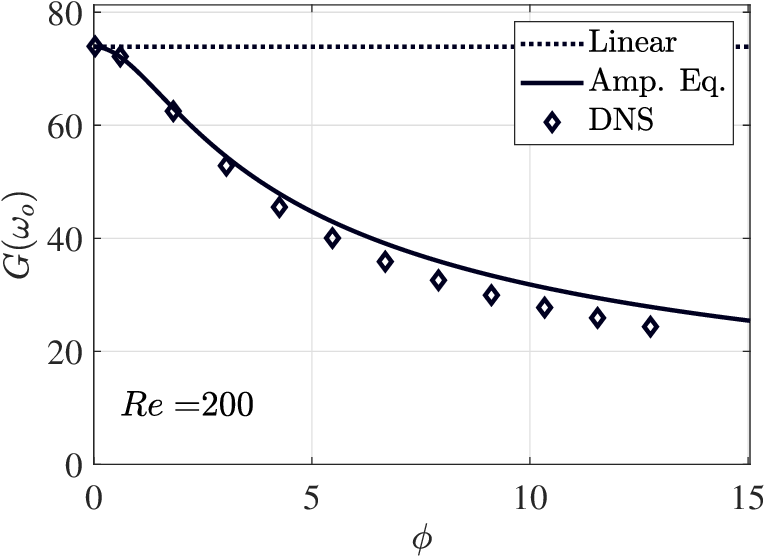}
  \caption{\label{fig:Hga}}
\end{subfigure}
\hfil
\begin{subfigure}{1\columnwidth}
  \includegraphics[trim={0cm 0cm 0cm 0cm},clip,width=1\linewidth]{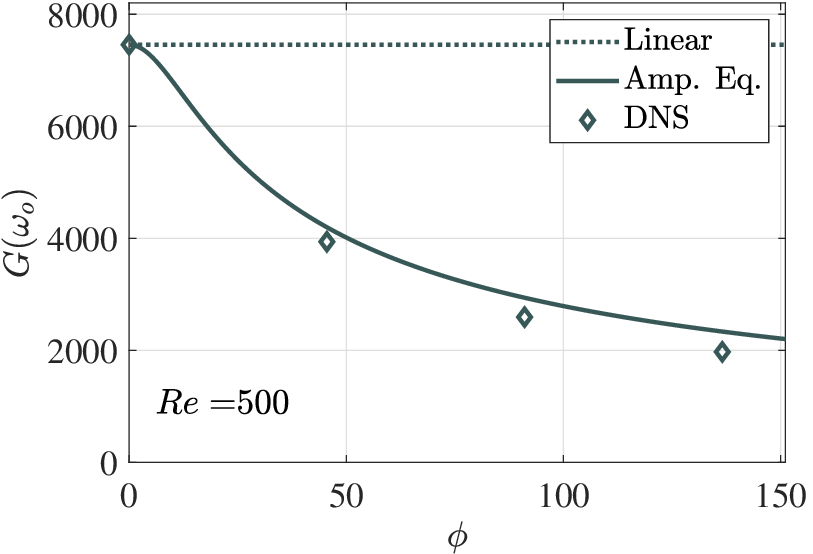}
  \caption{\label{fig:Hgb}}
\end{subfigure}
\begin{subfigure}{1\columnwidth}
  \includegraphics[trim={0cm 0cm 0cm 0cm},clip,width=0.95\linewidth]{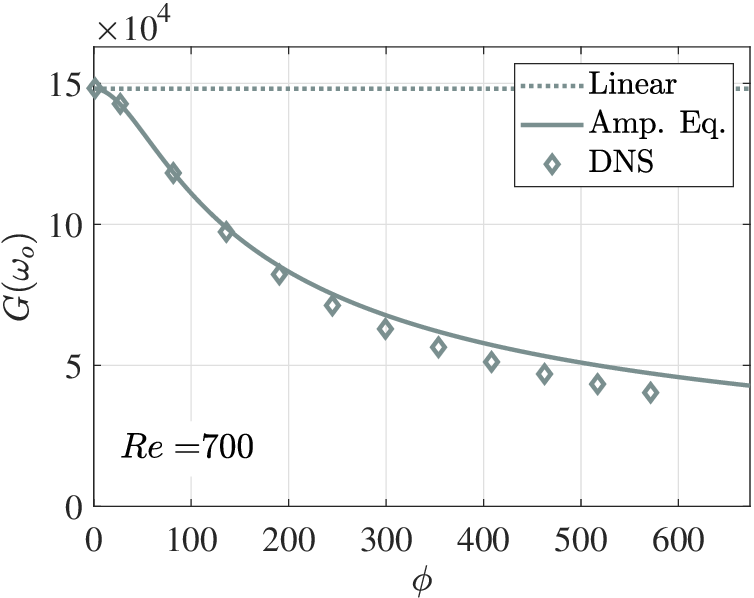}
  \caption{\label{fig:Hgc}}
\end{subfigure}
\hfil
\begin{subfigure}{1\columnwidth}
  \includegraphics[trim={0cm 0cm 0cm 0cm},clip,width=1\linewidth]{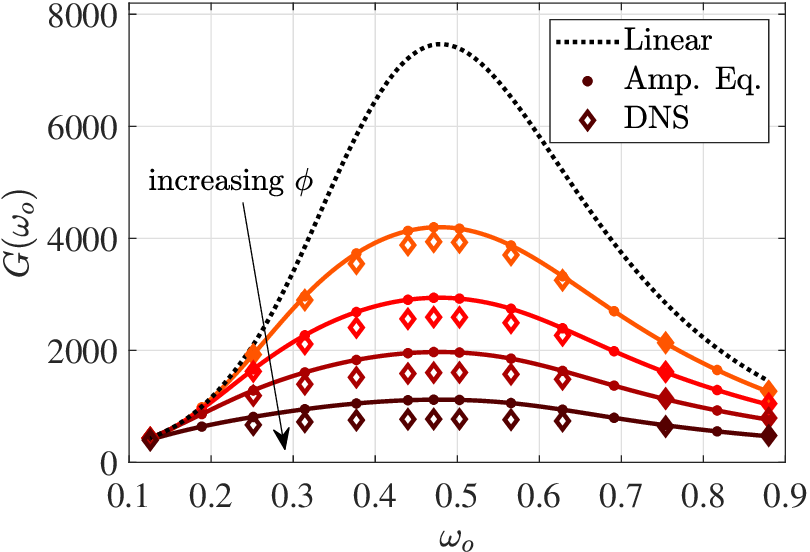}
  \caption{\label{fig:Hgd}}
\end{subfigure}
\caption{Harmonic gain for the BFS flow subjected to the (linear) optimal forcing structure, i.e., $\hbf_h=\cbf_1$. The gain is obtained from a linear ($=1/\ez$, dotted line), a leading-order weakly nonlinear ($=|A_1|/(\ez \phi)$ from Eq.~(\ref{eq:ghw}), continuous line) and a fully nonlinear (DNS, diamond markers) approach. In \subref{fig:Hga}-\subref{fig:Hgc}, the forcing frequency is set to $\wz=0.471$ and the gain is shown as a function of $\phi$, the rescaled harmonic forcing amplitude. It relates to a flow at $\Ren=200$ in \subref{fig:Hga}, at $\Ren=500$ in \subref{fig:Hgb}, and at $\Ren=700$ in \subref{fig:Hgc}. In \subref{fig:Hgd}, on the contrary, it is the Reynolds number which is fixed at $\Ren=500$, but the frequency is varied. Darker nuances correspond to larger forcing amplitudes $\ez^{3/2}\phi \in[1,2,4,10]/\sqrt{2}\times 10^{-4}$ ($\ez$ depending on the frequency). } \label{fig:HgvsF}
\end{figure*}

In Fig.~\ref{fig:HgvsF}, we show the evolution of harmonic gain of the BFS flow. The harmonic forcing structure is chosen as the optimal one, i.e. $\hbf_h=\cbf_1$.
Three approaches are compared: in (i) the linear one, the harmonic gain does not depend on the forcing amplitude. In (ii) the weakly nonlinear approach, the gain is given at leading order by $|A_1|/(\ez\phi)$ according to Eq.~(\ref{eq:ghw}), where $A_1$ solves the scalar, polynomial equation (\ref{eq:ampeq}). This approach is supposed to be valid as long as $\phi=O(1)$. In (iii) the fully nonlinear approach, the harmonic gain comes from a fully nonlinear direct numerical simulation (DNS), where the Fourier component oscillating at $\wz$ is extracted by post-processing (other frequencies, typically harmonics, are created by nonlinearities). Note that since we have chosen $\hbf_h = \cbf_1$, the sub-optimal structures are not linearly excited (although they are at a nonlinear level). This choice was made so that reducing the dynamics to the amplitude equation (\ref{eq:ampeq}) is exact in the linear regime, and the discrepancies between the weakly and fully nonlinear regimes can only stem from neglected nonlinearities. 

In Figs.~\ref{fig:Hga}-\ref{fig:Hgc}, the forcing frequency is fixed to $\wz = 0.471$, which yields the largest linear gain for $\Ren=500$, and the harmonic gain is shown as a function of the rescaled forcing amplitude $\phi$ for three different Reynolds numbers, one for each frame. For all considered Reynolds numbers, the agreement between the weakly and fully nonlinear gains is satisfactory. Although resulting from a simple polynomial equation whose coefficients can be computed from linear fields only, the weakly nonlinear approach yields a substantial improvement over the linear one in predicting fully nonlinear results. It predicts well the apparent decrease of the harmonic gain as the strength of the nonlinearities is increased. Nevertheless, the agreement between weakly and fully nonlinear results degrades progressively as $\phi$ is increased, presumably due to the ensuing increasing importance of the neglected, higher-order terms. The values of $\phi$ for which the discrepancies between the weakly and fully nonlinear results become visible are strikingly large, even more so as the Reynolds number is increased, and far above the postulated $O(1)$. It seems that $\phi$ can reach $O(\ez^{-1/2})$ values, and the weakly nonlinear results are still good. A possible rationale for this is that, due to the convective nature of the nonmodal mechanisms at work here, the optimal response $\cbu_1$ is located much further downstream as compared to the optimal forcing structure $\cbf_1$. This very weak spatial support between the forcing and its response implies that the inner product between the nonlinear interactions that ensue from the latter, and the former, is weak, and thus the coefficients $\mu$ and $\nu$ defined in Eqs.~(\ref{eq:coeffshmu}) and (\ref{eq:coeffshnu}) are very small. In other words, the retroaction of the nonlinearly-induced mean and harmonic fields, on $\cbu_1$, is weak, for $\cbu_1$ is triggered by a perturbation that is much further upstream. More quantitatively, the coefficients $\mu$ and $\nu$ are found to scale as $O(\ez)$ (please see Table~$1$ in Ref.~\cite{Ducimetiere22}), which decays exponentially with the Reynolds number \cite{Boujo15}. Consequently, for the nonlinear term $(\mu+\nu)A_1|A_1|^2$ to balance the linear terms $A_1$ and $\phi$ in Eq.~(\ref{eq:ampeq}), we must indeed increase $A_1$ and $\phi$ up to $O(\ez^{-1/2})$ values. We insist that this phenomenon is specific to a non-normality of convective nature, and, as we shall see, does not happen for component-wise non-normality (parallel flows) where forcing and response have a large spatial overlap. 

Figure~\ref{fig:Hgd} also depicts the evolution of the harmonic gain as the forcing amplitude is increased, but the Reynolds number is fixed to $\Ren=500$, and results corresponding to different forcing frequencies $\wz$ (considered independently) are reported. The same three approaches are compared. The amplitude equation (\ref{eq:ampeq}) yields good results for all the considered forcing frequencies, and predicts well the nonlinear attenuation of the whole harmonic gain curve over $\wz$. The good accuracy of the weakly nonlinear predictions is also valuable in the sense that they have only required a few minutes of computational time, whereas the fully nonlinear results have required a few days. 

Owing to its simplicity, the amplitude equation approach not only makes it possible to save computational resources but also furnishes a physical interpretation. Specifically, for all the considered frequencies and Reynolds numbers, the value of $|\mu|$ was found to be systematically larger than that of $|\nu|$. Thereby the nonlinear decrease of the harmonic gain can here be interpreted as being mostly due to a mean flow effect, and not related to a second harmonic effect, which corroborates the conclusions drawn in Ref.~\cite{Lugo16}.

In Fig.~\ref{fig:HgvsF} for the BFS flow, nonlinearities appear to systematically induce a decrease in the harmonic gain, as they saturate the energy level of the response. That is true for both the weakly and fully nonlinear approaches, the latter making no assumptions regarding the response structure. This saturating effect of the nonlinearities was shown in Ref.~\cite{Lugo16} (Fig.~$3$ therein) to maintain itself up to an arbitrarily large forcing amplitude, which was related to the nonlinearly-induced mean flow distortion reducing the length of the recirculation region, associated with the shortening of the shear layer. However, nonlinearities need not reduce the gain, and we have, for instance, shown in Ref.~\cite{Ducimetiere22} that the fully nonlinear harmonic gain of the two-dimensional plane Poiseuille flow, in a streamwise-periodic box, can increase or evolve non-monotonically with the forcing amplitude (please see Fig.~$10$ therein). The weakly nonlinear approach was also found to yield accurate predictions in these configurations. 

\section{Weakly nonlinear response to a stochastic forcing\label{sec:sto}}

In this second section, we subject the Navier-Stokes equations to a weak stochastic forcing. These latter equations are then reduced to a scalar one for the amplitude of the dominant component, optimal under some pre-selected gain measure, of the overall stochastic response. Again, this scalar amplitude equation incorporates leading-order nonlinearities, becoming increasingly important as the forcing intensity is increased. The general idea is the same as in the previous section, as it proceeds according to the projection of an asymptotic expansion in the subspace spanned by the dominant response. An implementation difference lies in the fact that the stochastic nature of the forcing and its response presently suggests considering a more general inner product. The latter here induces the variance norm, which includes a temporal and an ensemble average.

Let us stress that the present section, and the next one, are destined for readers specifically interested in the implementation details of the method when applied to stochastic forcing and transient growth problems, respectively. On the contrary, readers only interested in the global principle of the method may satisfy themselves with the developments presented in the previous section, and could directly jump to the conclusion of this long article for a summary.

\subsection{ Linear regime \label{sec:substo1}}

Anticipating the application case which we will consider in a moment, the base flow $\bUb$ chosen in this section is a three-dimensional plane channel flow, bounded in the $y$ (``crosswise") direction, but invariant in the $x$ (``streamwise") and $z$ (``spanwise") directions. Furthermore, the chosen external stochastic forcing will act on only one wavenumber $\al$ in the streamwise direction, and on only one wavenumber $\bet$ in the spanwise one. Specifically, the stochastic forcing reads
\begin{align}
\bff_s(x,y,z,t;\tet) = \pbf_s(y)\pxi(t;\tet)\eixz + \cc,
\label{eq:ssf}
\end{align}
where the complex-valued field $\pbf_s(y)$ is an arbitrary spatial structure, generically complex-valued. In the rest of this section, the prime accent emphasizes that the field is the spatial Fourier component associated with the $(\al,\bet)$ wavenumber pair. We have introduced the random (parametric) argument $\tet \in \Tet$, with $\Tet$ the sample space, i.e., the set of all possible elementary outcomes. The sample space $\Tet$ is associated with a given probability measure. To each outcome $\tet$, occurring with a certain probability given by the probability measure, is assigned a scalar and complex-valued function of time, $\pxi(t;\tet)$, according to a given rule. Such a particular $\pxi(t;\tet)$ constitutes a ``realization" of what is called in the following a ``stochastic process", thus defined as the collection of all realizations (one for each outcome $\tet$). Alternatively, for a fixed $t$, the scalar $\pxi(t;\tet)$ can be seen as a classical random variable (we refer to Ref.~\cite{Scott13} Chapter~$2.1$ for a more elaborate definition of a stochastic process). The stochastic process $\pxi(t;\tet)$ is typically chosen as a colored (including white) noise.

Owing to the invariance of the base flow along the $x$ and $z$ direction, the linear stochastic response $\bq$ will also oscillate only along the $(\al,\bet)$ wavenumber pair. This way, it is possible to write
\begin{align}
\bq(x,y,z,t;\tet) = \pbq(y,t;\tet)\eixz + \cc,
\label{eq:sperf}
\end{align}
and the field $\pbq(y,t;\tet)$ obeys the linear equation
\begin{align}
\bB\pa_t\pbq = \bL_{1} \pbq + \bP^T\pbf_s\pxi.
\label{eq:lsto}
\end{align}
The action of the linear operator $\bL_{n}$ over some $\pae{\bu_a,p_a}^{T}$ is computed as
\begin{align}
&\bL_{n}\ve{\bu_a}{p_a} \nonumber \\
&=\maa{-\bC_{n}(\bUb,\bu_a)-\bC_{0}(\bu_a,\bUb) +\Ren^{-1}\Delta_{n}\bu_a}{-\nab_{n} p_a}{\nab_{n} \cdot \bu_a}{0} 
\label{eq:Leab}
\end{align}
together with $\Delta_{n} = (n\al)^2 + \pa_{yy} +(n\bet)^2$, as well as $\nab_{n} = \pae{\rmi n\al,\pa_y, \rmi n\bet}^T$ and $n$ an integer. The nonlinear, bilinear operator $\bC_{n}$ was also defined as
\begin{align}
\bC_{n}(\bu_a,\bu_b) = (\bu_a \cdot \nab_{n})\bu_b,
\label{eq:cab}
\end{align}
The operator $\bC_{n}(\bu_a,\bu_b)$ computes the transport, by the velocity field $\bu_a$, of the velocity field $\bu_b$.

Note that the stochastic forcing in Eq.~(\ref{eq:ssf}), to which we subject the flow, is far from being general. That is first because, as said, it only acts over a single wavenumber pair $(\al,\bet)$. This constitutes a loss of generality in the nonlinear flow regime where the responses over different pair values cannot just be summed, but interact. Existing literature on parallel shear flows provides much evidence that such nonlinear interactions (or energy transfers) between different wavenumber pairs are an essential feature of both the transition and the self-sustaining some fully nonlinear states \cite{Pringle12, Duguet10, Cherubini11, Rigas21}. In addition, the forcing consists of a scalar noise process depending solely on time, multiplying a structure that is frozen in space. We refer to Ref.~\cite{Rolland24} (Eq.~$(3)$ therein) for an example of a more general stochastic forcing to which the flow can be subjected. 

Due to the stochastic nature of the present problem, the ensemble averaging operation over all realizations, or ``expected value", denoted $\ea{\bullet}$, is further introduced. It is important to realize that the expected value is a deterministic quantity and in this no longer depends on $\tet$. The temporal averaging operation over a finite-time signal of length $T\rightarrow\infty$, denoted $\ta{\bullet}$ with
\begin{align}
        \ta{\bullet} = \frac{1}{T}\int_{0}^{T} \bullet  \ \rmd t, 
\end{align}
is also introduced, as well as the finite-time Fourier transform and its inverse
\begin{subequations}
\begin{align}
        \hxi(\om;\tet) &= \Fo{\pxi(t;\tet)} = \frac{1}{\sqrt{T}}\int_{0}^{T}\pxi(t;\tet) e^{-\rmi \om t} \rmd t, \\
        \pxi(t;\tet) &= \Foi{\hxi(\om;\tet)} = \frac{\sqrt{T}}{2\pi}\int_{-\infty}^{\infty}\hxi(\om;\tet) e^{\rmi \om t} \rmd \om,
\end{align}
\end{subequations}
respectively. We verify that $\Foi{\Fo{\pxi}}=\pxi$ as well as $\Fo{\Foi{\hxi}}=\hxi$, if we adopt the following interpretation of the Dirac impulse $\de$
\begin{subequations}
\begin{align}
        &\int_{-\infty}^{\infty} \rme^{\rmi\om(t-b)} \rmd \om=2\pi\de(t-b), \label{eq:dirin} \\
        &\int_{0}^{T} \rme^{\rmi t(b-\om)} \rmd t=2\pi\de(b-\om), 
\end{align}
\end{subequations}
where $\de(x)$ has a diverging amplitude if $x=0$, and is null otherwise. 

Let us insist that the Fourier transform of the noise process, $\hxi(\om;\tet)$, is known \textit{a priori}. It is associated with a power spectral density computed as $\ea{|\hxi(\om;\tet)|^2}$. An important quantity associated with the noise is its overall intensity, named $\bga$, which is equal to the zero-frequency part of its power spectral density, i.e.,
\begin{align}
         \bga=\ea{|\hxi(0;\tet)|^2}.
        \label{eq:noint}
\end{align}
For instance, in case of white noise, $\bga$ corresponds to the constant value of $\ea{|\hxi(\om;\tet)|^2}$ over the frequencies.

The base flow is again assumed to be stable, such that the transients ensuing from switching on the forcing eventually fade away in the limit $t\rightarrow \infty$. In the statistically steady regime, Eq.~(\ref{eq:lsto}) can be solved in the Fourier domain for the velocity field $\hbu(y,\om;\tet) = \Fo{\pbu(y,t;\tet)}$ according to 
\begin{align}
    \hbu(y,\om;\tet) = \hxi(\om;\tet)\bR_{1}(\om)\pbf_s(y),
    \label{eq:lsst}
\end{align}
with
\begin{align}
     \bR_{n}(\om)=\bP(\rmi \om \bB - \bL_{n})^{-1}\bP^{T}.
    \label{eq:lsstr}
\end{align}
If $\bL_{1}$ is non-normal, then in line with the previous approach concerning the harmonic response, for a given stochastic process $\pxi$, it is relevant to construct an orthonormal basis of forcing structure onto which an arbitrary structure $\pbf_s$ can be projected. These structures are then ranked according to their contributions to some response-to-forcing amplification measure. As a consequence of the stochastic nature of the forcing prescribed in Eq.~(\ref{eq:lsto}), it is relevant to measure the response under the mean square norm of the velocity, which we call ``variance" in the following by abuse of language. We mathematically define the variance as being the norm $\npt{\bullet}^2$ induced by the inner product $\spt{\bullet}{\bullet}$.
The subscript ``$v$" stands for ``variance" whereas the subscript ``$\mathcal{T}$" emphasizes the inner product is applied to time-dependent fields, i.e., applies in the temporal domain (in opposition with that applying in the frequency domain, defined in a moment). Specifically, this inner product measures the resemblance between two stochastic velocity fields, say $\pbu_a(t;\tet)$ and $\pbu_b(t;\tet)$, according to 
\begin{align}
&\spt{\pbu_a(t;\tet)}{\pbu_b(t;\tet)} \nonumber \\
&= \ea{\ta{\ssp{\pbu_a(t;\tet)}{\pbu_b(t;\tet)}}} \nonumber \\
& = \ea{\frac{1}{T}\int_{0}^{T}\int_{\Omega}\pbu_a(t;\tet)^H\pbu_b(t;\tet)\rmd \Omega \rmd t}.
\label{eq:ipstocT}
\end{align}
This inner product takes the expected value of the temporal average of the $L^2$ inner product between two fields. Even if it is no longer mentioned explicitly from Eq.~(\ref{eq:ipstocT}), the only spatial coordinate over which the fields depend is the $y$ coordinate, thereby the $L^2$ inner product only integrates over the latter.

We can also define the equivalent inner product $\spf{\bullet}{\bullet}$, inducing the same variance norm but applying on the Fourier transforms (hence the subscript ``$\mathcal{F}$") as
\begin{align}
&\spf{\hbu_a(\om;\tet)}{\hbu_b(\om;\tet)} \nonumber \\
&= \ea{\frac{1}{2\pi}\iinf \ssp{\hbu_a(\om;\tet)}{\hbu_b(\om;\tet)}\rmd \om},
\label{eq:ipstocF}
\end{align}
and which is such that
\begin{align}
&\spt{\pbu_a(t;\tet)}{\pbu_b(t;\tet)} \nonumber \\
&= \ea{\frac{T}{4\pi^2}\iinf \iinf \ssp{\hbu_a(\om;\tet)}{\hbu_b(s;\tet)} \ta{e^{\rmi t(s-\om)}} \rmd \om \rmd s} \nonumber \\
        &= \ea{\frac{1}{2\pi}\iinf \iinf \ssp{\hbu_a(\om;\tet)}{\hbu_b(s;\tet)} \de(s-\om) \rmd \om \rmd s} \nonumber \\
        &= \spf{\hbu_a(\om;\tet)}{\hbu_b(\om;\tet)}.
\label{eq:ipstocFd}
\end{align}
In other words, computing $\spt{\bullet}{\bullet}$ between two fields in the temporal domain is equivalent to computing $\spf{\bullet}{\bullet}$ between their Fourier transforms. In Eq.~(\ref{eq:ipstocFd}), the interpretation of the Dirac impulse in Eq.~(\ref{eq:dirin})  was used, and the transformation from integrating over time to integrating over the frequencies amounts to using Parseval's theorem.

Optimizing over the forcing structure $\pbf_s$, the maximum ratio between the variance of the response, i.e. $\npt{\pbu}^2 = \spt{\pbu}{\pbu}$, divided by the forcing intensity, i.e. $\bga||\pbf_s||^2$, attainable by the system is defined as
\begin{align}
        G^2_{(\al,\bet)} = \max_{\pbf_s} \frac{\npt{\pbu(t;\tet)}^2}{\bga||\pbf_s||^2} = \max_{\pbf_s} \frac{\ea{\ta{||\pbu(t;\tet)||^2}}}{\bga||\pbf_s||^2}.
    \label{eq:ete}
\end{align}
This problem is similar to that in Eq.~(\ref{eq:hg1}), except that the response is here measured under an induced norm $\npt{\bullet}^2$, which is more general and more suited to the nature of the present problem, than simply taking the $L^2$ of a specific Fourier component. We insist that the variance of the response is divided by the intensity of the forcing, instead of also its variance (the latter being $\npt{\pxi}^2||\pbf_s||^2$ and not just $\bga||\pbf_s||^2$), for $\npt{\pxi}^2$ is possibly extremely large and diverges in case of a white noise (which would give a zero stochastic gain). The variance of the response, however, remains finite even in case of white noise excitation, as the large frequencies are naturally damped thanks to the presence of the temporal derivative in Eq.~(\ref{eq:lsto}).

By using Eq.~(\ref{eq:lsst}) the variance of the response, $\npt{\pbu(t;\tet)}^2 =\npf{\hbu(\om;\tet)}^2$, can be expressed as
\begin{align}
        &\npf{\hbu(\om;\tet)}^2\nonumber \\
        &= \frac{1}{2\pi}\iinf \ea{|\hxi(\om;\tet)|^2}\ssp{\pbf_s}{\bR_{1}^{\dag}(\om)\bR_{1}(\om)\pbf_s}  \rmd \om \nonumber \\
         &= \bga\ssp{\pbf_s}{\bBi\pbf_s},
        \label{eq:rvar}
\end{align}
and where the self-adjoint, strictly positive definite operator
\begin{align}
        \bBi =  \frac{1}{2\pi \bga}\iinf \ea{|\hxi(\om;\tet)|^2}\bR_{1}^{\dag}(\om)\bR_{1}(\om)  \rmd \om,
    \label{eq:ffr}
\end{align}
was defined. The superscript ``$\infty$" emphasizes the statistically steady nature of the operator. Note that if $\pxi$ is chosen as a white noise process, then $\ea{|\hxi(\om;\tet)|^2}=\bga$ for all frequencies, and the operator $\bBi$ in Eq.~(\ref{eq:ffr}) is equal to that of the same name derived in Refs.~\cite{Farrell93, Farrell96}. Eventually, by using Eq.~(\ref{eq:rvar}), the stochastic gain associated with the pair $(\al,\bet)$ becomes  
\begin{align}
        G^2_{(\al,\bet)} = \max_{\pbf_s} \frac{\ssp{\pbf_s}{\bBi\pbf_s}}{\ssp{\pbf_s}{\pbf_s}} = \frac{1}{\ez^2}, 
    \label{eq:ete2}
\end{align}
where we have also defined $\ez=1/G_{(\al,\bet)}$ as its inverse. 

The family of optimal and sub-optimal stochastic forcing structures, generically complex-valued, are then found as the eigenmodes $\set{\cbf_j}_{j\geq 1}$ of the operator $\bBi$. Each $\cbf_j$ is normalized as $||\cbf_j||=1$ and associated with a real, strictly positive eigenvalue $G^2_j$, such that 
\begin{align}
        \bBi\cbf_j = G^2_j\cbf_j, \quad j=1,2,...
\end{align}
The eigenvalues $\set{G^2_j}_{j\geq 1}$ are again sorted by decreasing value, and the stochastic forcing structure $\cbf_1$ leads to the largest gain $G_1 = G_{(\al,\bet)}$. The remaining modes $\set{\cbf_j}_{j\geq 2}$ constitute sub-optimal stochastic forcing structures. Ensuing from the fact that $\bBi$ is a self-adjoint operator, the $\set{\cbf_j}_{j\geq 1}$ family is orthonormal under the $L^2$ inner product, i.e. $\ssp{\cbf_j}{\cbf_k}=\de_{ij}$ (the Kronecker delta symbol). In what follows, without loss of generality let us consider a noise process with a unit intensity, i.e., $\bga=1$.

We further define $\cbu_j(\om;\tet)$ as the normalized linear response, in the Fourier domain, to the forcing $\hxi(\om;\tet)\cbf_j$, according to  
\begin{align}
        \cbu_j(\om;\tet) = G_j^{-1}\hxi(\om;\tet)\bR_{1}(\om)\cbf_j, \quad j\geq 1,
        \label{eq:stn}
\end{align}
and its equivalent $\bbu_j(t;\tet)$ in the temporal domain
\begin{align}
       \bbu_j(t;\tet)=\Foi{\cbu_j(\om;\tet)}.
        \label{eq:stn2}
\end{align}
This for instance makes $\cbu_1(\om;\tet)$ the normalized response to the optimal stochastic forcing structure (also called ``optimal response" in what follows by abuse of language), as well as $\set{\cbu_j(\om;\tet)}_{j\geq2}$ the normalized responses to the sub-optimal ones. The prefactor $G_j^{-1}$ in Eq.~(\ref{eq:stn}) ensures that each $\cbu_j(\om;\tet)$ possesses a unit variance. Indeed, computing the inner product $\spf{\bullet}{\bullet}$ between two responses $\cbu_j(\om;\tet)$ and $\cbu_k(\om;\tet)$ leads to
\begin{align}
        &\spf{\cbu_j}{\cbu_k} \nonumber \\
        &= \frac{G_j^{-1}G_k^{-1}}{2\pi}\iinf \ea{|\hxi(\om;\tet)|^2}\ssp{\cbf_j}{\bR_{1}(\om)^{\dag}\bR_{1}(\om)\cbf_k}\rmd \om \nonumber \\
        &= G_j^{-1}G_k^{-1}\ssp{\cbf_j}{\bBi\cbf_k} \nonumber \\
        &= G_j^{-1}G_k\ssp{\cbf_j}{\cbf_k} \nonumber \\
        &= \delta_{jk},
        \label{eq:orths}
\end{align}
which is equal to one if $j=k$ and zero otherwise (and where we have used the definition in Eq.~(\ref{eq:ffr}) with $\bga=1$). Thereby, the $\set{\cbu_j(\om;\tet)}_{j\geq 1}$ family of responses is orthonormal under the inner product $\spf{\bullet}{\bullet}$ (although it generically is not under the $L^2$ inner product). By definition, this implies the temporal equivalent $\set{\bbu_j(t;\tet)}_{j\geq 1}$ to be orthonormal under the inner product $\spt{\bullet}{\bullet}$ .

Decomposing the forcing structure $\pbf_s$ in the $\set{\cbf_j}_{j \geq 1}$ family, then applying $\hxi(\om;\tet)\bR_{1}(\om)$, leads to the rewriting of the linear stochastic response in Eq.~(\ref{eq:lsst}) according to
\begin{align}
    \hbu(\om;\tet) &= \sum_{j\geq 1}\ssp{\cbf_j}{\pbf_s}\hxi(\om;\tet)\bR_{1}(\om)\cbf_j \nonumber \\
    &= \sum_{j\geq 1}\ssp{\cbf_j}{\pbf_s}G_j\cbu_j(\om;\tet).
    \label{eq:lsst2}
\end{align}

From now on, we assume the optimal gain $G_1=1/\ez$ to be large and thereby $\ez$, its inverse, to be very small. Furthermore, $G_1$ is also assumed to be much larger than the sub-optimal ones, i.e. $G_1 \gg G_2 > G_3 > ...$. More quantitatively, the same specific scaling as in Eq.~(\ref{eq:scg}) is assumed, which we recall to be 
\begin{align}
& \set{G_1,G_2,...,G_N,G_{N+1},...} \nonumber \\
& = \set{\frac{1}{\ez},\frac{\ga_2}{\ez^{1/2}},...,\frac{\ga_N}{\ez^{1/2}},\ga_{N+1},...}, \quad \ga_j = O(1), \quad j\geq 2. 
\label{eq:scgbis}
\end{align}
The scalings in Eq.~(\ref{eq:scgbis}), together with the expression in Eq.~(\ref{eq:lsst2}), implies the linear stochastic response to rewrite
\begin{align}
    \hbu(\om;\tet) =& \frac{1}{\ez}\Biggl[ \ssp{\cbf_1}{\pbf_s}\cbu_1(\om;\tet) + \ez^{1/2}\sum_{j=2}^{N}\ssp{\cbf_j}{\pbf_s}\cbu_j(\om;\tet) \nonumber \\ 
    & + \ez\sum_{j>N}\ssp{\cbf_j}{\pbf_s}\cbu_j(\om;\tet) \Biggl].
    \label{eq:lsst3}
\end{align}
Equation (\ref{eq:lsst3}) indicates that the stochastic response to an arbitrary structure $\pbf_s$, is dominated by the response $\cbu_1(\om;\tet)$ to the optimal forcing structure $\cbf_1$, complemented by a smaller rest. This rest is contained in the orthogonal subspace spanned by the responses to the sub-optimal forcing structures $\set{\cbu_j(\om;\tet)}_{j\geq 2}$. It is in this sense in perfect analogy with Eq.~(\ref{eq:harr2}) for the harmonic response, again at the difference that the orthogonality property here refers to the inner product $\spf{\bullet}{\bullet}$ (or $\spt{\bullet}{\bullet}$ in the temporal domain) and not simply the $L^2$ one. 

\subsection{ Weakly nonlinear continuation \label{sec:substo2}}

Referring again to Fig.~\ref{fig:sketch} but keeping in mind this change in the inner product, we now describe the stochastic response in a weakly nonlinear regime. The fully nonlinear Navier-Stokes equations, subjected to a small stochastic forcing under the form given in Eq.~(\ref{eq:ssf}), read
\begin{align}
\bB \pa_t \ve{\bU}{P} =& \ve{-(\bU \cdot \nab)\bU + \Ren^{-1}\Delta \bU - \nab P}{\nab \cdot \bU} \nonumber \\
&+ \ez^{3/2}\ve{\phi\pbf_s\pxi \eixz +\cc}{0}.
\label{eq:NSf2}
\end{align}
Let us normalize $\pbf_s$ according to $||\pbf_s||^2=1$, such that the forcing intensity in Eq.~(\ref{eq:NSf2}) directly is $\phi \ez^{3/2}$ (please remember that we have chosen $\bga=1$). In other terms, the free scalar parameter $\phi = O(1)$ appears as the intensity of the stochastic forcing when rescaled by $\ez^{3/2}$. We insist that the prescribed stochastic forcing can contain infinitely many temporal frequencies but only acts on the $(\al, \bet)$-wavenumber pair in space. This assumption is made to lighten the calculations, as it precludes the apparition of terms corresponding to the interaction between different wavenumber pairs. However, the proposed method would also apply to more involved forcing terms.

We again seek a solution $\pae{\bU,P}^T(t;\tet)$ in the continuity of the linear one in Eq.~(\ref{eq:lsst3}), implying, for the fundamental component, order separations in terms of powers of $\ez^{1/2}$ between the leading contribution of $\cbu_1(\om;\tet)$, and these of the sub-optimal responses. Practically, in the statistically steady regime, $\pae{\bU,P}^T(t;\tet)$ is sought under the form of an asymptotic expansion in terms of integer powers of $\ez^{1/2}$ according to 
\begin{widetext}
\begin{align}
\ve{\bU(t;\tet)}{P(t;\tet)} =& \ve{\bUb}{\pb}  + \ez^{1/2}\sae{A_1\ve{\bbu_1(t;\tet)}{\bp_1(t;\tet)}\eixz + \cc} + \ez\sae{\sum_{j=2}^{N}A_j\ve{\bbu_j(t;\tet)}{\bp_j(t;\tet)}\eixz + \cc + \ve{\obu_2(t;\tet)}{\op_2(t;\tet)}} \nonumber \\
& + \ez^{3/2}\sae{\sum_{j>N}A_j\ve{\bbu_j(t;\tet)}{\bp_j(t;\tet)}\eixz + \cc + \ve{\obu_3(t;\tet)}{\op_3(t;\tet)}} + \underbrace{\sum_{m\geq 4} \ez^{m/2}\ve{\obu_m(t;\tet)}{\op_m(t;\tet)}}_{=O\pae{\ez^{2}}}.
\label{eq:asy1s}
\end{align}
\end{widetext}
The complex-valued, deterministic, constant scalar amplitudes $\set{A_j}_{j \geq 1}$, where $A_j=O(1)$ for all $j\geq 1$, are the unknowns of interest. In Eq.~(\ref{eq:asy1s}) we have introduced the nonlinearly-induced small (they are multiplied by $\ez^{m/2}$) corrections terms $\obu_m(t;\tet)$ for $m \geq 2$. 
Each is described as a Fourier series in space 
\begin{align}
\obu_m(t;\tet) =& \obu_{m,0}(t;\tet)  \nonumber \\
& + \pae{\sum_{ n \geq 1}\obu_{m,n}(t;\tet)\eixzm{n} + \cc},
\label{eq:fs2}
\end{align}
for $m\geq 2$ which, unlike in Eq.~(\ref{eq:fs}), this time must include a contribution from the fundamental wavenumber pair $(\al,\bet)$ at $n=1$. That is because the family $\set{\bbu_j(t;\tet)}_{j\geq 1}$ contains as many members as the number of spatial degrees of freedom. Thereby, a linear combination of the $\set{\bbu_j(t;\tet)}_{j\geq 1}$, weighed by the amplitudes $\set{A_j}_{j\geq 1}$ that depends neither on time nor on the stochastic argument, cannot describe any kind of temporal and stochastic variations. By construction, only is such a decomposition complete when describing the response to a specific forcing of the form $\pxi(t;\tet)$ times an arbitrary spatial structure (as expressed in Eq.~(\ref{eq:lsst2})). However, the higher-order nonlinear interactions forcing terms have no reason to be written under this specific form. Thereby, the description of the fundamental field must be completed at a nonlinear level, to account for additional sorts of stochastic and temporal variations that do not belong to the span of $\set{\bbu_j(t;\tet)}_{j\geq 1}$ (by ``span" we mean all linear combinations with constant coefficients). From Eq.~(\ref{eq:fs2}), the component of $\bU(t;\tet)$ in Eq.~(\ref{eq:asy1s}) which oscillates in space along $(\al,\bet)$, say $\pbU(t;\tet)$, is then
\begin{align}
\pbU(t;\tet) =& \ez^{1/2}A_1\bbu_1(t;\tet) \nonumber \\
&+ \ez\pae{\obu_{2,1}(t;\tet) + \sum_{j=2}^{N}A_j\bbu_j(t;\tet)} \nonumber \\
&+ \ez^{3/2}\pae{\obu_{3,1}(t;\tet) + \sum_{j>N}A_j\bbu_j(t;\tet)} \nonumber \\ 
&+ O\pae{\ez^{2}}.\label{eq:sb1}
\end{align}
Nevertheless, in order to avoid redundancy in the description of $\pbU(t;\tet)$ in Eq.~(\ref{eq:sb1}), for each $m\geq 2$ we exclude $\obu_{m,1}(t;\tet)$ from the subspace spanned by the $\set{\bbu_j(t;\tet)}_{j\geq 1}$ family. Specifically, for each $m\geq 2$, we impose  $\obu_{m,1}(t;\tet)$ to be orthogonal to all the members of the $\set{\bbu_j(t;\tet)}_{j\geq 1}$ family as
\begin{align}
\spt{\bbu_j(t;\tet)}{\obu_{m,1}(t;\tet)}=0, \quad j\geq 1, m\geq 2
\label{eq:nrc}
\end{align}
and rename $\obu_{m,1}=\obup_{m,1}$ to emphasize this important orthogonality property. This way, the part of the weakly nonlinear response $\pbU(t;\tet)$ which is spanned by the $\set{\cbu_j}_{j\geq 1}$ family (complete at linear level), can be reconstituted solely from the knowledge of the amplitudes $\set{A_j}_{j\geq 1}$. The $\obup_{m,1}$ contributions, for all $m \geq 2$, only describe a nonlinearly-induced orthogonal part. Consequently, as in the previous section, the weakly nonlinear expansion in Eq.~(\ref{eq:sb1}) is here too constructed in such a way as to be consistent with its linear counterpart in Eq.~(\ref{eq:lsst3}). This implies that, in the linear regime, the fields $\obup_{m,1}$ must vanish (not only their projections) and that each amplitude $A_j$ must reduce to $\phi \ga_j\ssp{\cbf_j}{\pbf_s}$ (with $\ga_1=1$) for all $j\geq 1$. Additionally, the amplitudes $\set{A_j}_{j \geq 1}$ incorporate leading-order nonlinear effects, which progressively come into play as $\phi$ is increased towards order one values, making the $\set{A_j}_{j \geq 1}$ typically depart from their linear values.

Each amplitude $A_k$ can be directly extracted by projecting Eq.~(\ref{eq:sb1}) onto $\bbu_k(t;\tet)$ with the inner product $\spt{\bullet}{\bullet}$. Indeed, 
\begin{align}
\spt{\bbu_k}{\pbU} =& \ez^{1/2}A_1\spt{\bbu_k}{\bbu_1} + \ez\sum_{j=2}^{N}A_j\spt{\bbu_k}{\bbu_j}
\nonumber \\ 
& + \ez^{3/2}\sum_{j>N}A_j\spt{\bbu_k}{\bbu_j},
\label{eq:sb2}
\end{align}
(which does not contain $O(\ez^2)$ terms). Relying on the orthonormality property of the $\set{\bbu_j(t;\tet)}_{j \geq 1}$ family, $A_k$ is the only remaining amplitude in Eq.~(\ref{eq:sb2}). Importantly, that is also why the $\set{A_j}_{j \geq 1}$ are chosen as being deterministic and constant: without this, they could not go out of the inner product in Eq.~(\ref{eq:sb2}) and thus could not bear this specific, immediate meaning.    

In Eq.~(\ref{eq:asy1s}) the pressure field $\bp_j$, associated with the response $\bbu_j$, is such that, in the Fourier domain
\begin{align}
\ve{\cbu_j(\om;\tet)}{\ccp_j(\om;\tet)} =  G^{-1}_j\hxi(\om;\tet)\pae{\rmi \om\bB-\bL_{1}}^{-1} \bP^{T}\cbf_j, \quad j \geq 1.
\label{eq:dep}
\end{align}

Inserting Eqs.~(\ref{eq:asy1s}) and (\ref{eq:fs2}) in the Navier-Stokes equations (\ref{eq:NSf2}), leads to an expansion for each of the spatial Fourier component. The one for the component oscillating at $(0,0)$ reads
\begin{align}
\bz =& \ez \Biggl[-(\bB\pa_t-\bL_{0})\ve{\obu_{2,0}}{\op_{2,0}} \nonumber\\
&- |A_1|^2\bP^{T}\pae{\bC_{1}(\bbu^*_1,\bbu_1) +\cc}\Biggl]+ O(\ez^{3/2}).
\label{eq:aFs0}
\end{align}
As for it, the component oscillating at $(2\al,2\bet)$ is governed by
\begin{align}
\bz =&  \ez \sae{-\pae{\bB\pa_t-\bL_{2}}\ve{\obu_{2,2}}{\op_{2,2}} - A_1^2\bP^{T}\bC_{1}(\bbu_1,\bbu_1)} \nonumber\\
&+ O(\ez^{3/2}).
\label{eq:aFs2}
\end{align}
Collecting terms at $O(\ez)$ in Eqs.~(\ref{eq:aFs0}) and (\ref{eq:aFs2}) leads to equations whose solutions are expressed in the Fourier domain (in time) as
\begin{align}
&\hobu_{2,0}(\om;\tet) = |A_1|^2\hobu_{2,0}^{|A_1|^2}(\om;\tet), \nonumber \\
&\hobu_{2,0}^{|A_1|^2}(\om;\tet) = -\bR_{0}(\om)\Fo{\bC_{1}(\bbu^*_1(t;\tet),\bbu_1(t;\tet)) + \cc }, \label{eq:so20s}
\end{align}
for the nonlinearly induced (spatial) mean flow distortion, as well as
\begin{align}
&\hobu_{2,2}(\om;\tet) = A_1^2\hobu_{2,2}^{A_1^2}(\om;\tet), \nonumber \\
&\hobu_{2,2}^{A_1^2}(\om;\tet) = -\bR_{2}(\om)\Fo{\bC_{1}(\bbu_1(t;\tet),\bbu_1(t;\tet)) }. \label{eq:so22s}
\end{align}
for the nonlinearly induced second harmonic field.
\begin{widetext}
The expansion for the fundamental component oscillating at $(\al,\bet)$ is
\begin{align}
 \bz =& \ez^{1/2}\sae{-A_1 \pae{\bB\pa_t-\bL_{1}} \ve{\bbu_1}{\bp_1}} +\ez \sae{-\pae{\bB\pa_t-\bL_{1}} \ve{\obup_{2,1}}{\opp_{2,1}} -\sum_{j=2}^{N}A_j \pae{\bB\pa_t-\bL_{1}} \ve{\bbu_j}{\bp_j} }\nonumber  \\
& +\ez^{3/2} \Biggr[-\pae{\bB\pa_t-\bL_{1}} \ve{\obup_{3,1}}{\opp_{3,1}} -\sum_{j > N}A_j \pae{\bB\pa_t-\bL_{1}} \ve{\bbu_j}{\bp_j} - A_1\bP^{T}\Big(\bC_{0}(\bbu_1,\obu_{2,0})  + \bC_{1}(\obu_{2,0},\bbu_1) \Big)
 \nonumber \\
& - A^*_1\bP^{T}\Big(\bC_{2}(\bbu_1^*,\obu_{2,2}) + \bC_{-1}(\obu_{2,2},\bbu_1^*)\Big) + \phi \bP^{T} \pxi\pbf_s \Biggr] + O(\ez^2).
\label{eq:aFs1}
\end{align}
Eq.~(\ref{eq:aFs1}) is sent in the Fourier domain (in time) and simplified by using the definition in Eq.~(\ref{eq:dep}). This leads to 
\begin{align}
\bz =& \ez^{1/2}\sae{-A_1 G_1^{-1}\hxi \bP^{T}\cbf_1} +\ez \sae{-\bP^T\obfp_{2,1}-\sum_{j=2}^{N}A_j G_j^{-1}\hxi \bP^{T}\cbf_j} \nonumber  \\
& +\ez^{3/2} \Biggr[-\bP^T\obfp_{3,1} -\sum_{j>N}A_j G_j^{-1}\hxi \bP^{T}\cbf_j - A_1|A_1|^2\bP^{T}\fnl + \phi \bP^{T} \hxi\pbf_s \Biggr] + O(\ez^2),
\label{eq:aFs12}
\end{align}
where, by using Eq.~(\ref{eq:so20s}) and Eq.~(\ref{eq:so22s}), we have introduced the forcing term $\fnl(\om;\tet)$ defined as
\begin{align}
\fnl(\om;\tet) &= \fnlm(\om;\tet) + \fnlh(\om;\tet),
\label{eq:defnl}
\end{align}
with
\begin{subequations}
\begin{align}
\fnlm(\om;\tet) &= \Fo{\bC_{0}(\bbu_1(t;\tet),\obu^{|A_1|^2}_{2,0}(t;\tet))  + \bC_{1}(\obu^{|A_1|^2}_{2,0}(t;\tet),\bbu_1(t;\tet))}, \label{eq:defnl0} \\
\fnlh(\om;\tet) &= \Fo{\bC_{2}(\bbu_1^*(t;\tet),\obu^{A_1^2}_{2,2}(t;\tet)) + \bC_{-1}(\obu^{A_1^2}_{2,2}(t;\tet),\bbu_1^*(t;\tet))}.
\label{eq:defnl2}
\end{align}
\end{subequations}
The forcing $\fnlm$ is induced by nonlinear interactions only, hence the subscript ``$nl$". More precisely, the forcing $\fnlm(\om;\tet)$ (resp. $\fnlh(\om;\tet)$) is the nonlinear interaction between the mean flow distortion (resp. the second harmonic) and the leading-order response. Between Eq.~(\ref{eq:aFs1}) and Eq.~(\ref{eq:aFs12}), we have also defined the forcing terms $\obfp_{m,1}(\om;\tet)$ according to $(\hobu^{\perp}_{m,1},\hat{\op}^{\perp}_{m,1})
^T=(\rmi \om \bB-\bL_{1})^{-1}\bP^T\obfp_{m,1}$ for $m\geq 2$. The latter expression is without loss of generality despite $\obfp_{m,1}$ forcing the momentum equations only, since the velocity field $\hobu^{\perp}_{m,1}$ must be divergence-free. Eventually, by using the scalings of the gains given in Eq.~(\ref{eq:scgbis}) and applying $\bP$, it is possible to transform Eq.~(\ref{eq:aFs12}) into
\begin{align}
& \bz = \ez \sae{-\obfp_{2,1}(\om;\tet)} +  \ez^{3/2}\sae{ -\obfp_{3,1}(\om;\tet)-A_1\hxi(\om;\tet) \cbf_1 - \sum_{j\geq 2}\ga^{-1}_jA_j \hxi(\om;\tet) \cbf_j - A_1|A_1|^2\fnl(\om;\tet) + \phi \hxi(\om;\tet)\pbf_s}.
\label{eq:aFs123}
\end{align}
\end{widetext} 
Once again, the terms cannot be collected at each order in Eq.~(\ref{eq:aFs123}), for it would over-determinate the values of the $\set{A_j}_{j\geq 1}$. This is true despite the presence of the unknown $\obfp_{m,1}$ terms, with $m\geq 2$, precisely because these latter are contained in a subspace orthogonal to that spanned by the $\set{\hxi\cbf_j}_{j\geq 1}$. The scalar equation for the amplitude $A_j$ can be freed from contributions in $\set{\hxi\cbf_k}_{k\geq 1, k\neq j}$ and in $\obfp_{m,1}$, by taking the inner product $\spf{\bullet}{\bullet}$ of Eq.~(\ref{eq:aFs123}) with the $j$th ``adjoint" stochastic response $\cbu^{\dag}_j(\om;\tet)$, defined as 
\begin{align}
\cbu^{\dag}_j(\om;\tet) = G_j^{-1}\bR^{\dag}_{1}(\om)\cbu_j(\om;\tet), \quad j\geq 1.
\label{eq:ars}
\end{align}
Taking the $\spf{\bullet}{\bullet}$ inner product of Eq.~(\ref{eq:aFs123})  with $\cbu^{\dag}_j(\om;\tet)$, amounts to pre-multiplying Eq.~(\ref{eq:aFs123}) by $\bR_{1}(\om)$ and then taking the inner product with $G_j^{-1}\cbu_j(\om;\tet)$. The prefactor $G_j^{-1}$ in Eq.~(\ref{eq:ars}) ensures the convenient normalization 
\begin{align}
\spf{\cbu^{\dag}_j}{\hxi\cbf_k} &= G_j^{-1}\spf{\cbu_j}{\hxi\bR_{1}(\om)\cbf_k} \nonumber \\
&= G_j^{-1}G_k\spf{\cbu_j}{\cbu_k}\nonumber \\
&=\de_{jk},
\label{eq:arsn}
\end{align}
which is equal to one if $j=k$. 
In particular, projecting Eq.~(\ref{eq:aFs123}) over $\cbu_1^{\dag}(\om;\tet)$ leads to an amplitude equation for $A_1$, given by 
\begin{align}
\boxed{0 = -A_1 - (\mu + \nu)A_1|A_1|^2 + \phi \ssp{\cbf_1}{\pbf_s}} + O(\ez^{1/2}), 
\label{eq:astoc}
\end{align}
with
\begin{subequations}
\begin{align}
\mu &= \spf{\cbu^{\dag}_1(\om;\tet)}{\fnlm(\om;\tet)} \nonumber \\
&= \frac{\ez}{2\pi} \ea{\iinf \ssp{\cbu_1(\om;\tet)}{\bR_{1}(\om)\fnlm(\om;\tet)} \rmd \om}, \label{eq:cofmus} \\
\nu &= \spf{\cbu^{\dag}_1(\om;\tet)}{\fnlh(\om;\tet)} \nonumber \\
&= \frac{\ez}{2\pi} \ea{\iinf \ssp{\cbu_1(\om;\tet)}{\bR_{1}(\om)\fnlh(\om;\tet)} \rmd \om}. \label{eq:cofnus}
\end{align}
\end{subequations}
In deriving Eq.~(\ref{eq:astoc}), we have used Eq.~(\ref{eq:arsn}) together with the fact that $\spf{\cbu_j}{\cbu_k}=0$ if $j\neq k$. In addition,
\begin{align}
\spf{\cbu^{\dag}_j}{\hxi\pbf_s} &= G_j^{-1}\spf{\cbu_j}{\hxi\bR_{1}(\om)\pbf_s} \nonumber \\
& = G_j^{-1}\spf{\cbu_j}{\sum_{k\geq 1}\ssp{\cbf_k}{\pbf_s}G_k\cbu_k} \nonumber \\
& = \ssp{\cbf_j}{\pbf_s}, \quad  j\geq 1.
\end{align}
The terms in $\obfp_{m,1}$ do not contribute to any of the equations for the $\set{A_j}_{j \geq 1}$ since
\begin{align}
\spf{\cbu^{\dag}_j}{\obfp_{m,1}} &= G_j^{-1}\spf{\cbu_j}{\hobu^{\perp}_{m,1}} = 0,
\end{align}
for every $m\geq 2$ by the orthogonality property expressed in Eq.~(\ref{eq:nrc}). 

Equation (\ref{eq:astoc}), for the amplitude of the response to the optimal stochastic forcing structure, is found to possess a similar form as that for the amplitude of the optimal harmonic response in Eq.~(\ref{eq:ampeq}). In particular, in both cases, the leading-order nonlinear term is $\propto A_1|A_1|^2$. It is a consequence of the fact that the excitation is periodic, in space along the wavenumber pair $(\al,\bet)$ for the present section, and in time in the previous section. Indeed, the quadratic terms $A_1^2$ and $|A_1|^2$ cannot oscillate along the fundamental wavenumber or frequency and, thereby cannot consistently appear in the amplitude equations. However, the nonlinear coefficients in Eqs.~(\ref{eq:cofmus})-(\ref{eq:cofnus}) are more computationally costly to evaluate than that in Eq.~(\ref{eq:ampeq}), for they also involve an integration over the frequencies and an ensemble average.

In the linear limit where the forcing intensity vanishes $\phi \rightarrow 0$ (while keeping $\ez$ fixed), Eq.~(\ref{eq:astoc}) reduces to $A_1 = \phi \ssp{\cbf_1}{\pbf_s}$, which is the expected linear solution. The nonlinear term $(\mu+\nu)A_1|A_1|^2$ captures leading order nonlinear effects on $A_1$, making it depart from its linear solution as $\phi$ is increased. The term $\mu A_1|A_1|^2$ quantifies how the nonlinear spatial mean flow distortion, $|A_1|^2\obu_{2,0}^{|A_1|^2}(t;\tet)$, interacts with the stochastic response $A_1\bbu_1(t;\tet)$, to, on average, result in a modification of its amplitude $A_1$. The term $\nu A_1|A_1|^2$, for its part, embeds how the second harmonic field, $A_1^2\obu_{2,2}^{A_1^2}(t;\tet)$, interacts with $A^*_1\bbu^*_1(t;\tet)$, to, on average, also bring about a modification of the response amplitude. 

Similarly, for each $A_j$ with $j \geq 2$, the corresponding leading-order equation can be obtained  by projecting Eq.~(\ref{eq:aFs123}) onto $\cbu_j^{\dag}(\om;\tet)$. As for the harmonic forcing case, these equations are linear at leading order and slaved to $A_1$. This is a result of the assumed large gap between $G_1$ and the sub-optimal gains, which led to introducing $A_1$ at dominant order in the asymptotic expansion in Eq.~(\ref{eq:asy1s}), and the sub-optimal amplitudes only at the next orders. These equations for the $\set{A_j}_{j \geq 2}$ are not written down, since only $A_1$ is necessary for determining the leading-order stochastic gain.

Indeed, the weakly nonlinear stochastic response in Eq.~(\ref{eq:sb1}) is associated with the gain
\begin{align}
&G_{(\al,\bet)} \nonumber \\
&= \frac{\npt{\pbU(t;\tet)}}{\ez^{3/2}\phi ||\pbf_s||} \nonumber \\
&= \frac{\pae{ \ez|A_1|^2 + \ez^2\sum_{j=2}^{N}|A_j|^2 + \ez^2\npt{\obup_{2,1}}^2 + O(\ez^3) }^{1/2} }{\ez^{3/2}\phi} \nonumber \\
&= \frac{|A_1|}{\ez\phi}\pae{1 + O(\ez)},
\label{eq:gsw}
\end{align}
where the orthonormality of the $\set{\bbu_j(t;\tet)}_{j \geq 1}$ basis under the inner product $\spt{\bullet}{\bullet}$ was invoked, together with Eq.~(\ref{eq:nrc}).  

Eventually, let us insist on the fact that the expansion in Eq.~(\ref{eq:asy1s}) is a particular form under which to seek the weakly nonlinear response, but another could have been chosen, perhaps leading to a more accurate description. For instance, as already mentioned, the $\set{A_j}_{j \geq 1}$ could represent amplitudes according to some other basis. But even in keeping the same basis, the $\set{A_j}_{j \geq 1}$ could also have been chosen depending on the frequency and/or the stochastic argument. This would have considerably complicated the calculations. However, these $\set{A_j(\om;\tet)}_{j \geq 1}$ would have carried more information than the constant, deterministic one considered in the present section. In particular, the $A_1(\om;\tet)$ would not have given only the weakly nonlinear evolution of the scalar measure $G_{(\al,\bet)}$, but that of the whole Fourier spectrum of the leading-order response according to $A_1(\om;\tet)\cbu_1(\om;\tet)$. Furthermore, a linear combination of the $\set{\cbu_j(\om;\tet)}_{j\geq1}$ with amplitudes depending on both the frequency and the stochastic argument would have made it possible to describe any kind of stochastic and spatio-temporal variations, without the need to add the complementary terms $\obup_{m,1}$. In this sense, the approach would have shared some similarities with that proposed in Ref.~\cite{Sapsis09}. A difference is that, since we restrict ourselves to weakly nonlinear regimes, the basis elements and the hierarchization of their respective contributions are determined \textit{a priori} from purely linear considerations.

Deriving an equation for such $A_1(\om;\tet)$ was performed in Ref.~\cite{Ducimetiere22b}, and indeed found to give excellent results at the cost of heavier calculations. The method used in Ref.~\cite{Ducimetiere22b} was different from that presented here, and relied on an operator perturbation technique and the introduction of multiple time scales. However, we believe that making the amplitudes in Eq.~(\ref{eq:asy1s}) depend on the frequency (or time) and the stochastic argument, then proceeding with the projection technique proposed here, would lead to the same equation as in Ref.~\cite{Ducimetiere22b}. 

\subsection{Application case: the Couette-Poiseuille flow \label{sec:substo3}}

We now test the performance of the amplitude equation (\ref{eq:astoc}) on the Couette-Poiseuille flow. The latter was numerically shown in Refs.~\cite{Bergstrom04, Shuai22} to support substantial nonmodal linear responses, which we understand from Ref.~\cite{Trefethen93} can interact with nonlinear mechanisms to bring about a transition to a turbulent state despite linear stability. Such subcritical transition to turbulence was found to occur in the laboratory experiment reported in Refs.~\cite{Klotz17, Klotz21, Liup21}. The subcritical nature of the Couette-Poiseuille flow makes it a challenging application case, which is also the reason why it is here selected. 

The Couette-Poiseuille flow, varying over the $y \in[-1;1]$ coordinate within a wall moving at velocity $1$ at $y=1$, and a fixed wall (i.e., moving at velocity $0$) at $y=-1$, reads
\begin{align}
\bUb(y) = \pae{\frac{3}{4}(y^2-1)+\frac{1}{2}(y-1)+1,0,0}^T.
\label{eq:CPflow}
\end{align}
This flow is purely along the streamwise direction and is invariant along the latter, as well as along the spanwise one. 

The stochastic process $\pxi(t;\tet)$ is chosen as being a band-limited ``white" noise, Gaussian distributed, and with zero average. It is characterized in the Fourier domain according to
\begin{align}
|\hxi(\om;\tet)|^2 = 1 \ \mbox{for} \ |\om|\leq \om_c,
\label{eq:appstoc}
\end{align}
and $\hxi(\om;\tet)=0$ otherwise. The symbol $\om_c$ denotes the band-limiting cut-off frequency (hence the subscript ``$c$"). We set the band-limiting frequency to $\om_c=\pi$, sufficiently large to encompass the dominant dynamics, and yet sufficiently small to make it possible to choose a large time step in the numerical simulations (the latter having to be smaller or equal to $\pi/\om_c$). 
\begin{figure}
\centering
\includegraphics[trim={0.0cm 0.0cm 0.0cm 0.0cm},clip,width=0.975\linewidth]{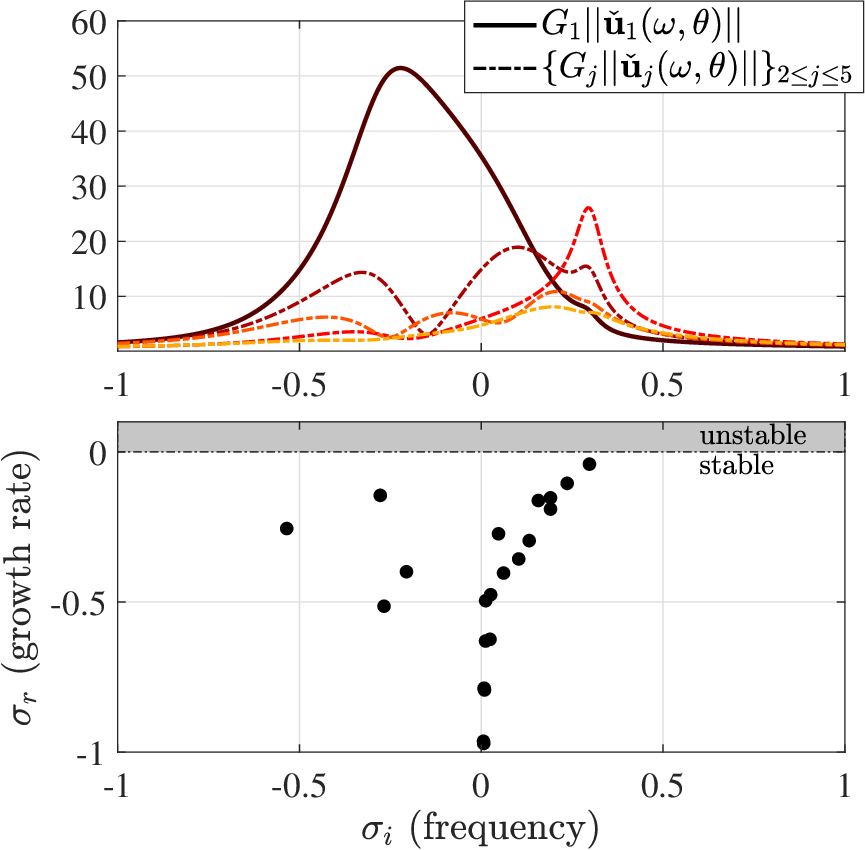}
\caption{For the Couette-Poiseuille flow with $(\Ren,\al,\bet)=(300,1,1)$. Top: $L^2$ norm of the unnormalized responses $\set{G_j\cbu_j(\om,\tet)}_{5\geq j \geq 1}$ to the optimal (full line) and first four sub-optimal (dashed-dotted line, lighter nuance lower in the hierarchy) stochastic forcing structures, respectively, with the noise specified in Eq.~(\ref{eq:appstoc}). Bottom: spectrum of this same flow in the complex plane (each dot an eigenvalue). The frequency ranges on the $x$-axis are the same for both the top and the bottom frames. }
\label{fig:Srr2}
\end{figure}
In practice, for each considered $\om$, the corresponding $\hxi(\om;\tet)$ is a complex number with a unit module but a phase randomly drawn according to a uniform law between $0$ and $2\pi$. The noise defined in Eq.~(\ref{eq:appstoc}) indeed possesses a unit intensity $\bga=1$. Associated with such $\hxi(\om;\tet)$, and for the parameters $(\Ren,\al,\bet)=(300,1,1)$, we illustrate in Fig.~\ref{fig:Srr2} the $L^2$ norm of the unnormalized response to the optimal stochastic forcing, i.e., we show $||G_1\cbu_1(\om,\tet)||=||\bR_{1}(\om)\cbf_1|||\hxi(\om;\tet)|=||\bR_{1}(\om)\cbf_1||$ for $|\om|\leq \om_c$. The unnormalized responses to the first four sub-optimal forcing structures are also shown. The flow responds mostly to frequencies $\om\in[-1,0.5]$, suggesting that, at least for these parameters, $\om_c=\pi$ is a sufficiently large cut-off frequency indeed. It can be increased to arbitrarily large values without this to change the results presented below, and the introduction of a cut-off frequency is unnecessary since the noise is measured by its intensity and not by its variance.

We note that the noise defined in Eq.~(\ref{eq:appstoc}) is very specific and is, for instance, a poor turbulence model. 
Nevertheless, this choice can be justified physically by the fact that the system is by assumption strongly non-normal, and thus the structure of the responses will be inherent to the operator in itself. Consequently, the stochastic excitation need not be carefully crafted for the conclusions to remain general. However, it is perhaps more reasonable to justify this choice by recalling that our interests lie in the methodological aspects more than in doing a detailed physical analysis, and, in this sense, we have no strong reasons to privilege another noise over that in Eq.~(\ref{eq:appstoc}).

\begin{figure*}
\centering
\begin{subfigure}{1\columnwidth}
  \includegraphics[trim={0cm 0cm 0cm 0cm},clip,width=0.95\linewidth]{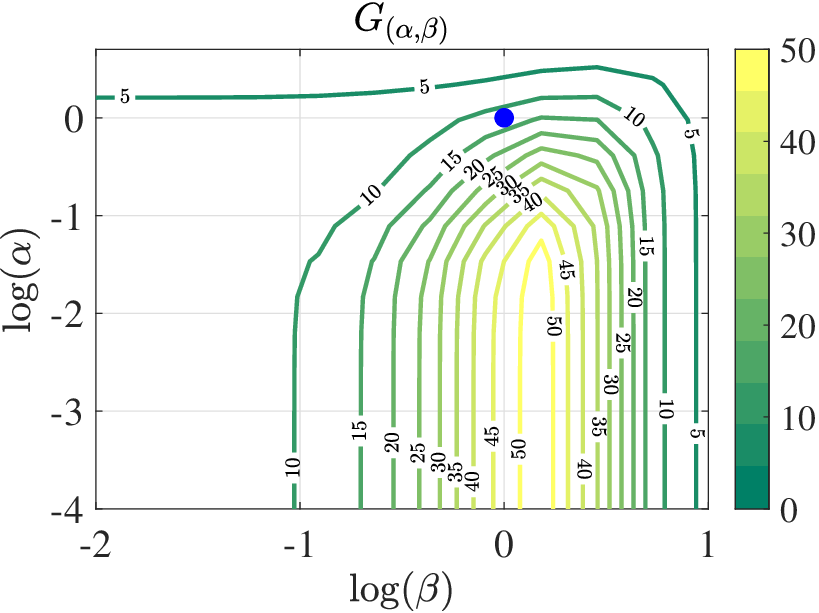}
  \caption{ \label{fig:Saba}}
\end{subfigure}
\hfil
\begin{subfigure}{1\columnwidth}
  \includegraphics[trim={0cm 0cm 0cm 0cm},clip,width=0.95\linewidth]{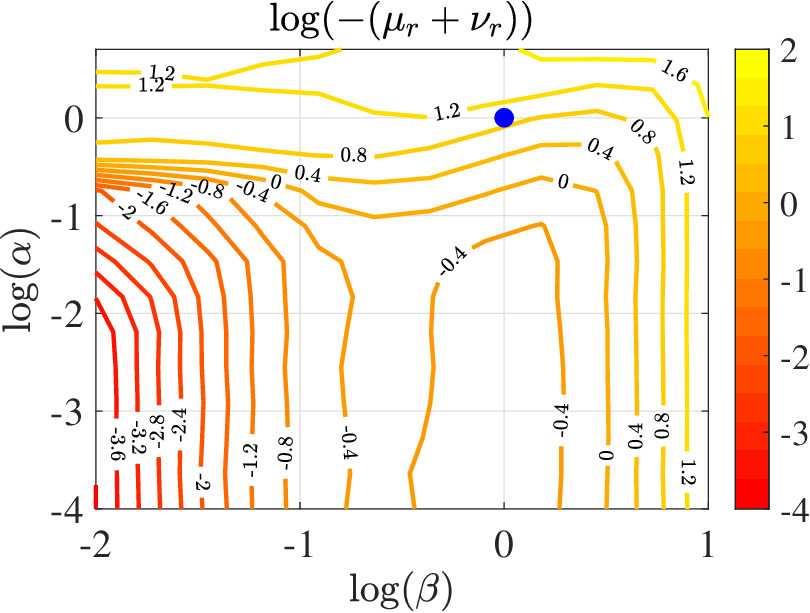}
   \caption{ \label{fig:Sabb}}
\end{subfigure}
\caption{\subref{fig:Saba} Isocontours of the optimal linear stochastic gain, as defined in Eq.~(\ref{eq:ete}), for the Couette-Poiseuille flow at $\Ren=300$ and under the specific noise in Eq.~(\ref{eq:appstoc}). The gain is shown as a function of the spanwise wavenumber $\bet$ and the streamwise one $\al$ (both in log scale). \subref{fig:Sabb} In the same wavenumbers plane, isocontours of the logarithm of $-(\mu_r+\nu_r)$, where $\mu$ and $\nu$ are the nonlinear coefficients defined in Eqs.~(\ref{eq:cofmus}) and (\ref{eq:cofnus}), respectively. The logarithm occurs to be always defined. In both frames, lighter nuances correspond to larger values, and a blue dot is placed at $(\al,\bet)=(1,1)$, the wavenumber pair chosen for a detailed nonlinear analysis.} \label{fig:Sab}
\end{figure*}

In Fig.~\ref{fig:Sab}(a) we report the linear optimal stochastic gain $G_{(\al,\bet)}$, as defined in Eq.~(\ref{eq:ete2}), for the Couette-Poiseuille flow at $\Ren=300$ and corresponding to the choice of noise in Eq.~(\ref{eq:appstoc}). We refer to Appendix~\ref{app:stocnum} for some numerical details. The gain is plotted as a function of the streamwise wavenumber $\al$ as well as the spanwise one $\beta$ (in log scales).
The largest gain is found for $\al=0$ and $\bet \approx 1.5$, where the forcing structure consists of streamwise invariant vortices amplified through the lift-up mechanism. We note that the same conclusion was drawn in Ref.~\cite{Farrell93} for the stochastically forced plane Poiseuille flow at $\Ren=200$. 

Still for $\Ren=300$, the wavenumber pair $(\al,\bet)=(1,1)$, corresponding to an oblique structure that is not the linearly most amplified (please see the dot in Fig.~\ref{fig:Saba}), is further selected for a detailed nonlinear analysis. At this parameters, it is true that the first sub-optimal gain $G_2=5.15 \approx 1.46 \sqrt{12.46}$ scales like the square root of the optimal one $G_1=12.46$, and that there is a reasonably large gap between them. We also understand from Fig.~\ref{fig:Srr2} that the optimal response indeed results from nonmodal mechanisms, as the corresponding $L^2$-norm profile over the frequencies is much wider and flatter than Lorentzian. It reaches its most prominent norm over $\om\in[-0.5,0]$, which seems hardly related to a single eigenvalue but contributes from many of them.

\begin{figure}
\centering
\includegraphics[trim={0.0cm 0.0cm 0.0cm 0.0cm},clip,width=0.85\linewidth]{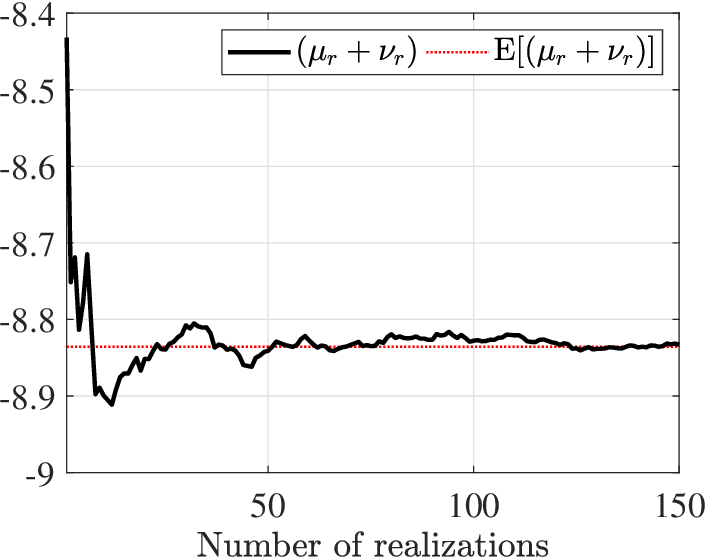}
\caption{For the Couette-Poiseuille flow with $(\Ren,\al,\bet)=(300,1,1)$. Convergence curve (continuous line) of the real part of the weakly nonlinear coefficient, i.e., $\mu_r+\nu_r$, as a function of the number of realizations. The dash-dotted line highlights the value of $\mu_r+\nu_r$ after a thousand realizations (not shown), virtually equal to the expected value.}
\label{fig:Sc}
\end{figure}

The weakly nonlinear approach requires the computation of the weakly nonlinear coefficients defined in Eqs.~(\ref{eq:cofmus}) and (\ref{eq:cofnus}). Since their expressions involve an ensemble average, several realizations of the quantity inside the expected value must be performed until a satisfactory convergence. We show in Fig.~\ref{fig:Sc} the convergence curve of $\mu_r+\nu_r$, the real part of the weakly nonlinear coefficient $\mu+\nu$, as denoted by the subscript ``$r$", with the number of realizations. 
Appendix~\ref{app:stocnum} also provides some numerical clarifications regarding the computation of the weakly nonlinear coefficients. The value of $\mu_r+\nu_r$ seems to converge with an acceptable tolerance after approximately a hundred realizations. The fact that it converges to a negative value is important, for it is $\mu_r+\nu_r$ only which governs the weakly nonlinear evolution of $|A_1|$ as $\phi$ is increased. The imaginary part $\mu_i+\nu_i$, in contrast, determines only the phase of $A_1$. More precisely, the negativity of $\mu_r+\nu_r$ implies that $|A_1|$ remains smaller than its linear solution, and thus the gain will decrease while increasing $\phi$. In other terms, for this particular wavenumber pair and Reynolds number, nonlinearities, at least in their weak expression, have a saturating effect on the amplitude of the response.

To assess this saturating effect of weak nonlinearities, we report in Fig.~\ref{fig:Sabb} the value of the logarithm of $-(\mu_r+\nu_r)$ in the wavenumbers plane. To save computational time, in producing Fig.~\ref{fig:Sabb} only twenty realizations have been performed for each wavenumber pair. The value of $\mu_r+\nu_r$ is found to be systematically negative for all the wavenumbers considered, which is why the logarithm is well defined everywhere in Fig.~\ref{fig:Sabb}. The same conclusion holds for $\Ren=500$, although not shown in the figure. Thereby, we did not find a wavenumber pair where weak nonlinearities could enhance the stochastic gain as $\phi$ increases. We also observed that larger values of $\al$ and/or $\bet$ seem to yield larger values of $-(\mu_r+\nu_r)$. This is presumably due to the increasing size of the nonlinear interactions forcing terms and their responses, since computing these former requires applying the gradient operator $\nab_{n} = [\rmi n\al,\pa_y, \rmi n\bet]^T$, linear in both $\al$ and $\bet$. For this reason, structures oscillating faster in space are predicted to experience a more severe weakly nonlinear saturation.

\begin{figure*}
\centering
\includegraphics[trim={0.0cm 0.0cm 0.0cm 0.0cm},clip,width=0.85\linewidth]{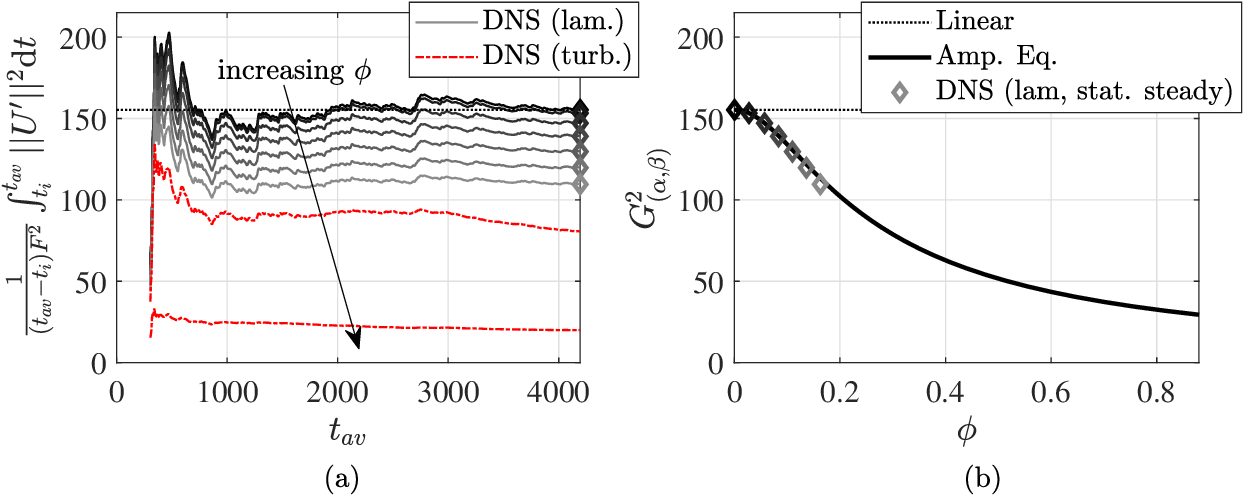}
\caption{The Couette-Poiseuille flow is considered with $(\Ren,\al,\bet)=(300,1,1)$ and where the stochastic forcing structure is chosen as the (linear) optimal one, i.e., $\pbf_s=\cbf_1$. (a) Temporal average of $||\pbU(t)||/F^2$ as a function of the averaging time $t_{av} \in [t_i=300;T=4200]$, defined in Eq.~(\ref{eq:gaicon}), and for different values of $\phi$ (lighter nuance larger $\phi$). 
Continuous lines denote laminar flow responses, whereas dash-dotted lines indicate the flow to have become turbulent (see text) for $\phi=0.22$ and $\phi=0.88$. 
For the laminar signals, a diamond is placed at $t_{av}=T$ to denote a converged temporal average; under the ergodic hypothesis, it corresponds to the fully nonlinear stochastic gain, directly reported in the right frame. (b) Stochastic gain as a function of $\phi$, as obtained from a linear ($=1/\ez$, dotted line), a leading-order weakly nonlinear ($=|A_1|/(\ez \phi)$ from Eq.~(\ref{eq:gsw}), continuous line) and a fully nonlinear (DNS laminar signals only, diamond markers) approach. }
\label{fig:GpRe300}
\end{figure*}

Going back to the specific analysis of the pair $(\al,\bet)=(1,1)$, we show in Fig.~\ref{fig:GpRe300}(b) the evolution of stochastic gain as a function of the rescaled stochastic forcing intensity $\phi$. As in the previous section, the stochastic forcing structure is chosen as the optimal one, i.e. $\pbf_s = \cbf_1$, such that the sub-optimal responses are only excited nonlinearly. 
This curve is analogous to these proposed in Figs.~\ref{fig:Hga}-\ref{fig:Hgc} in the harmonic case. Three approaches are again compared. In (i) the linear approach, the gain is equal to $=1/\ez$ and does not change with the forcing intensity. In (ii) the weakly nonlinear approach, the gain is $|A_1|/(\ez\phi)$ at leading-order according to Eq.~(\ref{eq:gsw}), where $A_1$ solves the polynomial equation (\ref{eq:astoc}). We also show with diamonds the (iii) fully nonlinear results extracted from DNS. These latter have been performed in a three-dimensional computational domain with coordinates
\begin{align}
x\in[0,2\pi/\al], \quad y\in[-1;1], \quad z \in[-\pi/\bet;\pi/\bet],
\label{eq:Nekbox}
\end{align}
and periodic boundary conditions in the $x$ and $z$ directions (please see Appendix~\ref{app:stocnum} for details concerning the numerical method). This makes possible the nonlinear production of wavenumber pairs ($n\al,p\bet$) with $n$ and $p$ two independent integers. Indeed, let us insist on the fact that, while the forcing solely excites the pair $(\al,\bet)=(1,1)$, nonlinearities generically excite different wavenumbers, typically harmonic pairs, but not only. However, by definition, the fully stochastic gain reported in Fig.~\ref{fig:GpRe300}(b) only refers to the $(\al,\bet)=(1,1)$ component of the stochastic response, denoted $\pbU$ in the calculations. Therefore, the latter must be extracted from the DNS by a spatial Fourier decomposition of the data beforehand.

Due to the norm to which the stochastic gain refers, the temporal and the ensemble average of $||\pbU||^2$ must be performed. The DNS lasted $T =4200$ units of time, to which we have discarded the first $300=t_i$ to remove the transient part of the response and only average in the presumably statistically steady regime. To assess the convergence of the temporal average of $||\pbU||^2$ with the averaging time $t_{av}$, where $t_i \leq t_{av} \leq T$, we show in Fig.~\ref{fig:GpRe300}(a) the evolution of the quantity
\begin{align}
\frac{1}{(t_{av}-t_i)} \int_{t_i}^{t_{av}}\frac{||\pbU(t)||^2}{F^2}\rmd t, \quad F=\phi \ez^{3/2}.
\label{eq:gaicon}
\end{align}
It should converge to a fixed value as $t_{av}$ is increased towards $T$, where it is equal to the temporal average of $||\pbU||^2/F^2$. For all considered values of $\phi$ but the two last ones, i.e., for the continuous line curves in Fig.~\ref{fig:GpRe300}(a), we observe that the quantity defined in Eq.~(\ref{eq:gaicon}) indeed converges towards some fixed value as $t_{av}=T$, highlighted by a diamond marker. The qualitatively similar evolution of all continuous line curves with $t_{av}$ is to be noticed. It is first made possible by the fact that they all have been subjected to the same noise process, only the value of $\phi$ being different between the two different curves. Another important element is that, for the corresponding $\phi$ values, the flow remains laminar. We further assume that these stochastic responses are ergodic processes, such that the last values of the continuous lines in Fig.~\ref{fig:GpRe300}(a), apparently converged and denoted by a diamond marker, directly give the stochastic gain without the need to additionally perform an ensemble average. This assumption was made for practical reasons only, since each curve in Fig.~\ref{fig:GpRe300}(a) takes about one week of computational time to be produced. Accordingly, the diamond marker in Fig.~\ref{fig:GpRe300}(a) are directly reported in Fig.~\ref{fig:GpRe300}(b).

For the two largest considered forcing intensities, corresponding to the dash-dotted continuous lines in Fig.~\ref{fig:GpRe300}(a), the flow becomes turbulent at a time smaller the larger $\phi$ is. The flow is declared turbulent by visual inspection of snapshots corresponding to different times and their corresponding Fourier spectra in space. Indeed, the turbulent regime is characterized by a substantially richer wavenumber content than the laminar regime. In this latter, the spectrum is dominated by the fundamental excited pair and its harmonics, i.e., by $n(\al,\bet)$ with $n=0,1,2,3,..$. This has no reason to remain true in a turbulent state and the latter is structurally fundamentally different from the one approached with the weakly nonlinear expansion in Eq.~(\ref{eq:asy1s}). Therefore, to only compare flow states of the same (laminar) nature and not mislead the reader, we did not report any corresponding stochastic gain value in Fig.~\ref{fig:GpRe300}(b) when the flow has transited to turbulence. 

Let us now comment on the agreement, in Fig.~\ref{fig:GpRe300}(b), between the weakly and fully nonlinear approaches. As long as the flow remains laminar for sufficiently small $\phi$, the weakly nonlinear approach performs well in predicting the nonlinear decrease of the stochastic gain as $\phi$ becomes larger. It brings a substantial improvement over the linear approach, and at a numerical cost that is considerably reduced as compared to that of the DNS. In addition, by comparing $\mu_r=-7.381$ with $\nu_r=-1.454$, we understand that this initial stochastic gain decrease is mostly due to a (spatial) mean flow distortion effect. The stochastic forcing appears to be less amplified by the spatial mean flow than by the base flow. This initial gain decrease, found in both approaches, may appear surprising given the subcritical nature of the flow and the subsequent transition to turbulence for larger forcing, although there is no contradiction there. It is possible that a more general forcing, exciting several wavenumber pairs, would lead to a gain increase even for small forcing values through nonlinear (triadic) interactions between these latter. Furthermore, the applied forcing structure is only optimal at the linear level, and need not resemble the nonlinear optimal (the latter being parameterized by the forcing intensity). The coefficients $\mu_r$ and $\nu_r$, being relative to the forcing structure considered, they could as well change sign if the nonlinear optimal was applied instead of the linear one.

However, the weakly nonlinear approach fails to predict the flow transition to turbulence, in the sense that the weakly nonlinear branch in Fig.~\ref{fig:GpRe300}(b) continues to exist and be unique for arbitrarily large $\phi$. This result could appear disappointing in the sense that the flow transition to another state could have corresponded to a loss of solution of Eq.~(\ref{eq:astoc}), or a loss of its stability. Going back to the harmonic forcing problem, this was for instance found to be the case in Ref.~\cite{Ducimetiere22} for the Poiseuille flow (please see Fig.~$10$ therein). Instead, the fact that the amplitude equation here gives no information about the flow transition to turbulence above some value $\phi$, seems to indicate that such transition is due to a leading-order nonlinear interaction between the stochastic response, and some ``secondary" mode appearing on top (and as a consequence) of the latter. Indeed, such nonlinear coupling between the stochastic response and some other structure(s) is inherently absent in the weakly nonlinear approach. That is because the weakly nonlinear approach postulates in Eq.~(\ref{eq:asy1s}) that the stochastic response is along the linear optimal one at the leading order. Thus, it only consists of a single amplitude equation for linear optimal response, and cannot take into account a leading-order structural enrichment that could arise, for instance, from a secondary instability. Thereby, it cannot take into account the ensuing nonlinear interactions that can lead to another flow state (here, turbulent). By contrast, in the weakly nonmodal harmonically forced Poiseuille flow in Ref.~\cite{Ducimetiere22}, Fig.~$10$, the amplitude equation in itself could predict the subcritical transition, since the latter was due to the effect of nonlinearities directly on the eigenvalue largely responsible for the harmonic response.

\begin{figure*}
\centering
\includegraphics[trim={0.0cm 0.0cm 0.0cm 0.0cm},clip,width=0.85\linewidth]{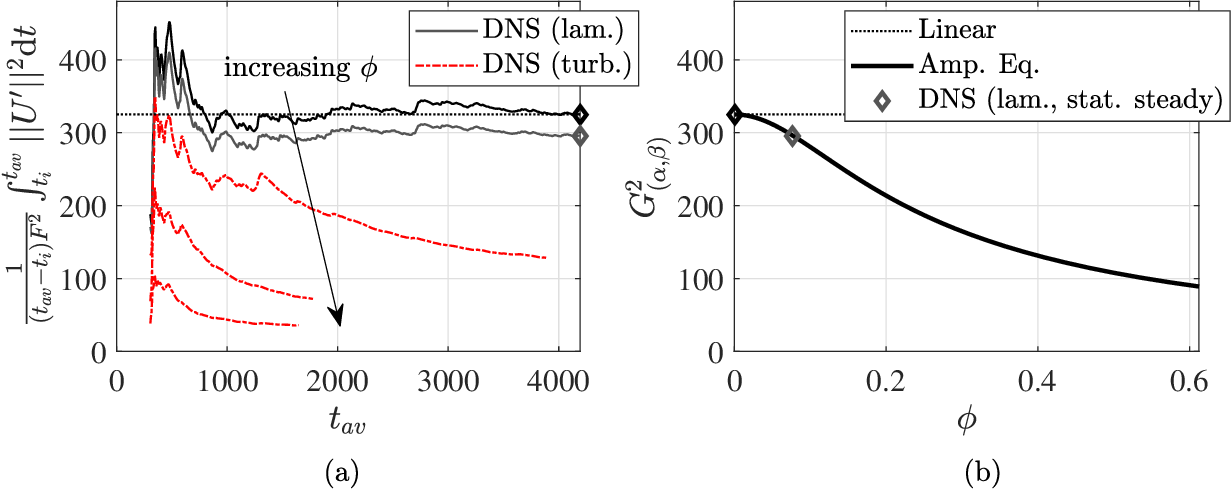}
\caption{Same as Fig.~\ref{fig:GpRe300} but the Reynolds number was increased to $\Ren=500$. For the three largest considered $\phi \in [0.15,0.31,0.61]$, the flow became turbulent. }
\label{fig:GpRe500}
\end{figure*}

The same figure as Fig.~\ref{fig:GpRe300}, but by increasing the Reynolds number to $\Ren=500$, is shown in Fig.~\ref{fig:GpRe500}.
The results are qualitatively similar to those for $\Ren=300$. Quantitatively, the threshold value of the rescaled stochastic forcing intensity $\phi$ above which the flow becomes turbulent is lower for $\Ren=500$ as compared to $\Ren=300$. Again, the weakly nonlinear approach is relevant for sufficiently small values of $\phi$ for the flow to remain laminar and structurally close to the linear response. Yet, these values of $\phi$ are sufficiently large for the weakly nonlinear approach to bring a significant correction over the linear gain value.  

Eventually, note that in Figs.~\ref{fig:GpRe300} and \ref{fig:GpRe500}, the value of $\phi$ for which nonlinearities become important indeed is of $O(1)$. This is in stark contrast with Fig.~\ref{fig:HgvsF} where it was seen to be of $O(\ez^{-1/2})$. That is because in Fig.~\ref{fig:HgvsF} the non-normality resulted from a Kelvin-Helmholtz convective mechanism, spatially developing in the non-parallel BFS base-flow. Figures~\ref{fig:GpRe300} and \ref{fig:GpRe500}, however, are concerned with a parallel flow and thus the non-normality is component-wise. Thereby, the optimal forcing and its response generically have substantial spatial support, which results in $O(1)$ nonlinear coefficients.

\section{Weakly nonlinear response to an initial perturbation\label{sec:tra}}

In this third and last section, we complete our survey by considering the nonmodal response to an initial perturbation, in a weakly nonlinear regime. Proceeding according to the same general idea as in the two previous sections, the Navier-Stokes equations are presently reduced to a scalar one for the amplitude of response to the optimal initial condition. While the nature of the problem suggests using the same $L^2$ inner product as in the first section, a particularity of the reduction procedure lies here in the fact that the amplitude must here depend on time. 

Once again, we invite readers who are not interested in the technical details of the method, when specifically applied to the transient growth problem, to skip directly to Sec.~\ref{sec:sp} for concluding remarks.

\subsection{Linear regime \label{sec:tra1}}

As in the previous section, the base flow $\bUb$ is also chosen here as a three-dimensional plane channel flow, bounded in the $y$ direction, but invariant in the $x$ and $z$ directions. In addition, an infinitesimal initial perturbation $\bu_0$ is assumed to be spatially periodic with wavenumbers $\al$ and $\bet$ along the $x$ and $z$ coordinates, respectively, such that
\begin{align}
\bu_0(x,y,z) = \pbu_0(y)\eixz + \cc.
\label{eq:u02}
\end{align}
The field $\pbu_0(y)$ is an arbitrary spatial velocity structure, generically complex-valued. Owing to the invariance of the base flow along the $x$ and $z$ directions, the linear response $\bq$ to the initial perturbation in Eq.~(\ref{eq:u02}) will also oscillate only along the $(\al,\bet)$ pair. It is thus possible to write
\begin{align}
\bq(x,y,z,t) = \pbq(y,t)\eixz + \cc,
\label{eq:tperf}
\end{align}
and the field $\pbq(y,t)$ obeys the linear equation
\begin{align}
\bB\pa_t\pbq = \bL_{1} \pbq, \quad \mbox{with} \quad \pbq(y,0)=\bP^T\pbu_0(y).
\label{eq:ltg}
\end{align}
We refer to Eq.~(\ref{eq:Leab}) for the definition of the linear operator $\bL_{1}$. Note that in this particular section, no assumption is made regarding the stability of $\bL_{1}$, which could as well be unstable. As shown in Sec.~\ref{sec:ifoptg}, Eq.~(\ref{eq:ltg}) can be formally solved by using the propagator operator $\bPhi(t,0)$. The latter is such that its application onto $\pbu_0$, the velocity field at time $t=0$, directly gives the velocity field $\pbu$ at some time $t$, i.e.
\begin{align}
\pbu(y,t) = \bPhi(t,0)\pbu_0(y).
\label{eq:ltg2}
\end{align}
From now on, the dependence of the field on the spatial coordinate $y$ will not be made explicit anymore. As argued in Sec.~\ref{sec:ifoptg}, it is relevant to seek the initial perturbation structure $\pbu_0$ leading to the largest amplification under the norm induced by the $L^2$ inner product, and evaluated at some $t_o$ called ``temporal horizon". Specifically, it is relevant to compute the transient gain
\begin{align}
\Gt &= \max_{\pbu_0}\frac{||\pbu(t_o)||}{||\pbu_0||} = \max_{\pbu_0}\frac{||\bPhi(t_o,0)\pbu_0||}{||\pbu_0||} \nonumber \\
&= \frac{1}{\ez},
\label{eq:tgg}
\end{align}
where the subscript $``(\al,\bet)"$ emphasizes that the gain is parameterized by the wavenumber pair $(\al,\bet)$. We have again defined $\ez$ as the gain inverse. The maximization problem in Eq.~(\ref{eq:tgg}) is solved by computing the right and left singular modes of the propagator evaluated at some temporal horizon $t_o$, i.e., $\bPhi(t_o,0)$, which yield the $\set{\bbu_{0,j}}_{j\geq 1}$ and $\set{\bbu_{j}(t_o)}_{j\geq 1}$ families, respectively. Each family is orthonormal under the $L^2$ inner product and the normalization $||\bbu_{0,j}||=||\bbu_{j}(t_o)||=1$ applies for every $j \geq 1$. Each pair of singular vectors is associated with the singular value $G_j$, such that the relations
\begin{align}
G_j\bbu_{j}(t_o) = \bPhi(t_o,0)\bbu_{0,j}, \quad G_j\bbu_{0,j} = \bPhi(t_o,0)^{\dag}\bbu_{j}(t_o),
\label{eq:svtg}
\end{align}
are true for every $j \geq 1$. The singular values $\set{G_j}_{j\geq 1}$ are sorted by decreasing values. This makes $G_1$ the optimal gain, i.e., $\Gt = G_1$, and $\bbu_{0,1}$ the associated optimal initial condition. Accordingly, $\bbu_1(t_o)$ is the normalized response to the optimal initial condition, evaluated at $t=t_o$. All the $j \geq 2$ correspond to sub-optimal gains and pairs.

While $\bbu_{j}(t_o)$ denote the response to $G_j^{-1}\bbu_{0,j}$ evaluated at $t=t_o$, the latter response can also be computed for all times $0\leq t\leq t_o$ according to
\begin{align}
\bbu_j(t) = G_j^{-1}\bPhi(t,0)\bbu_{0,j},
\label{eq:tgg2}
\end{align}
where the first relation in Eq.~(\ref{eq:svtg}) is recovered at $t=t_o$. In other terms, $\bbu_j(t)$ is the linear trajectory seeded by $G_j^{-1}\bbu_{0,j}$.

From now on, let us normalize $\pbu_0$, an arbitrary initial condition, according to $||\pbu_0||=1$. By decomposing $\pbu_0$ in the orthonormal basis $\set{\bbu_{0,j}}_{j \geq 1}$, we are led to the following decomposition of the transient response in Eq.~(\ref{eq:ltg2}),
\begin{align}
\pbu(t) = \bPhi(t,0)\pbu_0 &= \sum_{j\geq 1} \ssp{\bbu_{0,j}}{\pbu_0}G_j\bbu_j(t).
\label{eq:ltg2dec}
\end{align}
This decomposition, however, has to be interpreted with more caution than its equivalent in Eq.~(\ref{eq:harr}) for the harmonic forcing problem, or in Eq.~(\ref{eq:lsst2}) for the stochastic one. Indeed, the $\set{\bbu_j(t)}_{j\geq 1}$ family is generically not orthonormal for $t<t_o$. That is both because it is not orthogonal for $0 < t < t_o$, and because, for all $j\geq 1$, $||\bbu_j(t)||$ is not equal to one for $0\leq t<t_o$. Consequently, to remain consistent with the previous sections, we would rather be interested in describing the transient response $\pbu(t)$ as some component along $\bbu_1(t)$, plus another term $\bbup(t)$, defined as being orthogonal to the latter for all times, rather than only at $t=t_o$ and $t=0$ (i.e., $\ssp{\bbu_1(t)}{\bbup(t)}=0$ for all $t$). This gives 
\begin{align}
\pbu(t) = \frac{\ssp{\bbu_1(t)}{\pbu(t)}}{\ssp{\bbu_1(t)}{\bbu_1(t)}}\bbu_1(t) + \bbup(t),
\label{eq:dec4}
\end{align}
which, by identification with Eq.~(\ref{eq:ltg2dec}), leads to 
\begin{align}
\bbup(t) = \sum_{j\geq 2}\ssp{\bbu_{0,j}}{\pbu_0}G_j\bbup_j(t), \label{eq:dec6}
\end{align}
where we have defined
\begin{align}
\bbup_j(t) = \bbu_j(t)-\frac{\ssp{\bbu_1(t)}{\bbu_j(t)}}{\ssp{\bbu_1(t)}{\bbu_1(t)}}\bbu_1(t), \quad j \geq 2.
\label{eq:dec7}
\end{align}
The field $\bbup_j(t)$ is the component of $\bbu_j(t)$ which remains orthogonal to $\bbu_1(t)$, for all times and $j \geq 2$. We note that $\bbup_j(t_o)=\bbu(t_o)$ of unit norm, for all $j\geq 1$. 

In what follows, the optimal gain $G_1=1/\ez$ is again assumed to be large and thus $\ez$ to be small. Furthermore, we also assume a large gap between the value of the optimal gain and that of the sub-optimal ones, i.e., $G_1 \gg G_2 > G_3 > ...$ and, for the third time throughout this article, we postulate the scalings
\begin{align}
&\set{G_1,G_2,...,G_N,G_{N+1},...} \nonumber \\
&= \set{\frac{1}{\ez},\frac{\ga_2}{\ez^{1/2}},...,\frac{\ga_N}{\ez^{1/2}},\ga_{N+1},...}, \quad \ga_j = O(1), \quad j\geq 2.
\label{eq:scgtr}
\end{align}
This scaling implies that the transient response $\pbu(t)$ can be rewritten as
\begin{widetext}
\begin{align}
\pbu(t) =& \frac{1}{\ez}\Biggl(\frac{\ssp{\bbu_1(t)}{\ez \pbu(t)}}{\ssp{\bbu_1(t)}{\bbu_1(t)}}\bbu_1(t) +\underbrace{\ez^{1/2}\sum_{j=2}^{N}\ssp{\bbu_{0,j}}{\pbu_0}\ga_j \bbup_j(t) + \ez\sum_{j>N}\ssp{\bbu_{0,j}}{\pbu_0}\ga_j \bbup_j(t)}_{=\bbup(t)}\Biggl),
\label{eq:dec44}
\end{align}
\end{widetext}
where Eqs.~(\ref{eq:dec4}) and (\ref{eq:dec6}) have been used. Since for $t \simeq t_o$ we have $||\ez \pbu(t)||=O(1)$ as well as $||\bbup_j(t)||=O(1)$ for $j \geq 2 $, in Eq.~(\ref{eq:dec44}) the component along $\bbu_1(t)$ has a contribution which is $\ez^{-1/2} \gg 1$ larger than the term $\bbup(t)$ for $t$ around $t_o$. In other terms, for $t$ around $t_o$, it is possible to interpret the transient response as being predominantly along $\bbu_1(t)$ plus a smaller rest contained in the orthogonal subspace, i.e.,
\begin{align}
\pbu(t) = \frac{1}{\ez}\pae{\frac{\ssp{\bbu_1(t)}{\ez\pbu(t)}}{\ssp{\bbu_1(t)}{\bbu_1(t)}}\bbu_1(t) + \underbrace{O(\ez^{1/2})}_{=\bbup(t)}}, \quad t \simeq t_o.
\label{eq:dec5}
\end{align}
Let us, however, insist that this order separation, between the component along $\bbu_1(t)$ and the rest, only holds for $t$ around $t_o$. For times around $t=0$ in particular, by definition in Eq.~(\ref{eq:tgg2}) the norm $||\bbu_j(t)||$ is of order $G_j^{-1}$ for $j\geq 1$. Thereby, we deduce from Eq.~(\ref{eq:ltg2dec}) that the respective components along each of the $\set{\bbu_j(t)}_{j\geq 1}$ all contribute to the same (order one) amount to $\pbu(t)$. Since the $\set{\bbu_j(t)}_{j\geq 1}$ family is orthonormal at $t=0$, it is thus not possible to conclude that, at short times, the component along $\bbu_1(t)$ dominates over its orthogonal counterpart.
 
\begin{figure*}
\centering
 \begin{subfigure}{1\columnwidth}
  \includegraphics[trim={0cm 0cm 0cm 0cm},clip,width=0.95\linewidth]{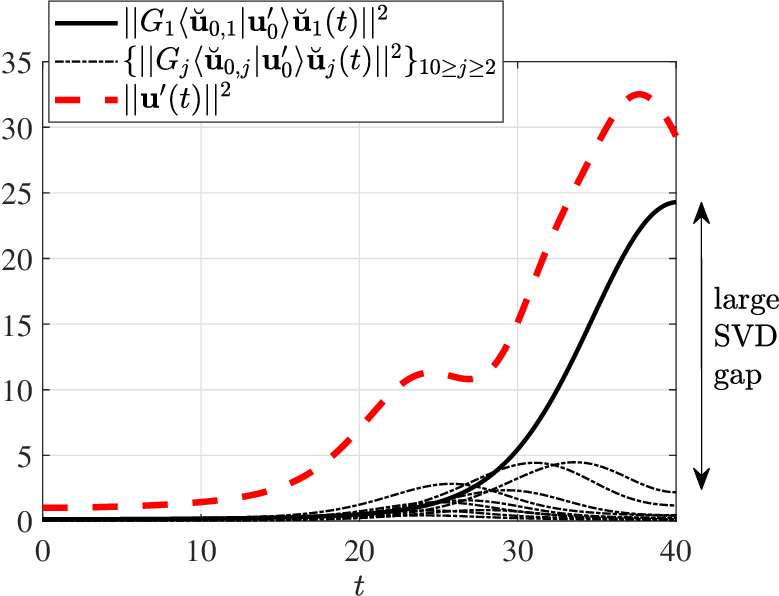}
  \caption{\label{fig:tgapa}}
\end{subfigure}
\hfil
\begin{subfigure}{1\columnwidth}
  \includegraphics[trim={0cm 0cm 0cm 0cm},clip,width=0.95\linewidth]{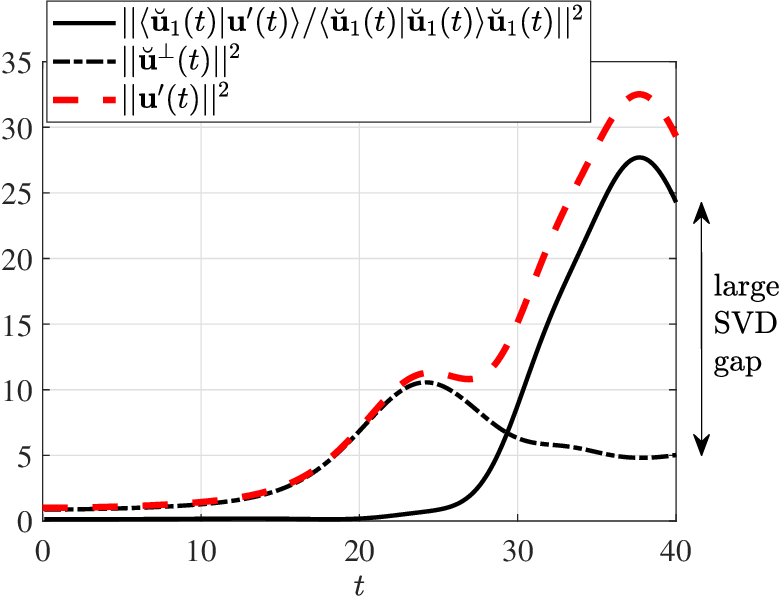}
  \caption{\label{fig:tgapb}}
\end{subfigure}
\caption{\subref{fig:tgapa} For the two-dimensional Lamb-Oseen vortex flow with azimuthal wavenumber $\al=2$, temporal horizon $t_o=40$ and $\Ren=10^4$. The dotted line shows the $L^2$-norm squared of some linear trajectory $\pbu(t)=\bPhi(t,0)\pbu_0$, as a function of $t\in[0;t_o]$, where $\pbu_0$ (with $||\pbu_0||^2=1$) is the sum over the first ten optimal and sub-optimal initial conditions, ponderated by random coefficients. The continuous line shows the norm of the contribution along the response to the optimal initial condition, i.e. $\ssp{\bbu_{0,1}}{\pbu_0}G_1\bbu_1(t)$ (see Eq.~(\ref{eq:ltg2dec})). Similarly, the thin dashed-dotted lines are for the contributions along the response to the first nine sub-optimal initial conditions. Beware that, since $\set{\bbu_j(t)}_{j\geq 1}$ is not an orthogonal family for $0<t<t_o$, the dotted line is the sum of all the other curves only at $t=0$ and $t=t_o$. \subref{fig:tgapb} Alternative decomposition, as expressed in Eq.~(\ref{eq:dec4}), of the linear trajectory $\pbu(t)$ into a component along $\bbu_1(t)$ (its norm squared the continuous line) plus a rest $\bbu^{\perp}(t)$ orthogonal to $\bbu_1(t)$ for all times (its norm squared the dashed-dotted line). This time, by construction, the dotted line is the sum of the two other curves for all times.} \label{fig:tgap}
\end{figure*}

Figure~\ref{fig:tgap} is meant to illustrate this. It shows the linear transient response of the two-dimensional Lamb-Oseen vortex flow at $\Ren=10^4$, to an initial condition with azimuthal wavenumber $\al=2$. There, the initial condition $\pbu_0$ (with $||\pbu_0||^2=1$) is chosen as the sum of the first ten optimal and sub-optimal initial conditions, associated with the temporal horizon $t_o=40$, and ponderated by random coefficients (whose sum of squares is equal to one, following from the normalization of $\pbu_0$ and of the $\set{\bbu_{0,j}}_{j\geq 1}$). We indeed observe in Fig.~\ref{fig:tgap} that the gap between $G_1=1/\ez=13.44$ and the sub-optimal gains (e.g., $G_2=5.22=1.42\sqrt{13.44}$), makes the contribution along $\bbu_1(t)$ dominant in the transient response for times around $t_o$ only, but generically not for smaller ones.

Nevertheless, in what follows, once again we take advantage of this large gap to reduce the Navier-Stokes equations to a scalar weakly nonlinear equation for the amplitude of $\bbu_1(t)$.

\subsection{Weakly nonlinear continuation \label{sec:tra2}}

Let us consider the fully nonlinear, unforced Navier-Stokes equations
\begin{align}
\bB \pa_t \ve{\bU}{P}= \ve{-(\bU \cdot \nab)\bU + \Ren^{-1}\Delta \bU - \nab P}{\nab \cdot \bU},
\label{eq:NSf3INI}
\end{align}
with the initial condition
\begin{align}
\ve{\bU(0)}{P(0)} = \ve{\bUb}{\pb} + \ez^{3/2}\bP^T\pae{\psi\pbu_0\eixz +\cc},
\label{eq:NSf3}
\end{align}
where recall the initial condition structure, $\pbu_0$, to be normalized as $||\pbu_0||=1$. It is scaled to appear at $O(\ez^{3/2})$ with an amplitude $\psi = O(1)$, the latter being a complex-valued free parameter. We emphasize that the initial condition oscillates in space purely along the pair $(\al,\bet)$. In the nonlinear regime, this constitutes a loss of generality which is purposely made to lighten the ensuing calculations. The method, however, is not fundamentally limited to spatially harmonic initial conditions.

We aim to determine the scalar, complex-valued amplitude $A_1(\ez^{1/2} t)=O(1)$ corresponding to the following postulated form for the weakly nonlinear transient response 
\begin{widetext}
\begin{align}
\ve{\bU(t)}{P(t)}  = & \ve{\bUb}{\pb} + \ez^{1/2}\sae{A_1(\ez^{1/2} t)\ve{\bbu_1(t)}{\bp_1(t)}\eixz + \cc} + \ez\sae{\sum_{j=2}^{N} A_j(\ez^{1/2} t) \ve{\bbup_j(t)}{\bpp_j(t)}\eixz + \cc + \ve{\obu_2(t)}{\op_2(t)}} \nonumber \\
& + \ez^{3/2}\sae{\sum_{j>N} A_j(\ez^{1/2} t) \ve{\bbup_j(t)}{\bpp_j(t)}\eixz + \cc + \ve{\obu_3(t)}{\op_3(t)}} + \underbrace{\sum_{m\geq 4} \ez^{m/2}\ve{\obu_m(t)}{\op_m(t)}}_{=O\pae{\ez^{2}}},
\label{eq:asytg}
\end{align}
where all the others $A_j = O(1)$ for $j\geq 2$ are also unknowns. 
\end{widetext}
The symbols $\obu_m(t)$ with $m \geq 2$ denote spatial harmonics fields without including the fundamental, such that 
\begin{align}
\obu_m(t) = \obu_{m,0}(t) + \pae{\sum_{ n \geq 2}\obu_{m,n}(t)\eixzm{n} + \cc},
\label{eq:fs3}
\end{align}
for $m \geq 2$. As in the two previous sections, expansion in Eq.~(\ref{eq:asytg}) is such that the component of the weakly nonlinear transient response along the pair $(\al,\bet)$, reading 
\begin{align}
\pbU(t) = & \ez^{1/2} A_1(\ez^{1/2} t) \bbu_1(t) +\ez \sum_{j=2}^{N} A_j(\ez^{1/2} t) \bbup_j(t) \nonumber \\
& + \ez^{3/2} \sum_{j>N} A_j(\ez^{1/2} t)\bbup_j(t),
\label{eq:asytgU}
\end{align}
is similar to its solution in the linear regime given in Eq.~(\ref{eq:dec44}). This is at the exception that, at each order, the prefactors in front of the fields $\bbu_1$ and $\set{\bbup_{j}}_{j\geq 2}$ in Eq.~(\ref{eq:dec44}) have been replaced by the unknown amplitudes $\set{A_j}_{j\geq1}$ which also incorporate weakly nonlinear effects. Again, this was purposely made to ensure the continuity between the linear and weakly nonlinear regimes. 

We insist, however, that the separation of the orders in Eq.~(\ref{eq:asytg}) holds only for $t \simeq t_o$. On the contrary, for small times $t\simeq 0$, the terms in $\set{\bbup_j(t)}_{j\geq 2}$, as well as the one in $\bbu_1(t)$, all appear at $O(\ez^{3/2})$ due to the scaling of the initial condition. The latter fact will become clear in a moment.   

Contrary to the previous sections, the amplitude $A_1(\ez^{1/2} t)$ must here depend on time. It is still for reasons of consistency with the linear regime, where the time-dependent solution $A_1(\ez^{1/2} t) = \psi\ssp{\bbu_1(t)}{\ez \bPhi(t,0)\pbu_0}/\ssp{\bbu_1(t)}{\bbu_1(t)}$ should be recovered in the limit $\psi \rightarrow 0$. Indeed, the latter expression follows from the decomposition in Eq.~(\ref{eq:dec4}), where $\pbu(t)=\ez^{3/2}\psi\bPhi(t,0)\pbu_0$ and recalling that $A_1$ is pre-multiplied by $\ez^{1/2}$ in the expansion of the response. We further assume the amplitude $A_1$ to vary slowly with time, in the sense that $\rmd_t A_1 =O(\ez^{1/2})$, as we anticipate that it will simplify the temporal integrals involving powers of $A_1$. While this hypothesis presumably degrades the accuracy of the prediction, as the strength of the nonlinearities is increased, it also makes the resulting amplitude equation substantially easier to solve. It would also have been possible to make $A_1$ depending on $t$ directly, which could presumably capture faster nonlinear modification of $A_1$ at the cost of a more complicated amplitude equation. 

The initial condition associated with each field contribution in Eq.~(\ref{eq:asytg}) is obtained by evaluating the latter expansion at $t=0$, applying $\bP$, and then identifying with the externally applied initial condition in Eq.~(\ref{eq:NSf3}). This gives
\begin{align}
&\ez^{1/2}A_1(0)\ez\bbu_{0,1} + \ez\sum_{j=2}^{N}A_j(0)G_j^{-1}\bbu_{0,j} \nonumber \\ 
&+ \ez^{3/2}\sum_{j > N}A_j(0)G_j^{-1}\bbu_{0,j} = \ez^{3/2}\psi \pbu_0, 
\label{eq:asytg0}
\end{align}
where we have used that $\bbup_j(0)=G_j^{-1}\bbu_{0,j}$. Equivalently, 
\begin{align}
A_1(0)\bbu_{0,1} + \sum_{j \geq 2 }A_j(0)\ga_j^{-1}\bbu_{0,j} = \psi \pbu_0.
\label{eq:asytg04}
\end{align}
The orthonormality of the $\set{\bbu_{0,j}}_{j \geq 1}$ family further implies
\begin{align}
A_1(0)=\psi \ssp{\bbu_{0,1}}{\pbu_0}, \quad A_j(0)=\ga_j \psi \ssp{\bbu_{0,j}}{\pbu_0}, \quad j\geq 2.
\label{eq:asytg02}
\end{align}
In addition, since only the $(\al,\bet)$ pair is initialized, all harmonics fields must initially be null, i.e., $\obu_{m,n}(0) = \bz$ for all considered $m$ and $n$.

Injecting Eqs.~(\ref{eq:asytg}) and (\ref{eq:fs3}) in the Navier-Stokes equations leads to 
\begin{align}
\bz =& \ez \Biggl[-(\bB\pa_t-\bL_{0})\ve{\obu_{2,0}}{\op_{2,0}} \nonumber \\
&- |A_1|^2\bP^{T}\pae{\bC_{1}(\bbu^*_1,\bbu_1) +\cc}\Biggl] + O(\ez^{3/2}),
\label{eq:atg0}
\end{align}
for the component oscillating at the wavenumber pair $(0,0)$, together with
\begin{align}
\bz =&  \ez \Biggl[-\pae{\bB\pa_t-\bL_{2}}\ve{\obu_{2,2}}{\op_{2,2}} \nonumber \\
& - A_1^2\bP^{T}\bC_{1}(\bbu_1,\bbu_1)\Biggl]+ O(\ez^{3/2}),
\label{eq:atg2}
\end{align}
for the component oscillating at $(2\al,2\bet)$. By collectings terms of Eq.~(\ref{eq:atg0}) at $O(\ez)$, we show in Appendix~\ref{app:tgA} that the zero harmonics velocity field $\obu_{2,0}$ reads
\begin{align}
\obu_{2,0}(t) = |A_1(\ez^{1/2} t)|^2 \obu^{|A_1|^2}_{2,0}(t) + O(\ez^{1/2}),
\label{eq:u201c}
\end{align}
where $\obu^{|A_1|^2}_{2,0}(t)$ solves
\begin{align}
\bB\pa_t\ve{\obu^{|A_1|^2}_{2,0}}{\op^{|A_1|^2}_{2,0}} = \bL_{0}\ve{\obu^{|A_1|^2}_{2,0}}{\op^{|A_1|^2}_{2,0}} - \bP^{T}\pae{\bC_{1}(\bbu^*_1,\bbu_1) +\cc},
\label{eq:u201cBIS}
\end{align}
with $\obu^{|A_1|^2}_{2,0}(0)=\bz$. Similarly, by collecting terms of Eq.~(\ref{eq:atg2}) at $O(\ez)$, it is also shown in Appendix~\ref{app:tgA} that the second harmonics velocity field $\obu_{2,2}$ is
\begin{align}
\obu_{2,2}(t) = A_1(\ez^{1/2} t)^2 \obu^{A_1^2}_{2,2}(t) + O(\ez^{1/2}),
\label{eq:u201d}
\end{align}
where $\obu^{A_1^2}_{2,2}(t)$ solves
\begin{align}
\bB\pa_t\ve{\obu^{A_1^2}_{2,2}}{\op^{A_1^2}_{2,2}} = \bL_{2}\ve{\obu^{A_1^2}_{2,2}}{\op^{A_1^2}_{2,2}} - \bP^{T}\bC_{1}(\bbu_1,\bbu_1),
\label{eq:u201dBIS}
\end{align}
with $\obu^{A_1^2}_{2,2}(0)=\bz$. 

\begin{widetext}
The expansion for the component oscillating at the fundamental wavenumber pair $(\al,\bet)$ reads
\begin{align}
\bz =& \ez^{1/2}\sae{-\pae{\bB\pa_t-\bL_{1}} \pae{A_1\ve{\bbu_1}{\bp_1}}} +\ez \sae{-\sum_{j=2}^{N} \pae{\bB\pa_t-\bL_{1}} \pae{A_j\ve{\bbup_j}{\bpp_j}}} +\ez^{3/2} \Biggr[ -\sum_{j > N} \pae{\bB\pa_t-\bL_{1}} \pae{A_j\ve{\bbup_j}{\bpp_j}} \nonumber  \\
&- A_1\bP^{T}\Big(\bC_{0}(\bbu_1,\obu_{2,0})  + \bC_{1}(\obu_{2,0},\bbu_1)\Big) - A^*_1\bP^{T}\Big(\bC_{2}(\bbu_1^*,\obu_{2,2}) + \bC_{-1}(\obu_{2,2},\bbu_1^*)\Big) \Biggr].
\label{eq:atg1}
\end{align}
Further substituting Eq.~(\ref{eq:u201c}) and Eq.~(\ref{eq:u201d}) in Eq.~(\ref{eq:atg1}), and then proceeding along the calculations lines proposed in Appendix~\ref{app:tgB}, makes it possible to simplify Eq.~(\ref{eq:atg1}) into
\begin{align}
\bz =& \ez^{3/2} \Big[ -A_1\bbu_{0,1}-\sum_{j \geq 2}\ga_j^{-1}A_j\pae{\bbu_{0,j} - \ssp{\bbu^{\dag}_1(t)}{\bbu_{0,j}}\bbu_{0,1}} + \psi\pbu_0 - A_1|A_1|^2\bPhi(0,t)\pae{\punlm + \punlh} \Big] + O(\ez^2).\label{eq:atgj}
\end{align}
In Eq.~(\ref{eq:atgj}), we have introduced the ``adjoint to the optimal transient response", $\bbu^{\dag}_1(t)$, as
\begin{align}
\bbu^{\dag}_1(t) = \frac{\ez\bPhi(t,0)^{\dag}\bbu_1(t)}{\ssp{\bbu_1(t)}{\bbu_1(t)}}, 
\label{eq:unld3}
\end{align}
where the scalar prefactor $\ez/\ssp{\bbu_1(t)}{\bbu_1(t)}$ ensures the convenient normalization
\begin{align}
\ssp{\bbu^{\dag}_1(t)}{\bbu_{0,1}} = \frac{\ssp{\bbu_1(t)}{\ez \bPhi(t,0)\bbu_{0,1}}}{\ssp{\bbu_1(t)}{\bbu_1(t)}} = \frac{\ssp{\bbu_1(t)}{\bbu_1(t)}}{\ssp{\bbu_1(t)}{\bbu_1(t)}}=1.
\label{eq:unld3n}
\end{align}
In Eq.~(\ref{eq:atgj}) we have also defined the third-order velocity fields $\punlm(t)$ and $\punlh(t)$. The former results from the nonlinear interaction of the optimal response $\bbu_1$ with the mean flow correction $\obu^{|A_1|^2}_{2,0}$, whereas the latter results from the interaction of $\bbu_1$ with the second harmonic field $\obu^{A_1^2}_{2,2}$. They are solutions of 
\begin{align}
(\bB\pa_t - \bL_{1})\ve{\punlm}{\pnlm} = \bP^T\pfnlm = \bP^T\pae{\bC_{0}(\bbu_1,\obu^{|A_1|^2}_{2,0})  + \bC_{1}(\obu^{|A_1|^2}_{2,0},\bbu_1)}, 
\label{eq:ati1}
\end{align}
as well as
\begin{align}
(\bB\pa_t - \bL_{1})\ve{\punlh}{\pnlh} = \bP^T\pfnlh =\bP^T\pae{\bC_{2}(\bbu_1^*,\obu^{A_1^2}_{2,2}) + \bC_{-1}(\obu^{A_1^2}_{2,2},\bbu_1^*)},  
\label{eq:ati2}
\end{align}
\end{widetext}
respectively, with $\punlh(0)=\punlm(0)=\bz$.
In Eq.~(\ref{eq:atgj}), the term $\pae{\bbu_{0,j}-\ssp{\bbu_1^{\dag}}{\bbu_{0,j}}\bbu_{0,1}}$ can easily be shown to be orthogonal to $\bbu_1^{\dag}$ for every $j\geq 1$, as a result of Eq.~(\ref{eq:unld3n}). Thereby, the simplest leading-order equation for $A_1$ can be extracted by taking the inner product between $\bbu^{\dag}_1(t)$ and Eq.~(\ref{eq:atgj}). This gives  
\begin{align}
\boxed{0  = -A_1 +\psi\ssp{\bbu^{\dag}_1}{\pbu_0} - A_1|A_1|^2(\mu+\nu)} + O(\ez^{1/2}), \label{eq:ampeqtg}
\end{align}
where we insist that the terms at $O(\ez^{1/2})$ can be neglected generally only for $t\simeq t_o$, and at the condition that the temporal integration performed in transforming Eq.~(\ref{eq:atg1}) into Eq.~(\ref{eq:atgj}) did not alter the asymptotic hierarchy at these times. We have defined the coefficients 
\begin{subequations}
\begin{align}
&\mu(t)=\ssp{\bbu^{\dag}_1(t)}{\bPhi(0,t)\punlm(t)} = \ez\frac{\ssp{\bbu_1(t)}{\punlm(t)}}{\ssp{\bbu_1(t)}{\bbu_1(t)}}, \label{eq:ampeqtgcmu} \\
&\nu(t)=\ssp{\bbu^{\dag}_1(t)}{\bPhi(0,t)\punlh(t)} = \ez\frac{\ssp{\bbu_1(t)}{\punlh(t)}}{\ssp{\bbu_1(t)}{\bbu_1(t)}}. \label{eq:ampeqtgcnu}
\end{align}
\end{subequations}

By evaluating Eq.~(\ref{eq:ampeqtg}) at $t=0$ and noticing that $\mu(0)=\nu(0)=0$ since $\punlh(0)=\punlm(0)=\bz$, we check that $A_1(0)=\psi\ssp{\bbu_{0,1}}{\pbu_0}$ in accordance with Eq.~(\ref{eq:asytg02}). Furthermore, by letting the rescaled amplitude of the initial condition $\psi$ tend towards zero, the nonlinear term in Eq.~(\ref{eq:ampeqtg}) becomes negligible such that
\begin{align}
A_1(\ez^{1/2} t) \rightarrow \psi \ssp{\bbu^{\dag}_1(t)}{\pbu_0} = \psi\frac{\ssp{\bbu_1(t)}{\ez\bPhi(t,0)\pbu_0}}{\ssp{\bbu_1(t)}{\bbu_1(t)}},
\label{eq:aeqtglin}
\end{align}
as $\psi \rightarrow 0$. We indeed recover the correct expression for $A_1$ in the linear regime.

In addition, the nonlinear term $(\mu+\nu)A_1|A_1|^2$ in Eq.~(\ref{eq:ampeqtg}) captures the leading-order nonlinear effects, making $A_1$ depart from Eq.~(\ref{eq:aeqtglin}) as the amplitude of the initial condition is increased through $\psi$. The coefficient $\mu(t)$ takes into account the fact that the transient response to the optimal initial condition, $A_1\bbu_1$, evolves over a base flow which is corrected by a time-dependent nonlinear (spatial) mean flow, equal to $|A_1|^2\obu^{|A_1|^2}_{2}$ at leading-order. Such a nonlinear correction to the base flow feeds back on the transient response and modifies its amplitude $A_1$. We note that this nonlinear coupling between the perturbation, and the mean flow carrying it, was also accounted for in Ref.~\cite{Pralits15}, and the basis for finding weakly nonlinear optimal initial conditions. The fact that the mean flow correction is simply proportional to $|A_1|^2$ despite resulting from the temporal integration of the interaction of the time-dependent $A_1\bbu_1$ with its complex conjugate. This would not have been true if $A_1$ had depended directly on $t$ instead of $\ez^{1/2} t$, for such $|A_1(t)|^2$ could not have got out of the temporal integral in computing the mean flow correction. Instead, the assumed time scale separation between $A_1(\ez^{1/2} t)$ and $\bbu_1(t)$ here guarantees that $|A_1(\ez^{1/2} t)|^2$ can get out of the integral and give the leading-order response (please see Eq.~(\ref{eq:u201aA})). In other words, the nonlinear term in Eq.~(\ref{eq:ampeqtg}) can be seen as resulting from a ``quasi-static" approximation, where the modulation of the mean flow distortion is sufficiently slow for the latter to be instantaneously proportional to $|A_1|^2$. The coefficient $\nu$ incorporates the fact that (spatial) leading-order, second harmonic field $A_1^2\obu^{A_1^2}_{2}$ interacts with $A_1^*\bbu^*_1$ to also bring about a nonlinear modification of $A_1$.    

We note that differentiating Eq.~(\ref{eq:ampeqtg}) with respect to time $t$, then keeping the linear term $\rmd_t A_1 = \ez^{1/2}\dot{A}_1$ at leading-order while absorbing in $O(\ez^{1/2})$ the nonlinear contribution in $\ez^{1/2}(\mu+\nu)(\dot{A}_1|A_1|^2 + ...)$, which results from the chain rule derivative, gives
\begin{align}
\rmd_t A_1 = \psi\rmd_t\pae{\ssp{\bbu^{\dag}_1}{\pbu_0}} - A_1|A_1|^2(\dot{\mu}+\dot{\nu}) + O(\ez^{1/2}), \label{eq:loduc}
\end{align}
where the $O(\ez^{1/2})$ term only contains higher-order nonlinear corrections. By neglecting these latter, Eq.~(\ref{eq:loduc}) is similar to that derived in our previous work in Ref.~\cite{Ducimetiere23}, Eq.~$(3.27)$ therein. In the latter article, we resorted to a perturbation of the inverse propagator to make it singular, embedded in a multiple time scales expansion closed with the Fredholm alternative. The resulting expression of the amplitude equation under the form $\rmd_t A_1= ...$ is inherent to the multiple-time scale method used in the latter article. Because of the term in $\ez^{1/2}(\mu+\nu)(\dot{A}_1|A_1|^2 + ...)$, absorbed at $O(\ez^{1/2})$ then neglected in Eq.~(\ref{eq:loduc}), but still present at leading order in Eq.~(\ref{eq:ampeqtg}), solving these two equations should thus lead to a $O(\ez^{1/2})$ difference between the results. 

It would have been possible to proceed with a weakly nonlinear expansion that follows directly from Eq.~(\ref{eq:ltg2dec}), and not from Eq.~(\ref{eq:dec44}). Specifically, in the expansion in Eq.~(\ref{eq:asytgU}), the $\set{\bbup_j(t)}_{j\geq 2}$ would have been replaced more simply by the $\set{\bbu_j(t)}_{j\geq 2}$, so as to become
\begin{align}
\pbU(t) =& \ez^{1/2} A_1(\ez^{1/2} t) \bbu_1(t) +\ez \sum_{j=2}^{N} A_j(\ez^{1/2} t) \bbu_j(t) \nonumber \\
& + \ez^{3/2} \sum_{j>N} A_j(\ez^{1/2} t)\bbu_j(t).
\label{eq:asytgUbis}
\end{align}
By performing the same calculation sequence, this would have led to the same expansion as in Eq.~(\ref{eq:atgj}), but where the summation term in Eq.~(\ref{eq:atgj}) would have become more simply $\sum_{j\geq2}\ga^{-1}_jA_j\bbu_{0,j}$. From here, an equation for $A_j$, linearly decoupled from all the others, could have been extracted upon taking the inner product simply with $\bbu_{0,j}$. This would have led to the equation
\begin{align}
0=&-A_1+\psi\ssp{\bbu_{0,1}}{\pbu_0}\nonumber\\
&-A_1|A_1|^2\ssp{\bbu_{0,1}}{\int_{0}^{t}\bPhi(0,s)\pae{\pfnlm(s)+\pfnlh(s)}\rmd s} \nonumber \\
&+O(\ez^{1/2}),
\label{eq:atgbis}
\end{align}
for $A_1$, perhaps more intuitive than Eq.~(\ref{eq:ampeqtg}). However, the coefficient multiplying the nonlinear term in Eq.~(\ref{eq:atgbis}) requires a backward integration of the direct problem (which is not the case for the coefficients defined in Eqs.~(\ref{eq:ampeqtgcmu}) and (\ref{eq:ampeqtgcnu})). This is notoriously known to lead to numerical difficulties, due to the existence of eigenvalues with very negative real parts, whose corresponding eigenmodes contribute to the backward integration in an exponentially diverging manner. In addition, we recall that the family $\set{\bbu_j(t)}_{j\geq 1}$ is not orthogonal for $0\leq t <t_o$, such that, for these times, the $A_1$ solving Eq.~(\ref{eq:atgbis}) would not have corresponded to the amplitude the orthogonal projection of the weakly nonlinear response onto $\bbu_1(t)$, whereas it is in the method developed above and leading to Eq.~(\ref{eq:ampeqtg}). A relative advantage of defining the $A_1$ as the amplitude of the orthogonal projection, is to relegate the error made in computing the transient gain to the next order for all $t\simeq t_o$ and not only at $t=t_o$. Indeed, the transient gain of the $(\al,\bet)$-component corresponding to the expansion in Eq.~(\ref{eq:asytg}) is
\begin{align}
&G_{(\al,\bet)}(t) \nonumber \\
&= \frac{||\pbU(t)||}{\ez^{3/2}\psi ||\pbu_0||} \nonumber \\
&= \frac{||\ez^{1/2} A_1\bbu_1 + \ez \sum_{j=2}^{N}A_j\bbup_j + O(\ez^{3/2})||}{\ez^{3/2}\psi ||\pbu_0||} \nonumber \\
&= \frac{\pae{ \ez|A_1|^2||\bbu_1||^2 + \ez^2 \sum_{j=2}^{N}|A_j|^2||\bbup_j||^2 + O(\ez^{3/2}) }^{1/2} }{\ez^{3/2}\psi} \nonumber \\
&= \frac{|A_1|||\bbu_1||}{\ez\psi}(1 + O(\ez)),
\label{eq:gtgw}
\end{align}
where the terms inside the $O(\ez)$ are veritably negligible for $t \simeq t_o$.  Now, if in the second equality in Eq.~(\ref{eq:gtgw}) the correction of $O(\ez)$ had not been orthogonal to $\bbu_1$, then it would have produced, in the third equality, a term of $O(\ez^{3/2})$ instead of $O(\ez^2)$. Thereby, the transient gain would have been $(\ez \psi)^{-1}|A_1|||\bbu_1||(1 + O(\ez^{1/2}))$ and the error made in approaching the gain with $(\ez \psi)^{-1}|A_1|||\bbu_1||$ would have been larger. That is one of the reasons for which the amplitude $A_1$ was defined as the amplitude of an orthogonal projection in this section: to minimize the error made in approaching the gain. Other reasons are that, as said, this makes easier the computation of nonlinear coefficients and guarantees a conceptual consistency with the two previous sections, where the amplitudes $A_1$ here too resulted from orthogonal projections.

\subsection{Application case: the Lamb-Oseen vortex flow \label{sec:tra3}}

Let us now appraise the validity of the amplitude equation (\ref{eq:ampeqtg}) on an application case. 
To this end, we show below some results for the same two-dimensional, axisymmetric, Lamb-Oseen vortex flow as considered in Ref.~\cite{Ducimetiere23}. The results below are obtained by solving Eq.~(\ref{eq:ampeqtg}) directly, whereas were obtained by solving its version under the form $\rmd_t A_1=...$ in Ref.~\cite{Ducimetiere23}. In both cases, only the leading order was considered in solving the amplitude equations. As said, there is thus a $O(\ez^{1/2})$ difference between the results below and that in Ref.~\cite{Ducimetiere23}, which is unimportant. The calculations were made in radial coordinates, and the base flow invariance was along the azimuthal direction, instead of being along the streamwise and spanwise directions. This, however, does not invalidate the calculations proposed above, and the same amplitude equation as Eq.~(\ref{eq:ampeqtg}) can be derived. Care should simply be brought in the definition of the inner product, as the radial coordinate typically implies including a Jacobian in the integral. The Lamb-Oseen vortex flow was shown numerically in Ref.~\cite{Rossi97} and experimentally in Refs.~\cite{Nolan99, Denoix94, VanHeijst89, VanHeijst91, Kloosterziel91}, to experience a subcritical transition toward a tripolar state. This makes it a sufficiently rich application case to evaluate both the strengths and limitations of the weakly nonlinear approach.  

In the rest of this section, the considered initial perturbation oscillates in the azimuthal direction with a wavenumber $\al=2$. Thereby, in the linear and weakly nonlinear paradigm, only the radial coordinate is resolved. The numerical simulations resolve the flow over a two-dimensional numerical grid. To compute the fully nonlinear gain, the flow component oscillating at $\al=2$ is then explicitly extracted using a Fourier series. We refer to Ref.~\cite{Ducimetiere23} for more information on the numerical procedure.

\begin{figure*}
\centering 
 \begin{subfigure}{1\columnwidth}
  \includegraphics[trim={0cm 0cm 0cm 0cm},clip,width=0.95\linewidth]{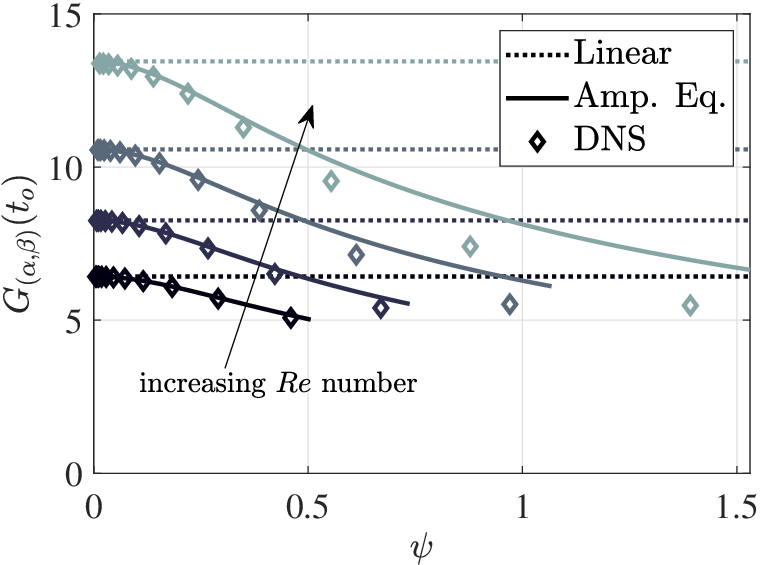}
  \caption{\label{fig:Taba}}
\end{subfigure}
\hfil
\begin{subfigure}{1\columnwidth}
  \includegraphics[trim={0cm 0cm 0cm 0cm},clip,width=0.95\linewidth]{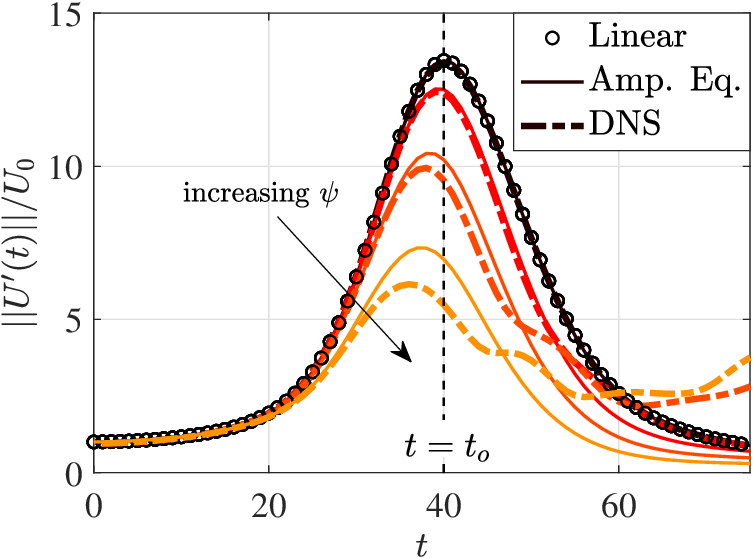}
  \caption{\label{fig:Tabb}}
\end{subfigure}
\caption{\subref{fig:Taba} Transient gain (at $t=t_o$) for the two-dimensional Lamb-Oseen vortex flow and a perturbation with azimuthal wavenumber $\al=2$. The initial condition structure is chosen as the (linear) optimal one, i.e., $\pbu_0=\bbu_{0,1}$. The gain is shown as a function of $\psi$, the rescaled amplitude of the initial condition. The gain is obtained by (i) a linear approach ($=1/\ez$, dotted line), (ii) a leading-order weakly nonlinear one ($=|A_1(\ez^{1/2} t_o)|/(\ez \psi)$ from Eq.~(\ref{eq:gtgw}), continuous line) and (iii) a fully nonlinear one (DNS, diamond markers). The temporal horizons are $t_o \in [25,30,35,40]$, corresponding to $\Ren \in [1.25,2.5,5,10] \times 10^3$, respectively. Larger Reynolds numbers correspond to lighter colors. \subref{fig:Tabb} For $\Ren=10^4$ only. Norm of the transient response $\pbU(t)$, divided by the amplitude of the initial condition $U_0 = \psi\ez^{3/2}$, as a function of time. This ratio is equal to the transient gain for $t=t_o$, highlighted by a vertical dashed line. Again it is obtained by (i) a linear approach ($=||\bbu_1(t)||/\ez$, circle markers), (ii) a leading-order weakly nonlinear one ($=||A_1(\ez^{1/2} t)\bbu_1(t)||/(\ez\psi)$, continuous lines), and (iii) a fully nonlinear one (DNS, dashed-dotted line). Larger values of $\psi\in[0.014, 0.22, 0.55, 1.40]$ correspond to lighter colors.}\label{fig:Tab}
\end{figure*}

We report in Fig.~\ref{fig:Taba} the evolution of the transient gain at $t=t_o$ as a function of $\psi$, the rescaled amplitude of the initial perturbation. 
We insist that all results refer to the azimuthal wavenumber $\al=2$. The considered temporal horizons are $t_o \in[25,30,35,40]$ for $\Ren \in [1.25,2.5,5,10]\times10^3$, respectively. Each temporal horizon is the one that leads to the largest optimal transient gain for its corresponding Reynolds number. Since it is an important assumption to the derivation of the amplitude equation (\ref{eq:ampeqtg}), we first checked that, for each considered set of parameters, the first sub-optimal gain is indeed of the order of the square root of the optimal one, and that the latter is large indeed. For instance, for $t_o=40$ and $\Ren=10^4$, we obtain indeed $G_2=5.22\approx 1.42\sqrt{13.44}$ and $G_1=13.44$. Furthermore, the initial condition is systematically chosen as the optimal one, i.e. $\pbu_0=\bbu_{0,1}$.  The three approaches for obtaining the gain, (i) linear ($=1/\ez$), (ii) leading-order weakly nonlinear ($=|A_1(\ez^{1/2} t_o)|/(\ez\psi)$ according to Eq.~(\ref{eq:gtgw}) and where $A_1$ solves Eq.~(\ref{eq:ampeqtg})) and (iii) fully nonlinear, are again compared in Fig.~\ref{fig:Taba}. It is the transient growth analogous to Figs.~\ref{fig:Hga}-\ref{fig:Hgc} and Fig.~\ref{fig:GpRe300}(b) for the stochastic and harmonic forcing problem, respectively. 

For all considered Reynolds numbers (larger $\Ren$ lighter nuances), the weakly nonlinear approach performs well in predicting the evolution of the fully nonlinear transient gains (at $t=t_o$). In particular, it accurately captures the decrease of the transient gain as the amplitude of the initial condition is increased. Further comparing the value of $\mu$ with that of $\nu$, indicates that this saturation effect is largely due to an (azimuthal) mean flow effect, more than due to a second harmonic effect (please see Fig.~$8$ in Ref.~\cite{Ducimetiere23}).

Figure~\ref{fig:Tabb} shows the evolution of $||\pbU(t)||/U_0$ over time, for a fixed $\Ren=10^4$ but different values of the amplitude of the initial condition. We recall $\pbU(t)$ to denote the transient response associated with the azimuthal wavenumber $\al=2$. To lighten the notation, we have also defined $U_0=\psi \ez^{3/2}$, the amplitude of the initial condition. This way, when evaluated at $t=t_o$, the ratio $||\pbU(t)||/U_0$ is directly the transient gain. In other terms, for $\Ren=10^4$ and at $t=t_o$, the information shown in Fig.~\ref{fig:Tabb} is redundant with that shown in Fig.~\ref{fig:Taba}. Figure~\ref{fig:Tabb}, however, also evaluates the relevance of the (leading-order) weakly nonlinear approach for different times than $t=t_o$. In the latter approach, $||\pbU(t)||/U_0$ reduces to $||A_1\bbu_1(t)||/(\psi \ez)$. For small times $t\leq t_o$, the weakly nonlinear approach performs well for all considered values of $\psi$. For times $t \simeq 0$, this good performance is simply explained by the fact that the response amplitude remains small there and thus nonlinearities do not become important. Indeed, the amplitude equation (\ref{eq:ampeqtg}) gives an exact result for $A_1$ in the linear regime, and since we have chosen $\pbu_0=\bbu_{0,1}$, no further sub-optimal trajectories are linearly excited. For larger times $t\geq t_o$ and large values of $\psi$, however, the fully nonlinear gains follow a temporal evolution that is completely different from that predicted by the weakly nonlinear approach. This corresponds in the DNS to the flow experiencing a subcritical transition toward a tripolar state, which is structurally very different. As argued in Ref.~\cite{Ducimetiere23}, this transition is triggered by the Reynolds stress divergence of the optimal response. The latter creates a very intense shear layer in the azimuthal mean flow, making the latter unstable to a shear instability. As mentioned in the previous section, the weakly nonlinear approach is inherently unable to predict such a transition, as it does not include an amplitude for a ``secondary" mode declaring on top of the optimal response. Thereby the predictions from the amplitude equation, for the optimal response only, depart from DNS results after the time needed for the shear-driven unstable mode to become dominant and yield leading-order nonlinear interactions with the optimal response.

In conclusion, here too, the amplitude equation performs well as long as the flow remains structurally close to the linear one, but fails to predict a subcritical transition and the associated complete restructuring of the flow (although we have shown in Ref.~\cite{Ducimetiere23} that the amplitude equation does predict a loss of solution at later times, indeed suggesting that the flow has transited there).

\section{Summary and perspective \label{sec:sp}}

\subsection{Summary}

In summary, we have considered strongly non-normal flow systems, governed by the Navier-Stokes equations, subjected to weak external excitations. From here, we have proposed a general method to reduce the Navier-Stokes equations to a weakly nonlinear one for the amplitude of the dominant nonmodal flow response. We have appraised independently three different types of external excitation: a harmonic forcing, a stochastic forcing, and eventually an initial perturbation. Despite being excitations of different natures, the reduction procedure followed from the same backbone principle, which we state as follows  
\begin{enumerate}
    \item  The nature of the excitation suggests the choice of a certain induced norm under which to measure the flow response. For instance, in the harmonic forcing problem, it is relevant to measure the component of the response oscillating at the forcing frequency under the $L^2$ norm.  In the stochastic forcing problem, however, the response is more appropriately measured under the variance norm, i.e. $\npt{\bullet}^2=\spt{\bullet}{\bullet}$.
    \item Once the induced norm for measuring the flow response has been selected, it is relevant to seek to optimize the flow response-to-forcing amplification. Note that this also requires selecting a norm under which to measure the forcing. It is generally chosen to be the same as for the response, but it need not be so (please see Eq.~(\ref{eq:ete}) for an example). Optimizing the response-to-forcing amplification amounts to constructing a family of excitation structures (e.g., forcing or initial condition) together with a family for their respective responses. The latter family of responses, in particular, is orthonormal under the inner product inducing the selected norm (for the response). For instance, for the stochastic forcing problem, solving Eq.~(\ref{eq:ete}) led to the family $\set{\bbu_j(t;\tet)}_{j \geq 1}$ of responses processes, orthonormal under the $\spt{\bullet}{\bullet}$ inner product. 
    \item Crucially, when measuring the response to some arbitrary structure under the selected induced norm, the contribution of the $j$th response constituting the family can be ranked according to the associated gain $G_j$. For strongly non-normal systems, it is often true that the contributions of only a few responses in the family dominate the linear response to some arbitrary structure. In other terms, it is often true that a small number of gains are much larger than all the others. It is sometimes called the ``low-rank" property. 
    \item If the ``low-rank" property holds, then projecting the Navier-Stokes equations in the low-dimensional subspace spanned by these few leading responses still gives the leading-order flow response, both in a linear and a weakly nonlinear regime at least. This projection is done under the inner product inducing the selected norm (for the response), e.g. $\spt{\bullet}{\bullet}$ for the stochastic response. In the phase space represented in the orthogonal basis formed by the responses set, as done in Fig.~\ref{fig:sketch}, the weakly nonlinear expansion we have proposed captures the curvature of the locus of the responses as the amplitude of the excitation is increased. Such curvature makes this locus depart from the linear response subspace, to which it is tangent in the linear regime.  
\end{enumerate}

To simplify the calculations and not obscure the general idea of the article, we have restricted the analysis to situations were, in the responses set, only the optimal response contributed the most to the overall linear response (i.e., $G_1 \gg G_2 >G_3>...$, equivalent to a ``unit-rank" assumption). In addition, in the stochastic forcing and transient growth problems, we have assumed the base flow to be parallel. Since no temporal symmetries generically exist in these two problems, this assumption was necessary for the base flow still to carry periodic perturbations, only in space instead of time. As an additional layer of simplification, only monochromatic perturbations (i.e., associated with a single wavenumber pair) have been considered at leading-order, which guaranteed that the first nonlinear interaction term to retroact on the fundamental pair was collected at the last order considered in the expansion, i.e., $O(\ez^{3/2})$, instead of the intermediate one, i.e., $O(\ez)$, thus making the calculations lighter. This spatial symmetry was not assumed in the harmonic forcing problem since the inherent temporal symmetry had the same effect on the calculations.

Made possible by these assumptions, we could formally derive two-degrees of freedom (since complex-valued) amplitude equations for the optimal harmonic response in Eq.~(\ref{eq:ampeq}), for the response to the optimal stochastic forcing in Eq.~(\ref{eq:astoc}), and for the response to the optimal initial condition in Eq.~(\ref{eq:ampeqtg}). In doing so, in stark contrast with classical techniques, we did not need to assume the existence of neutral or close-to-neutral eigenvalues of the linearized operator, and/or to assume the low-dimensionality of some master eigenspace in which to project the dynamics.   

For each of the three problems, we selected a different application case: a two-dimensional convectively unstable non-parallel flow for the harmonic forcing problem, a three-dimensional parallel one for the stochastic problem, and eventually a two-dimensional axisymmetric one for the transient growth problem. For sufficiently small excitation amplitudes, yet up to values large enough to depart from the linear regime, the amplitude equations systematically performed well in predicting the nonlinear evolution of the gain as the amplitude of the excitation was increased. This was found to be true whether the nonlinearities enhance, decrease, or induce a non-monotonic behavior of the gain. Owing to their simplicity of interpretation, they could also bring physical insights into the physical mechanisms responsible for such nonlinear evolutions of the gain. 

These good performances, however, did not systematically maintain themselves as the forcing amplitude was increased to large values, causing the nonlinear flow response to transit to a state structurally completely different from to the linear one. The amplitude equations mostly failed to predict such subcritical transitions of the flow. This was the case for the Couette-Poiseuille and the Lamb-Oseen vortex flows, subjected to a stochastic forcing and initial condition, respectively. Indeed, for these flows, the amplitude equation still predicted the branch stemming from the linear solution to exist and be unique, as the amplitude of the excitation was increased to arbitrarily large values. Meanwhile, the DNS showed that above some threshold value, the flow transited to another state. This defeat we interpreted as the subcritical transition being initiated by the apparition of a ``secondary" unstable mode on top of the nonmodal response, whose amplitude was not included in the weakly nonlinear expansion, whereas it must lead to leading-order interaction after a sufficient amount of time for the unstable mode to have grown. By contrast, in Fig.~$10$ of Ref.~\cite{Ducimetiere22}, the nonmodal amplitude equation in itself could predict the subcritical transition of the harmonically forced plane Poiseuille flow, by loss of stability of the solution or loss of the solution. The good performance of the amplitude equation in predicting there the subcritical behaviors was inherent to the fact that the response was mostly modal and linked to the presence of a weakly damped eigenvalue at the same frequency, and that this is precisely this eigenvalue which, by crossing the neutral axis, is responsible for the subcritical transition of the flow. 

Since the amplitude equations missed the subcritical transitions in the application cases considered in Secs.~\ref{sec:sto} and \ref{sec:tra}, they also could not predict the flow structures past the transitions. Again, this is a consequence of seeking the weakly nonlinear response as an asymptotic expansion where only the optimal response is included at leading order. This, by definition, condemns the former response to remain structurally close to the latter, such that our approach, at least in its current implementation, has very little freedom on the spatial structure. Although the latter feature is what makes possible the reduction to a very low-dimensional system, it is problematic in the subcritical transition scenario. There, the flow may go into a nonlinear state that is spatially much richer than that around which the expansion is performed. This enrichment may involve many sub-optimal structures (which were not relevant around the original attractor), but also, in the parallel shear flow cases considered Secs.~\ref{sec:sto} and \ref{sec:tra}, many additional wavenumber pairs (and not just harmonics of the fundamental pair, already included here). Indeed, in the subcritical transition to turbulence, for instance, existing literature indicates that the latter often involves nonlinear energy transfers between different wavenumber pairs \cite{Pringle12, Pralits15, Rigas21}. Overall, for our approach to make predictions regarding subcritical transitions, we believe that it should incorporate more degrees of freedom in space. This could be done either by including more structures at leading-order (please see discussion below) or, possibly, by introducing a slow space scale $X=\ez x$ such that the amplitudes eventually depend on it $A_1=A_1(X,...)$ \cite{Knobloch15}.

\subsection{Perspectives}

Although we are aware of the limitations of the approach adopted in this article, for the reasons mentioned above, or more fundamentally by the fact that it relies on a weakly nonlinear expansion, we do believe that it offers numerous perspectives. We list some of them below.   
\begin{itemize}
    \item From a methodological point of view, the approach could be generalized to flows for which some sub-optimal gains scale like the optimal one. This was found to happen in Ref.~\cite{Butler92, Blackburn08}, among many other examples. Thereby, the associated sub-optimal responses would need to be included in the leading order of the expansion, together with the optimal one. When the base flow is taken parallel, including different wavenumber pairs at the leading order could constitute another relevant extension of the method. These two types of enrichment of the leading-order spatial structure, i.e., by including sub-optimal responses and/or additional spatial wavenumbers, would result in a system of nonlinearly coupled amplitude equations which could characterize much richer nonlinear regimes. Such a system of amplitude equations, as they involve more degrees of freedom in space, could yield better performances in trying to predict subcritical transitions and subsequent states. 
    \item Another possible extension of the method is motivated by the fact that non-parallel base flows are also subject to strong nonmodal stochastic and transient responses, as illustrated in Refs.~\cite{Dergham2013} and \cite{Blackburn08}, respectively. Thereby, it would be useful to extend the calculations proposed in Secs.~\ref{sec:sto} and \ref{sec:tra} to non-parallel base flows, where a spatial Fourier decomposition of the perturbation is not possible anymore. This would be at the cost of slightly more complex calculations, resulting in the apparition of a quadratic term $\propto A_1^2$ in the final amplitude equation together with the cubic one $\propto A_1^3$ (as shown in Ref.~\cite{Ducimetiere22}, Eq.~($3.25$) therein).        
    \item Another perspective is motivated by the question of the role played by non-normality in noise-induced transition in fluid flows, which we believe to be of central physical importance. Please consider, for instance, a flow past the onset of a supercritical pitchfork bifurcation, where two attractors exist. Subjecting this flow to a stochastic forcing makes noise-induced transitions between the two attractors possible, after unpredictable and possibly extremely long times. This motivated us to derive in Ref.~\cite{Ducimetiere24} a stochastically forced amplitude equation for the symmetry-breaking eigenmode, which could reproduce transition statistics at a meager numerical cost. 
    However, in Ref.~\cite{Ducimetiere24} we only considered a narrow-band, slowly varying noise. The latter only triggered the slow dynamics, which was reduced on the symmetry-breaking eigenmode. It would be of great interest now to consider the flow response to a noise that is not narrow-banded. This was recently done in Ref.~\cite{McMullen24} for the flow past a cylinder, where the noise was an intrinsic feature of real molecular fluids. In the latter article, the noise was decomposed as the sum of a slow plus a fast component. The latter triggers a fast dynamics, which, in our context, could be greatly amplified by non-normality. Thereby, it would be interesting to couple the modal approach in Ref.~\cite{Ducimetiere24} with the nonmodal one proposed in Sec.~\ref{sec:sto}, to be able to consider the flow response to an arbitrary noise and not only a narrow-band one. The ensuing low-dimensional system would couple an equation for the symmetry-breaking eigenmode, with an equation for the part of the response triggered by non-normality. In this, it would quantify the extent to which the nonlinear coupling between the modal and the nonmodal part of the stochastic response is an essential ingredient to noise-induced transitions. 
    \item From a more fundamental perspective, we believe that a precise link with the center manifold theory could possibly be drawn, at least for the harmonic forcing problem. Indeed, in Fig.~\ref{fig:sketch}, the locus of the weakly nonlinear response (dashed line) may be seen as analogous to a center manifold for the modal paradigm, whereas $V_h$ would be analogous to the center eigenspace (the former being tangent to the latter in the linear regime). Thereby, proceeding along the calculation lines proposed in Refs.~\cite{Carini15, Negi24}, but rewriting the Navier-Stokes equations in the singular mode basis instead of the eigenbasis, then seeking the part of the nonlinear correction of solution which is contained in the sub-optimals subspace (i.e. $V^{\perp}_{1}$), as a graph over the amplitude along the optimal subspace (i.e. $V_{1}$) and the forcing amplitude, could be another manner to derive Eq.~(\ref{eq:ampeq}).  
    \item Eventually, the method presented in this paper was applied, but not restricted to the Navier-Stokes equations. The scope of the proposed approach also covers other physical systems governed by other stochastic partial or ordinary differential equations, as long as the linearized operator presents a high degree of non-normality. This includes solid mechanics \cite{Ha21}, network science \cite{Asllani18, Asllani18b, Zachary20, Sornette23}, the modeling of music instruments \cite{Weinreich77, Politzer15}, pattern formation \cite{Biancalani17, Vaclav17}, astrophysics \cite{Jaramillo21}, population dynamics \cite{Neubert97}, and many others fields of physics. \\
\end{itemize}

\begin{acknowledgments}
The central idea that all our previous results could be derived without relying on an operator perturbation, but more simply by using a projection method, came from Richard Kerswell on the occasion of the PhD defense of the first author. Thereby, the authors greatly acknowledge Prof. Kerswell for his fruitful suggestion, without which the present paper would not have existed. \\  

This work has been supported by the Swiss National Science Foundation, F.G. under grant number $200341$ and Y.-M. D. under grant number $225429$ (Postdoc.Mobility scholarship).
\end{acknowledgments}

\appendix


\begin{widetext}
\section{Generalization of the calculations of Sec.~\ref{sec:subhar2} with asymptotically expanded amplitudes} \label{app:Fuji}
By expanding each amplitude $A_j$ for $j\geq 1$ according to Eq.~(\ref{eq:apw}), the expansion of the harmonic response in Eq.~(\ref{eq:asy1}) becomes
\begin{align}
\ve{\bU}{P}=& \ve{\bUb}{\pb} + \ez^{1/2}\sae{A^{(0)}_1\ve{\cbu_1}{\ccp_1}\ei{\wz} + \cc} + \ez\sae{A^{(1)}_1\ve{\cbu_1}{\ccp_1}\ei{\wz} + \sum_{j=2}^{N}A^{(0)}_j\ve{\cbu_j}{\ccp_j}\ei{\wz} + \cc + \ve{\obu_2}{\op_2}} \nonumber \\
& + \ez^{3/2}\sae{A^{(2)}_1\ve{\cbu_1}{\ccp_1}\ei{\wz} + \sum_{j=2}^{N}A^{(1)}_j\ve{\cbu_j}{\ccp_j}\ei{\wz} + \sum_{j>N}A^{(0)}_j\ve{\cbu_j}{\ccp_j}\ei{\wz} + \cc + \ve{\obu_3}{\op_3}} + O\pae{\ez^{2}}.
\label{eq:asy1APP}
\end{align}
By further injecting Eq.~(\ref{eq:asy1APP}) in the Navier-Stokes equations (\ref{eq:NSf}) and remembering the definition in Eq.~(\ref{eq:depp}), the expansion for the fundamental component (i.e. oscillating at $\wz$) is obtained as
\begin{align}
\bz =& \ez^{1/2}\sae{-A^{(0)}_1 G^{-1}_1 \bP^T \cbf_1} +\ez \sae{-A^{(1)}_1 G^{-1}_1 \bP^T \cbf_1 -\sum_{j=2}^{N}A^{(0)}_j G^{-1}_j \bP^T \cbf_j} +\ez^{3/2} \Biggl[-A^{(2)}_1 G^{-1}_1 \bP^T \cbf_1 \nonumber \\
&-\sum_{j=2}^{N}A^{(1)}_j G^{-1}_j \bP^T \cbf_j - \sum_{j > N}A^{(0)}_jG^{-1}_j \bP^T \cbf_j - 2\bP^{T}\sae{A^{(0)}_1\bC(\cbu_1,\obu_{2,0})  + A^{(0),*}_1\bC(\cbu_1^*,\obu_{2,2})} + \phi \bP^{T} \hbf_h \Biggr] +O(\ez^2).
\label{eq:aF12APP}
\end{align}
By then injecting in Eq.~(\ref{eq:aF12APP}) the scalings of the gains in Eq.~(\ref{eq:scg}), as well as the zero and second harmonics fields solutions in Eqs.~(\ref{eq:so20}) and (\ref{eq:so22}), and eventually applying $\bP$, the expansion becomes
\begin{align}
 \bz = \ez^{3/2} & \Bigg [ -A^{(0)}_1\cbf_1 -\sum_{j\geq 2} \ga^{-1}_j A^{(0)}_j \cbf_j - 2A^{(0)}_1|A^{(0)}_1|^2\sae{\bC(\cbu_1,\obu_{2,0}^{|A_1|^2})  + \bC(\cbu_1^*,\obu_{2,2}^{A_1^2})} + \phi \hbf_h \Bigg ] + O(\ez^2).
\label{eq:aF13APP}
\end{align}
Eq.~(\ref{eq:aF13APP}) is the same as Eq.~(\ref{eq:aF13}) where $A_j$ has been replaced by its leading-order contribution $A^{(0)}_j$, for every $j\geq 1$. Thereby, projecting Eq.~(\ref{eq:aF13APP}) over $\cbf_j$ to extract an equation for $A^{(0)}_j$, then using again the expansion in Eq.~(\ref{eq:apw}) to rewrite this equation in terms of $A_j$, produces at leading-order an equation for $A_j$ which is the same as Eq.~(\ref{eq:ampeq}) for $j=1$ and the same as Eq.~(\ref{eq:ampeqj}) for $j\geq 2$. 
\end{widetext}

\section{Response to a stochastic forcing: numerical details \label{app:stocnum}}

The linear and weakly nonlinear computations are performed on \texttt{Matlab} by resorting to a pseudospectral method with the Gauss-Lobatto (``endpoints") collocation points and $200$ Chebyshev polynomials. The frequency domain is discretized using $N$ uniformly distributed points within 
\begin{align}
\om \in \sae{-\frac{\om_{m}}{2}+\Delta \om;\frac{\om_{m}}{2}} \quad \mbox{with} \quad \Delta \om=\frac{\om_m}{N},
\label{eq:fint}
\end{align}
and $\om_{m}=2\pi/\Delta t$ the sampling frequency (by definition two times the cut-off, or ``Nyquist" frequency). All integrations over the frequencies are performed using the function \texttt{trapz} pre-implemented in \texttt{Matlab}. The leading eigenvalues of $\bBi$ and associated eigenmodes are computed by using the \texttt{eigs} algorithm of \texttt{Matlab} with a $10^{-10}$ tolerance. 

Computing the weakly nonlinear coefficients $\mu$ and $\nu$ requires applying the convection operator, whose quadratic nonlinearity corresponds to a convolution product in the Fourier domain. Thereby, the nonlinearity can excite frequencies that are not externally forced. For this reason, it is important to select $\om_m \gg \om_c$, where we recall $\om_c$ as the noise band-limiting cut-off frequency, chosen as $\om_c=\pi$. The coefficients are found upon implementing the following algorithm:    
\begin{enumerate}
    \item Choose the values for $N$ and $\om_m$, which set the discretization of the frequency interval according to Eq.~(\ref{eq:fint}). The values $N=2^{14}=16384$ and $\om_m = 8\pi$ have been found more than enough for the numerical convergence of the results.   
    \item Over the discrete set of frequencies, pre-compute once and for all the deterministic scalar $\ez$ (inverse of the optimal gain) and fields $\cbf_1$ (optimal forcing structure) and $\bR_{1}(\om)\cbf_1$.
    \item Draw randomly a band-limited white noise $\hxi(\om;\tet)$ and deduce $\cbu_1(\om;\tet) = \ez \hxi(\om;\tet)\bR_{1}(\om)\cbf_1$.\label{draw}
     \item Determine $\bbu_1(t;\tet)=\Foi{\cbu_1(\om;\tet)}$ and then compute in the Fourier domain the second-order responses $\hobu_{2,0}^{|A_1|^2}(\om;\tet)$ and $\hobu_{2,2}^{A_1^2}(\om;\tet)$ according to Eqs.~(\ref{eq:so20s}) and (\ref{eq:so22s}), respectively.
     \item Determine $\obu_{2,0}^{|A_1|^2}(t;\tet) = \Foi{\hobu_{2,0}^{|A_1|^2}(\om;\tet)}$ and $\obu_{2,2}^{A_1^2}(t;\tet) = \Foi{\hobu_{2,2}^{A_1^2}(\om;\tet)}$. In turns, compute $\fnlm(\om;\tet)$ and $\fnlh(\om;\tet)$ according to Eq.~(\ref{eq:defnl}), and then their linear responses in the Fourier domain $\bR_{1}(\om)\fnlm(\om;\tet)$ and $\bR_{1}(\om)\fnlh(\om;\tet)$. 
     \item Compute the argument of the expected values in Eqs.~(\ref{eq:cofmus}) and (\ref{eq:cofnus}), consisting of the frequency integral of the inner product of known fields (times a prefactor), and store their values.
     \item Update the expected values (i.e., the ensemble averages) in Eqs.~(\ref{eq:cofmus}) and (\ref{eq:cofnus}) and return to step~\ref{draw}. until convergence of both $\mu$ and $\nu$. 
\end{enumerate}
The direct and inverse Fourier transforms are performed by using the ``\texttt{fft}" and ``\texttt{ifft}" commands of \texttt{Matlab}, with the proper normalization to ensure the satisfaction of Parseval's theorem. 

Fully nonlinear simulations (DNS) were performed using the spectral element open source solver \texttt{Nek5000}. The numerical grid was three-dimensional with $N_x=16$ elements in the streamwise direction (with periodic boundary conditions), $N_y = 10$ elements in the crosswise direction (imposed null velocity at $y=-1$ and unit velocity at $y=1$), and $N_z=16$ elements in the spanwise direction (with periodic boundary conditions). The time step of the simulations was $\Delta t = 4\times 10^{-3}$. Convergence of the results by refining the mesh and reducing the time step has been checked.

\section{Response to an initial perturbation: second-order solutions for the velocity fields}\label{app:tgA}

The solution to Eq.~(\ref{eq:atg0}) for the velocity field $\obu_{2,0}$ can formally be written 
\begin{align}
&\obu_{2,0}(t)=- \bPhi_0(t,0) \nonumber  \\
&\times\int_{0}^{t} |A_1(\ez^{1/2} s)|^2 \underbrace{\bPhi_0(0,s)\pae{\bC_1(\bbu^*_1(s),\bbu_1(s)) +\cc}}_{=-\rmd\bg(s)/\rmd s}.
\label{eq:u201A}
\end{align}
where $\bPhi_0$ is the propagator associated with $\bL_0$.
In deriving Eq.~(\ref{eq:u201A}), we have used the classical result stipulating that the solution of some generic forced linear system 
\begin{align}
(\bB\pa_t - \bL_a)\ve{\bu}{p}=\bP^T\bff,
\label{eq:linrefA}
\end{align}
is
\begin{align}
\bu(t) = \bPhi_a(t,0)\bu(0) + \bPhi_a(t,0)\int_{0}^{t} \bPhi_a(0,s)\bff(s)\rmd s,
\label{eq:linref}
\end{align}
with $\bPhi_a$ the propagator associated with $\bL_a$ (``$a$" being an arbitrary index). We have also used that $\obu_{2,0}(0)=\bz$. Integrating Eq.~(\ref{eq:u201A}) by parts results in 
\begin{align}
&\obu_{2,0}(t) =  \bPhi_0(t,0)\sae{|A_1(\ez^{1/2} s)|^2\bg(s)}^{s=t}_{s=0} \nonumber \\
&- \ez^{1/2} \bPhi_0(t,0) \int_{0}^{t} (\dot{A}_1(\ez^{1/2} s)A^*_1(\ez^{1/2} s) +\cc)\bg(s)\rmd s \nonumber \\
&=|A_1(\ez^{1/2} t)|^2 \bPhi_0(t,0)\bg(t) + O(\ez^{1/2}),
\label{eq:u201aA}
\end{align}
where the dot denotes the derivative with respect to the natural variable, and where we have defined $\bg(y)$ as
\begin{align}
\bg(y) = -\int_{0}^{y}\bPhi_0(0,s)\pae{\bC_1(\bbu^*_1(s),\bbu_1(s)) +\cc}\rmd s.
\label{eq:u201bA}
\end{align}
In other terms, it is possible to write
\begin{align}
\obu_{2,0}(t) = |A_1(\ez^{1/2} t)|^2 \obu^{|A_1|^2}_{2,0}(t) + O(\ez^{1/2}),
\label{eq:u201cA}
\end{align}
where
\begin{align}
\obu^{|A_1|^2}_{2,0}(t) =&\bPhi_0(t,0) \bg(t)\\
=& -\bPhi_0(t,0) \nonumber \\
&\times \int_{0}^{t}\bPhi_0(0,s)\pae{\bC_1(\bbu^*_1(s),\bbu_1(s)) +\cc}\rmd s.
\label{eq:u201cATRIS}
\end{align}
Equivalently, according to Eq.~(\ref{eq:linref}), the field $\obu^{|A_1|^2}_{2,0}(t)$ solves
\begin{align}
\bB\pa_t\ve{\obu^{|A_1|^2}_{2,0}}{\op^{|A_1|^2}_{2,0}} = \bL_{0}\ve{\obu^{|A_1|^2}_{2,0}}{\op^{|A_1|^2}_{2,0}} - \bP^{T}\pae{\bC_{1}(\bbu^*_1,\bbu_1) +\cc}.
\label{eq:u201cABIS}
\end{align}
with $\obu^{|A_1|^2}_{2,0}(0)=\bz$. We find similarly
\begin{align}
\obu_{2,2}(t) = A_1(\ez^{1/2} t)^2 \obu^{A_1^2}_{2,2}(t) + O(\ez^{1/2}),
\label{eq:u201dA}
\end{align}
where $\obu^{A_1^2}_{2,2}(t)$ solves
\begin{align}
\bB\pa_t\ve{\obu^{A_1^2}_{2,2}}{\op^{A_1^2}_{2,2}} = \bL_{2}\ve{\obu^{A_1^2}_{2,2}}{\op^{A_1^2}_{2,2}} - \bP^{T}\bC_{1}(\bbu_1,\bbu_1).
\label{eq:u201dABIS}
\end{align}

\section{Response to an initial perturbation: rewriting of the asymptotic expansion} \label{app:tgB}
Further substituting Eqs.~(\ref{eq:u201c}) and (\ref{eq:u201d}) in Eq.~(\ref{eq:atg1}) leads to

\begin{widetext}
\begin{align}
\bz =& \ez^{1/2}\sae{-\pae{\bB\pa_t-\bL_{1}} \pae{A_1\ve{\bbu_1}{\bp_1}}} +\ez \sae{-\sum_{j=2}^{N} \pae{\bB\pa_t-\bL_{1}} \pae{A_j\ve{\bbup_j}{\bpp_j}}}\nonumber  \\
&+\ez^{3/2} \sae{ -\sum_{j > N} \pae{\bB\pa_t-\bL_{1}} \pae{A_j\ve{\bbup_j}{\bpp_j}} - A_1|A_1|^2\bP^{T}(\pfnlm + \pfnlh) } + O(\ez^2),
\label{eq:atg12A}
\end{align}
where we have defined the nonlinear interactions forcing terms
\begin{align}
\pfnlm(t) &= \bC_{0}(\bbu_1(t),\obu^{|A_1|^2}_{2,0}(t))  + \bC_{1}(\obu^{|A_1|^2}_{2,0}(t),\bbu_1(t)), \nonumber \\
\pfnlh(t) &= \bC_{2}(\bbu_1^*(t),\obu^{A_1^2}_{2,2}(t)) + \bC_{-1}(\obu^{A_1^2}_{2,2}(t),\bbu_1^*(t)).
\label{eq:atgcA}
\end{align}
It is then useful to get rid of the operator $(\bB\pa_t-\bL_1)$ in Eq.~(\ref{eq:atg12A}). To do this, let us first multiply Eq.~(\ref{eq:linrefA}) by $\bP$ and inject the resulting $\bff(t) = \bP(\bB \pa_t-\bL_a)(\bu(t),p(t))^T$ in Eq.~(\ref{eq:linref}). This gives, 
\begin{align}
\bu(t) &= \bPhi_a(t,0)\bu(0) + \bPhi_a(t,0)\int_{0}^{t} \bPhi_a(0,s)\bP(\bB \pa_s-\bL_a)\ve{\bu(s)}{p(s)}\rmd s \Rightarrow \nonumber \\
\bPhi_a(0,t)\bu(t) &= \bu(0) + \int_{0}^{t} \bPhi_a(0,s)\bP(\bB \pa_s-\bL_a)\ve{\bu(s)}{p(s)}\rmd s \Rightarrow \nonumber \\
\bPhi_a(t,0)\rmd_t\pae{\bPhi_a(0,t)\bu(t)} &= \bP(\bB \pa_t-\bL_a)\ve{\bu(t)}{p(t)},
\label{eq:invprop}
\end{align}
which is a well-known result in dynamical systems theory. Thereby, applying $\bP$ to Eq.~(\ref{eq:atg12A}) and then using Eq.~(\ref{eq:invprop}) leads to the following rewriting
\begin{align}
\bz & = \ez^{1/2} \sae{ -\bPhi(t,0)\rmd_t\pae{\bPhi(0,t)A_1\bbu_1}} + \ez \sae{-\sum_{j=2}^{N}  \bPhi(t,0)\rmd_t\pae{\bPhi(0,t)A_j\bbup_j}}\nonumber  \\
& +\ez^{3/2} \sae{ -\sum_{j > N}\bPhi(t,0)\rmd_t\pae{\bPhi(0,t)A_j\bbup_j} - A_1|A_1|^2\pae{\pfnlm + \pfnlh}} + O(\ez^2) \label{eq:atgaA}
\end{align}
where, as in the main text, $\bPhi$ designates the propagator associated with $\bL_1$. Subsequently applying $\bPhi(0,t)$ to Eq.~(\ref{eq:atgaA}) and then integrating the resulting expression between $0$ and $t$, leads to, terms by terms,
\begin{align}
\int_{0}^{t} \underbrace{\bPhi(0,s)\bPhi(s,0)}_{=\bI}\rmd_s\pae{\bPhi(0,s)A_1(\ez^{1/2} s)\bbu_1(s)}\rmd s &=\ez\bbu_{0,1}\int_{0}^{t}\rmd_s \pae{A_1(\ez^{1/2} s)}\rmd s =\ez\bbu_{0,1}\pae{ A_1(\ez^{1/2} t)- A_1(0)},
\label{eq:atgdA}
\end{align}
since $\bPhi(0,s)\bbu_1(s)=\bbu_1(0)=\ez\bbu_{0,1}$ by definition. We also obtain for $j \geq 2$,
\begin{align}
&\int_{0}^{t} \underbrace{\bPhi(0,s)\bPhi(s,0)}_{=\bI}\rmd_s\pae{\bPhi(0,s)A_j(\ez^{1/2} s)\bbup_j(s)}\rmd s \nonumber  \\
&=A_j(\ez^{1/2} t)\bPhi(0,t)\bbup_j(t)-A_j(0)G_j^{-1}\bbu_{0,j} \nonumber \\
&=A_j(\ez^{1/2} t)\pae{G_j^{-1}\bbu_{0,j}-\frac{\ssp{\bbu_1(t)}{\bbu_j(t)}}{\ssp{\bbu_1(t)}{\bbu_1(t)}}\ez\bbu_{0,1}}-A_j(0)G_j^{-1}\bbu_{0,j} \nonumber \\
&=A_j(\ez^{1/2} t)\pae{G_j^{-1}\bbu_{0,j}-\frac{\ssp{\bbu_1(t)}{G_j^{-1}\bPhi(t,0)\bbu_{0,j}}}{\ssp{\bbu_1(t)}{\bbu_1(t)}}\ez \bbu_{0,1}}-A_j(0)G_j^{-1}\bbu_{0,j} \nonumber \\
&=A_j(\ez^{1/2} t)G_j^{-1}\pae{\bbu_{0,j} - \ssp{\bbu^{\dag}_1(t)}{\bbu_{0,j}}\bbu_{0,1}} -A_j(0)G_j^{-1}\bbu_{0,j}, 
\label{eq:atgeA}
\end{align}
where we have used that $\bbup_j(0) = \bbu_j(0) = G_j^{-1}\bbu_{0,j}$ as well as the definition in Eq.~(\ref{eq:dec7}). We have also defined the adjoint response $\bbu_1^{\dag}(t)=\ez\bPhi(t,0)^{\dag}\bbu_{1}(t)/\ssp{\bbu_1(t)}{\bbu_1(t)}$, normalized such that $\ssp{\bbu_1^{\dag}(t)}{\bbu_{0,1}}=1$. Injecting Eqs.~(\ref{eq:atgdA}) and (\ref{eq:atgeA}) in Eq.~(\ref{eq:atgaA}) leads to
\begin{align}
\bz  =& \ez^{3/2} \Biggl[-A_1\bbu_{0,1}-\sum_{j \geq 2}\ga_j^{-1}A_j\pae{\bbu_{0,j} - \ssp{\bbu^{\dag}_1(t)}{\bbu_{0,j}}\bbu_{0,1}} + \underbrace{\pae{A_1(0)\bbu_{0,1} +\sum_{j\geq 2}A_j(0)\ga_j^{-1}\bbu_{0,j}}}_{\mbox{= $\psi\pbu_0 $ by Eq.~(\ref{eq:asytg04})} } \nonumber \\
& - \int_{0}^{t}A_1|A_1|^2\bPhi(0,s)\pae{\pfnlm + \pfnlh} \rmd s \Biggr]  + O(\ez^2). \label{eq:atgfA}
\end{align}
By following the same reasoning as in Eq.~(\ref{eq:u201aA}), it is possible to expand
\begin{align}
&\int_{0}^{t}A_1(\ez^{1/2} s)|A_1(\ez^{1/2} s)|^2\bPhi(0,s)\pae{\pfnlm(s) + \pfnlh(s)} \rmd s \nonumber \\
&=A_1(\ez^{1/2} t)|A_1(\ez^{1/2} t)|^2\pae{\int_{0}^{t}\bPhi(0,s)\pfnlm(s)\rmd s + \int_{0}^{t}\bPhi(0,s)\pfnlh(s)\rmd s} + O(\ez^{1/2}) \nonumber \\
&=A_1(\ez^{1/2} t)|A_1(\ez^{1/2} t)|^2\bPhi(0,t)\pae{\punlm(t) + \punlh(t)} + O(\ez^{1/2}),
\label{eq:atggA}
\end{align}
where we have introduced
\begin{align}
&\punlm(t) = \bPhi(t,0)\int_{0}^{t}\bPhi(0,s)\pfnlm(s)\rmd s, \quad \mbox{and} \quad \punlh(t) = \bPhi(t,0)\int_{0}^{t}\bPhi(0,s)\pfnlh(s)\rmd s.
\label{eq:athA}
\end{align}
Equivalently, $\punlm(t)$ and $\punlh(t)$ solve
\begin{align}
(\bB\pa_t - \bL_{1})\ve{\punlm}{\pnlm} = \bP^T\pfnlm, \quad \mbox{and} \quad (\bB\pa_t - \bL_{1})\ve{\punlh}{\pnlh} = \bP^T\pfnlh, 
\label{eq:atiA}
\end{align}
respectively. By further identifying with $\psi\pbu_0$ the term between parenthesis involving the $\set{A_j(0)}_{j\geq1}$ in Eq.~(\ref{eq:atgfA}), according to Eq.~(\ref{eq:asytg0}), the expansion is simplified as
\begin{align}
\bz & = \ez^{3/2} \Big[ -A_1\bbu_{0,1}-\sum_{j \geq 2}\ga_j^{-1}A_j\pae{\bbu_{0,j} - \ssp{\bbu^{\dag}_1(t)}{\bbu_{0,j}}\bbu_{0,1}} + \psi\pbu_0 - A_1|A_1|^2\bPhi(0,t)\pae{\punlm + \punlh} \Big] + O(\ez^2).\label{eq:atgjA}
\end{align}
\end{widetext}

\bibliography{apssamp}

\providecommand{\noopsort}[1]{}\providecommand{\singleletter}[1]{#1}%
\begin{thebibliography}{120}%
\makeatletter
\providecommand \@ifxundefined [1]{%
 \@ifx{#1\undefined}
}%
\providecommand \@ifnum [1]{%
 \ifnum #1\expandafter \@firstoftwo
 \else \expandafter \@secondoftwo
 \fi
}%
\providecommand \@ifx [1]{%
 \ifx #1\expandafter \@firstoftwo
 \else \expandafter \@secondoftwo
 \fi
}%
\providecommand \natexlab [1]{#1}%
\providecommand \enquote  [1]{``#1''}%
\providecommand \bibnamefont  [1]{#1}%
\providecommand \bibfnamefont [1]{#1}%
\providecommand \citenamefont [1]{#1}%
\providecommand \href@noop [0]{\@secondoftwo}%
\providecommand \href [0]{\begingroup \@sanitize@url \@href}%
\providecommand \@href[1]{\@@startlink{#1}\@@href}%
\providecommand \@@href[1]{\endgroup#1\@@endlink}%
\providecommand \@sanitize@url [0]{\catcode `\\12\catcode `\$12\catcode `\&12\catcode `\#12\catcode `\^12\catcode `\_12\catcode `\%12\relax}%
\providecommand \@@startlink[1]{}%
\providecommand \@@endlink[0]{}%
\providecommand \url  [0]{\begingroup\@sanitize@url \@url }%
\providecommand \@url [1]{\endgroup\@href {#1}{\urlprefix }}%
\providecommand \urlprefix  [0]{URL }%
\providecommand \Eprint [0]{\href }%
\providecommand \doibase [0]{https://doi.org/}%
\providecommand \selectlanguage [0]{\@gobble}%
\providecommand \bibinfo  [0]{\@secondoftwo}%
\providecommand \bibfield  [0]{\@secondoftwo}%
\providecommand \translation [1]{[#1]}%
\providecommand \BibitemOpen [0]{}%
\providecommand \bibitemStop [0]{}%
\providecommand \bibitemNoStop [0]{.\EOS\space}%
\providecommand \EOS [0]{\spacefactor3000\relax}%
\providecommand \BibitemShut  [1]{\csname bibitem#1\endcsname}%
\let\auto@bib@innerbib\@empty
\bibitem [{\citenamefont {Meliga}(2018)}]{MeligHDR}%
  \BibitemOpen
  \bibfield  {author} {\bibinfo {author} {\bibfnamefont {P.}~\bibnamefont {Meliga}},\ }\emph {\bibinfo {title} {Linear and semi-linear analysis of large-scale oscillations in laminar and turbulent open flows}},\ \href@noop {} {\bibinfo {type} {Habilitation à diriger des recherches}},\ \bibinfo  {school} {Aix-Marseille Université} (\bibinfo {year} {2018})\BibitemShut {NoStop}%
\bibitem [{\citenamefont {Schmid}\ and\ \citenamefont {Henningson}(2001)}]{SH01}%
  \BibitemOpen
  \bibfield  {author} {\bibinfo {author} {\bibfnamefont {P.}~\bibnamefont {Schmid}}\ and\ \bibinfo {author} {\bibfnamefont {D.}~\bibnamefont {Henningson}},\ }\href@noop {} {\emph {\bibinfo {title} {Stability and transition in shear flows}}}\ (\bibinfo  {publisher} {Springer},\ \bibinfo {year} {2001})\BibitemShut {NoStop}%
\bibitem [{\citenamefont {Åkervik}\ \emph {et~al.}(2007)\citenamefont {Åkervik}, \citenamefont {Hœpffner}, \citenamefont {Ehrenstein},\ and\ \citenamefont {Henningson}}]{Akervik07}%
  \BibitemOpen
  \bibfield  {author} {\bibinfo {author} {\bibfnamefont {E.}~\bibnamefont {Åkervik}}, \bibinfo {author} {\bibfnamefont {J.}~\bibnamefont {Hœpffner}}, \bibinfo {author} {\bibfnamefont {U.}~\bibnamefont {Ehrenstein}},\ and\ \bibinfo {author} {\bibfnamefont {D.}~\bibnamefont {Henningson}},\ }\bibfield  {title} {\bibinfo {title} {Optimal growth, model reduction and control in a separated boundary-layer flow using global eigenmodes},\ }\href@noop {} {\bibfield  {journal} {\bibinfo  {journal} {J. Fluid Mech.}\ }\textbf {\bibinfo {volume} {579}},\ \bibinfo {pages} {305–314} (\bibinfo {year} {2007})}\BibitemShut {NoStop}%
\bibitem [{\citenamefont {Chomaz}(2005)}]{Chomaz05}%
  \BibitemOpen
  \bibfield  {author} {\bibinfo {author} {\bibfnamefont {J.-M.}\ \bibnamefont {Chomaz}},\ }\bibfield  {title} {\bibinfo {title} {Global instabilities in spatially developing flows: non-normality and nonlinearity},\ }\href@noop {} {\bibfield  {journal} {\bibinfo  {journal} {Annu. Rev. Fluid Mech.}\ }\textbf {\bibinfo {volume} {37}},\ \bibinfo {pages} {357} (\bibinfo {year} {2005})}\BibitemShut {NoStop}%
\bibitem [{\citenamefont {Reddy}\ and\ \citenamefont {Henningson}(1993)}]{Reddy93}%
  \BibitemOpen
  \bibfield  {author} {\bibinfo {author} {\bibfnamefont {S.}~\bibnamefont {Reddy}}\ and\ \bibinfo {author} {\bibfnamefont {D.}~\bibnamefont {Henningson}},\ }\bibfield  {title} {\bibinfo {title} {Energy growth in viscous channel flows},\ }\href@noop {} {\bibfield  {journal} {\bibinfo  {journal} {J. Fluid Mech.}\ }\textbf {\bibinfo {volume} {252}},\ \bibinfo {pages} {209–238} (\bibinfo {year} {1993})}\BibitemShut {NoStop}%
\bibitem [{\citenamefont {Trefethen}\ \emph {et~al.}(1993)\citenamefont {Trefethen}, \citenamefont {Trefethen}, \citenamefont {Reddy},\ and\ \citenamefont {Driscoll}}]{Trefethen93}%
  \BibitemOpen
  \bibfield  {author} {\bibinfo {author} {\bibfnamefont {L.}~\bibnamefont {Trefethen}}, \bibinfo {author} {\bibfnamefont {A.}~\bibnamefont {Trefethen}}, \bibinfo {author} {\bibfnamefont {S.}~\bibnamefont {Reddy}},\ and\ \bibinfo {author} {\bibfnamefont {T.}~\bibnamefont {Driscoll}},\ }\bibfield  {title} {\bibinfo {title} {Hydrodynamic stability without eigenvalues},\ }\href@noop {} {\bibfield  {journal} {\bibinfo  {journal} {Science}\ }\textbf {\bibinfo {volume} {261}},\ \bibinfo {pages} {578} (\bibinfo {year} {1993})}\BibitemShut {NoStop}%
\bibitem [{\citenamefont {Kerswell}(2018)}]{KerswellAnnuRev2018}%
  \BibitemOpen
  \bibfield  {author} {\bibinfo {author} {\bibfnamefont {R.}~\bibnamefont {Kerswell}},\ }\bibfield  {title} {\bibinfo {title} {Nonlinear nonmodal stability theory},\ }\href@noop {} {\bibfield  {journal} {\bibinfo  {journal} {Annu. Rev. Fluid Mech.}\ }\textbf {\bibinfo {volume} {50}},\ \bibinfo {pages} {319} (\bibinfo {year} {2018})}\BibitemShut {NoStop}%
\bibitem [{\citenamefont {Heifetz}\ and\ \citenamefont {Methven}(2005)}]{Heifetz05}%
  \BibitemOpen
  \bibfield  {author} {\bibinfo {author} {\bibfnamefont {E.}~\bibnamefont {Heifetz}}\ and\ \bibinfo {author} {\bibfnamefont {J.}~\bibnamefont {Methven}},\ }\bibfield  {title} {\bibinfo {title} {{Relating optimal growth to counterpropagating Rossby waves in shear instability}},\ }\href@noop {} {\bibfield  {journal} {\bibinfo  {journal} {Phys. Fluids}\ }\textbf {\bibinfo {volume} {17}},\ \bibinfo {pages} {064107} (\bibinfo {year} {2005})}\BibitemShut {NoStop}%
\bibitem [{\citenamefont {Butler}\ and\ \citenamefont {Farrell}(1992)}]{Butler92}%
  \BibitemOpen
  \bibfield  {author} {\bibinfo {author} {\bibfnamefont {K.}~\bibnamefont {Butler}}\ and\ \bibinfo {author} {\bibfnamefont {B.}~\bibnamefont {Farrell}},\ }\bibfield  {title} {\bibinfo {title} {Three‐dimensional optimal perturbations in viscous shear flow},\ }\href@noop {} {\bibfield  {journal} {\bibinfo  {journal} {Phys. Fluids A}\ }\textbf {\bibinfo {volume} {4}},\ \bibinfo {pages} {1637} (\bibinfo {year} {1992})}\BibitemShut {NoStop}%
\bibitem [{\citenamefont {Schmid}\ and\ \citenamefont {Henningson}(1994)}]{Schmid94}%
  \BibitemOpen
  \bibfield  {author} {\bibinfo {author} {\bibfnamefont {P.}~\bibnamefont {Schmid}}\ and\ \bibinfo {author} {\bibfnamefont {D.}~\bibnamefont {Henningson}},\ }\bibfield  {title} {\bibinfo {title} {Optimal energy density growth in {H}agen–{P}oiseuille flow},\ }\href@noop {} {\bibfield  {journal} {\bibinfo  {journal} {J. Fluid Mech.}\ }\textbf {\bibinfo {volume} {277}},\ \bibinfo {pages} {197–225} (\bibinfo {year} {1994})}\BibitemShut {NoStop}%
\bibitem [{\citenamefont {Corbett}\ and\ \citenamefont {Bottaro}(2000)}]{Corbett00}%
  \BibitemOpen
  \bibfield  {author} {\bibinfo {author} {\bibfnamefont {P.}~\bibnamefont {Corbett}}\ and\ \bibinfo {author} {\bibfnamefont {A.}~\bibnamefont {Bottaro}},\ }\bibfield  {title} {\bibinfo {title} {Optimal perturbations for boundary layers subject to stream-wise pressure gradient},\ }\href@noop {} {\bibfield  {journal} {\bibinfo  {journal} {Phys. Fluids}\ }\textbf {\bibinfo {volume} {12}},\ \bibinfo {pages} {120} (\bibinfo {year} {2000})}\BibitemShut {NoStop}%
\bibitem [{\citenamefont {Cossu}\ and\ \citenamefont {Chomaz}(1997)}]{Cossu97}%
  \BibitemOpen
  \bibfield  {author} {\bibinfo {author} {\bibfnamefont {C.}~\bibnamefont {Cossu}}\ and\ \bibinfo {author} {\bibfnamefont {J.-M.}\ \bibnamefont {Chomaz}},\ }\bibfield  {title} {\bibinfo {title} {Global measures of local convective instabilities},\ }\href@noop {} {\bibfield  {journal} {\bibinfo  {journal} {Phys. Rev. Lett.}\ }\textbf {\bibinfo {volume} {78}},\ \bibinfo {pages} {4387} (\bibinfo {year} {1997})}\BibitemShut {NoStop}%
\bibitem [{\citenamefont {Blackburn}\ \emph {et~al.}(2008)\citenamefont {Blackburn}, \citenamefont {Barkley},\ and\ \citenamefont {Sherwin}}]{Blackburn08}%
  \BibitemOpen
  \bibfield  {author} {\bibinfo {author} {\bibfnamefont {H.}~\bibnamefont {Blackburn}}, \bibinfo {author} {\bibfnamefont {D.}~\bibnamefont {Barkley}},\ and\ \bibinfo {author} {\bibfnamefont {S.}~\bibnamefont {Sherwin}},\ }\bibfield  {title} {\bibinfo {title} {Convective instability and transient growth in flow over a backward-facing step},\ }\href@noop {} {\bibfield  {journal} {\bibinfo  {journal} {J. Fluid Mech.}\ }\textbf {\bibinfo {volume} {603}},\ \bibinfo {pages} {271–304} (\bibinfo {year} {2008})}\BibitemShut {NoStop}%
\bibitem [{\citenamefont {Marquet}\ \emph {et~al.}(2008)\citenamefont {Marquet}, \citenamefont {Sipp}, \citenamefont {Chomaz},\ and\ \citenamefont {Jacquin}}]{Marquet08}%
  \BibitemOpen
  \bibfield  {author} {\bibinfo {author} {\bibfnamefont {O.}~\bibnamefont {Marquet}}, \bibinfo {author} {\bibfnamefont {D.}~\bibnamefont {Sipp}}, \bibinfo {author} {\bibfnamefont {J.-M.}\ \bibnamefont {Chomaz}},\ and\ \bibinfo {author} {\bibfnamefont {L.}~\bibnamefont {Jacquin}},\ }\bibfield  {title} {\bibinfo {title} {Amplifier and resonator dynamics of a low-{R}eynolds-number recirculation bubble in a global framework},\ }\href@noop {} {\bibfield  {journal} {\bibinfo  {journal} {J. Fluid Mech.}\ }\textbf {\bibinfo {volume} {605}},\ \bibinfo {pages} {429–443} (\bibinfo {year} {2008})}\BibitemShut {NoStop}%
\bibitem [{\citenamefont {Marquet}\ \emph {et~al.}(2009)\citenamefont {Marquet}, \citenamefont {Lombardi}, \citenamefont {Chomaz}, \citenamefont {Sipp},\ and\ \citenamefont {Jacquin}}]{Marquet09}%
  \BibitemOpen
  \bibfield  {author} {\bibinfo {author} {\bibfnamefont {O.}~\bibnamefont {Marquet}}, \bibinfo {author} {\bibfnamefont {M.}~\bibnamefont {Lombardi}}, \bibinfo {author} {\bibfnamefont {J.-M.}\ \bibnamefont {Chomaz}}, \bibinfo {author} {\bibfnamefont {D.}~\bibnamefont {Sipp}},\ and\ \bibinfo {author} {\bibfnamefont {L.}~\bibnamefont {Jacquin}},\ }\bibfield  {title} {\bibinfo {title} {Direct and adjoint global modes of a recirculation bubble: lift-up and convective non-normalities},\ }\href@noop {} {\bibfield  {journal} {\bibinfo  {journal} {J. Fluid Mech.}\ }\textbf {\bibinfo {volume} {622}},\ \bibinfo {pages} {1–21} (\bibinfo {year} {2009})}\BibitemShut {NoStop}%
\bibitem [{\citenamefont {Trefethen}\ and\ \citenamefont {Embree}(2005)}]{Tref05}%
  \BibitemOpen
  \bibfield  {author} {\bibinfo {author} {\bibfnamefont {L.}~\bibnamefont {Trefethen}}\ and\ \bibinfo {author} {\bibfnamefont {M.}~\bibnamefont {Embree}},\ }\href@noop {} {\emph {\bibinfo {title} {Spectra and pseudo-spectra}}}\ (\bibinfo  {publisher} {Princeton University Press},\ \bibinfo {year} {2005})\BibitemShut {NoStop}%
\bibitem [{\citenamefont {Ehrenstein}\ and\ \citenamefont {Gallaire}(2005)}]{Ehrenstein05}%
  \BibitemOpen
  \bibfield  {author} {\bibinfo {author} {\bibfnamefont {U.}~\bibnamefont {Ehrenstein}}\ and\ \bibinfo {author} {\bibfnamefont {F.}~\bibnamefont {Gallaire}},\ }\bibfield  {title} {\bibinfo {title} {On two-dimensional temporal modes in spatially evolving open flows: the flat-plate boundary layer},\ }\href@noop {} {\bibfield  {journal} {\bibinfo  {journal} {J. Fluid Mech.}\ }\textbf {\bibinfo {volume} {536}},\ \bibinfo {pages} {209} (\bibinfo {year} {2005})}\BibitemShut {NoStop}%
\bibitem [{\citenamefont {Åkervik}\ \emph {et~al.}(2008)\citenamefont {Åkervik}, \citenamefont {Ehrenstein}, \citenamefont {Gallaire},\ and\ \citenamefont {Henningson}}]{Akervik08}%
  \BibitemOpen
  \bibfield  {author} {\bibinfo {author} {\bibfnamefont {E.}~\bibnamefont {Åkervik}}, \bibinfo {author} {\bibfnamefont {U.}~\bibnamefont {Ehrenstein}}, \bibinfo {author} {\bibfnamefont {F.}~\bibnamefont {Gallaire}},\ and\ \bibinfo {author} {\bibfnamefont {D.}~\bibnamefont {Henningson}},\ }\bibfield  {title} {\bibinfo {title} {Global two-dimensional stability measures of the flat plate boundary-layer flow},\ }\href@noop {} {\bibfield  {journal} {\bibinfo  {journal} {Eur. J. Mech. B/Fluids}\ }\textbf {\bibinfo {volume} {27}},\ \bibinfo {pages} {501} (\bibinfo {year} {2008})}\BibitemShut {NoStop}%
\bibitem [{\citenamefont {Monokrousos}\ \emph {et~al.}(2010)\citenamefont {Monokrousos}, \citenamefont {Åkervik}, \citenamefont {Brandt},\ and\ \citenamefont {Henningson}}]{Monkrousos10}%
  \BibitemOpen
  \bibfield  {author} {\bibinfo {author} {\bibfnamefont {A.}~\bibnamefont {Monokrousos}}, \bibinfo {author} {\bibfnamefont {E.}~\bibnamefont {Åkervik}}, \bibinfo {author} {\bibfnamefont {L.}~\bibnamefont {Brandt}},\ and\ \bibinfo {author} {\bibfnamefont {D.}~\bibnamefont {Henningson}},\ }\bibfield  {title} {\bibinfo {title} {Global three-dimensional optimal disturbances in the {B}lasius boundary-layer flow using time-steppers},\ }\href@noop {} {\bibfield  {journal} {\bibinfo  {journal} {J. Fluid Mech.}\ }\textbf {\bibinfo {volume} {650}},\ \bibinfo {pages} {181} (\bibinfo {year} {2010})}\BibitemShut {NoStop}%
\bibitem [{\citenamefont {Ehrenstein}\ and\ \citenamefont {Gallaire}(2008)}]{Ehrenstein08}%
  \BibitemOpen
  \bibfield  {author} {\bibinfo {author} {\bibfnamefont {U.}~\bibnamefont {Ehrenstein}}\ and\ \bibinfo {author} {\bibfnamefont {F.}~\bibnamefont {Gallaire}},\ }\bibfield  {title} {\bibinfo {title} {Two-dimensional global low-frequency oscillations in a separating boundary-layer flow},\ }\href@noop {} {\bibfield  {journal} {\bibinfo  {journal} {J. Fluid Mech.}\ }\textbf {\bibinfo {volume} {614}},\ \bibinfo {pages} {315} (\bibinfo {year} {2008})}\BibitemShut {NoStop}%
\bibitem [{\citenamefont {Alizard}\ \emph {et~al.}(2009)\citenamefont {Alizard}, \citenamefont {Cherubini},\ and\ \citenamefont {Robinet}}]{Alizard09}%
  \BibitemOpen
  \bibfield  {author} {\bibinfo {author} {\bibfnamefont {F.}~\bibnamefont {Alizard}}, \bibinfo {author} {\bibfnamefont {S.}~\bibnamefont {Cherubini}},\ and\ \bibinfo {author} {\bibfnamefont {J.-C.}\ \bibnamefont {Robinet}},\ }\bibfield  {title} {\bibinfo {title} {Sensitivity and optimal forcing response in separated boundary layer flows},\ }\href@noop {} {\bibfield  {journal} {\bibinfo  {journal} {Phys. Fluids}\ }\textbf {\bibinfo {volume} {21}},\ \bibinfo {pages} {064108} (\bibinfo {year} {2009})}\BibitemShut {NoStop}%
\bibitem [{\citenamefont {Pringle}\ \emph {et~al.}(2012)\citenamefont {Pringle}, \citenamefont {Willis},\ and\ \citenamefont {Kerswell}}]{Pringle12}%
  \BibitemOpen
  \bibfield  {author} {\bibinfo {author} {\bibfnamefont {C.}~\bibnamefont {Pringle}}, \bibinfo {author} {\bibfnamefont {A.}~\bibnamefont {Willis}},\ and\ \bibinfo {author} {\bibfnamefont {R.}~\bibnamefont {Kerswell}},\ }\bibfield  {title} {\bibinfo {title} {Minimal seeds for shear flow turbulence: using nonlinear transient growth to touch the edge of chaos},\ }\href@noop {} {\bibfield  {journal} {\bibinfo  {journal} {J. Fluid Mech.}\ }\textbf {\bibinfo {volume} {702}},\ \bibinfo {pages} {415} (\bibinfo {year} {2012})}\BibitemShut {NoStop}%
\bibitem [{\citenamefont {Farano}\ \emph {et~al.}(2015)\citenamefont {Farano}, \citenamefont {Cherubini}, \citenamefont {Robinet},\ and\ \citenamefont {De~Palma}}]{Farano15}%
  \BibitemOpen
  \bibfield  {author} {\bibinfo {author} {\bibfnamefont {M.}~\bibnamefont {Farano}}, \bibinfo {author} {\bibfnamefont {S.}~\bibnamefont {Cherubini}}, \bibinfo {author} {\bibfnamefont {J.-C.}\ \bibnamefont {Robinet}},\ and\ \bibinfo {author} {\bibfnamefont {P.}~\bibnamefont {De~Palma}},\ }\bibfield  {title} {\bibinfo {title} {{H}airpin-like optimal perturbations in plane {P}oiseuille flow},\ }\href@noop {} {\bibfield  {journal} {\bibinfo  {journal} {J. Fluid Mech.}\ }\textbf {\bibinfo {volume} {775}},\ \bibinfo {pages} {R2} (\bibinfo {year} {2015})}\BibitemShut {NoStop}%
\bibitem [{\citenamefont {Rossi}\ \emph {et~al.}(1997)\citenamefont {Rossi}, \citenamefont {Lingevitch},\ and\ \citenamefont {Bernoff}}]{Rossi97}%
  \BibitemOpen
  \bibfield  {author} {\bibinfo {author} {\bibfnamefont {L.}~\bibnamefont {Rossi}}, \bibinfo {author} {\bibfnamefont {J.}~\bibnamefont {Lingevitch}},\ and\ \bibinfo {author} {\bibfnamefont {A.}~\bibnamefont {Bernoff}},\ }\bibfield  {title} {\bibinfo {title} {Quasi-steady monopole and tripole attractors for relaxing vortices},\ }\href@noop {} {\bibfield  {journal} {\bibinfo  {journal} {Phys. Fluids}\ }\textbf {\bibinfo {volume} {9}},\ \bibinfo {pages} {2329} (\bibinfo {year} {1997})}\BibitemShut {NoStop}%
\bibitem [{\citenamefont {Ducimeti\`ere}\ and\ \citenamefont {Gallaire}(2023)}]{Ducimetiere23}%
  \BibitemOpen
  \bibfield  {author} {\bibinfo {author} {\bibfnamefont {Y.-M.}\ \bibnamefont {Ducimeti\`ere}}\ and\ \bibinfo {author} {\bibfnamefont {F.}~\bibnamefont {Gallaire}},\ }\bibfield  {title} {\bibinfo {title} {A weakly nonlinear amplitude equation approach to the bypass transition in the two-dimensional {L}amb–{O}seen vortex},\ }\href@noop {} {\bibfield  {journal} {\bibinfo  {journal} {J. Fluid Mech.}\ }\textbf {\bibinfo {volume} {976}},\ \bibinfo {pages} {A10} (\bibinfo {year} {2023})}\BibitemShut {NoStop}%
\bibitem [{\citenamefont {Matsubara}\ and\ \citenamefont {Alfredsson}(2001)}]{Matsubara01}%
  \BibitemOpen
  \bibfield  {author} {\bibinfo {author} {\bibfnamefont {M.}~\bibnamefont {Matsubara}}\ and\ \bibinfo {author} {\bibfnamefont {P.}~\bibnamefont {Alfredsson}},\ }\bibfield  {title} {\bibinfo {title} {Disturbance growth in boundary layers subjected to free-stream turbulence},\ }\href@noop {} {\bibfield  {journal} {\bibinfo  {journal} {J. Fluid Mech.}\ }\textbf {\bibinfo {volume} {430}},\ \bibinfo {pages} {149–168} (\bibinfo {year} {2001})}\BibitemShut {NoStop}%
\bibitem [{\citenamefont {Andersson}\ \emph {et~al.}(1999)\citenamefont {Andersson}, \citenamefont {Berggren},\ and\ \citenamefont {Henningson}}]{Andersson99}%
  \BibitemOpen
  \bibfield  {author} {\bibinfo {author} {\bibfnamefont {P.}~\bibnamefont {Andersson}}, \bibinfo {author} {\bibfnamefont {M.}~\bibnamefont {Berggren}},\ and\ \bibinfo {author} {\bibfnamefont {D.}~\bibnamefont {Henningson}},\ }\bibfield  {title} {\bibinfo {title} {{Optimal disturbances and bypass transition in boundary layers}},\ }\href@noop {} {\bibfield  {journal} {\bibinfo  {journal} {Phys. Fluids}\ }\textbf {\bibinfo {volume} {11}},\ \bibinfo {pages} {134} (\bibinfo {year} {1999})}\BibitemShut {NoStop}%
\bibitem [{\citenamefont {Brandt}\ \emph {et~al.}(2004)\citenamefont {Brandt}, \citenamefont {Schlatter},\ and\ \citenamefont {Henningson}}]{Brandt04}%
  \BibitemOpen
  \bibfield  {author} {\bibinfo {author} {\bibfnamefont {L.}~\bibnamefont {Brandt}}, \bibinfo {author} {\bibfnamefont {P.}~\bibnamefont {Schlatter}},\ and\ \bibinfo {author} {\bibfnamefont {D.}~\bibnamefont {Henningson}},\ }\bibfield  {title} {\bibinfo {title} {Transition in boundary layers subject to free-stream turbulence},\ }\href@noop {} {\bibfield  {journal} {\bibinfo  {journal} {J. Fluid Mech.}\ }\textbf {\bibinfo {volume} {517}},\ \bibinfo {pages} {167–198} (\bibinfo {year} {2004})}\BibitemShut {NoStop}%
\bibitem [{\citenamefont {Song}\ \emph {et~al.}(2024)\citenamefont {Song}, \citenamefont {Dong},\ and\ \citenamefont {Zhao}}]{Song24}%
  \BibitemOpen
  \bibfield  {author} {\bibinfo {author} {\bibfnamefont {R.}~\bibnamefont {Song}}, \bibinfo {author} {\bibfnamefont {M.}~\bibnamefont {Dong}},\ and\ \bibinfo {author} {\bibfnamefont {L.}~\bibnamefont {Zhao}},\ }\bibfield  {title} {\bibinfo {title} {Principle of fundamental resonance in hypersonic boundary layers: an asymptotic viewpoint},\ }\href {https://doi.org/10.1017/jfm.2023.1043} {\bibfield  {journal} {\bibinfo  {journal} {J. Fluid Mech.}\ }\textbf {\bibinfo {volume} {978}},\ \bibinfo {pages} {A30} (\bibinfo {year} {2024})}\BibitemShut {NoStop}%
\bibitem [{\citenamefont {Hamilton}\ \emph {et~al.}(1995)\citenamefont {Hamilton}, \citenamefont {Kim},\ and\ \citenamefont {Waleffe}}]{Hamilton95}%
  \BibitemOpen
  \bibfield  {author} {\bibinfo {author} {\bibfnamefont {J.}~\bibnamefont {Hamilton}}, \bibinfo {author} {\bibfnamefont {J.}~\bibnamefont {Kim}},\ and\ \bibinfo {author} {\bibfnamefont {F.}~\bibnamefont {Waleffe}},\ }\bibfield  {title} {\bibinfo {title} {Regeneration mechanisms of near-wall turbulence structures},\ }\href@noop {} {\bibfield  {journal} {\bibinfo  {journal} {J. Fluid Mech.}\ }\textbf {\bibinfo {volume} {287}},\ \bibinfo {pages} {317–348} (\bibinfo {year} {1995})}\BibitemShut {NoStop}%
\bibitem [{\citenamefont {Waleffe}(1997)}]{Waleffe97}%
  \BibitemOpen
  \bibfield  {author} {\bibinfo {author} {\bibfnamefont {F.}~\bibnamefont {Waleffe}},\ }\bibfield  {title} {\bibinfo {title} {{On a self-sustaining process in shear flows}},\ }\href@noop {} {\bibfield  {journal} {\bibinfo  {journal} {Phys. Fluids}\ }\textbf {\bibinfo {volume} {9}},\ \bibinfo {pages} {883} (\bibinfo {year} {1997})}\BibitemShut {NoStop}%
\bibitem [{\citenamefont {Cherubini}\ \emph {et~al.}(2010)\citenamefont {Cherubini}, \citenamefont {De~Palma}, \citenamefont {Robinet},\ and\ \citenamefont {Bottaro}}]{Cherubini10}%
  \BibitemOpen
  \bibfield  {author} {\bibinfo {author} {\bibfnamefont {S.}~\bibnamefont {Cherubini}}, \bibinfo {author} {\bibfnamefont {P.}~\bibnamefont {De~Palma}}, \bibinfo {author} {\bibfnamefont {J.-C.}\ \bibnamefont {Robinet}},\ and\ \bibinfo {author} {\bibfnamefont {A.}~\bibnamefont {Bottaro}},\ }\bibfield  {title} {\bibinfo {title} {Rapid path to transition via nonlinear localized optimal perturbations in a boundary-layer flow},\ }\href@noop {} {\bibfield  {journal} {\bibinfo  {journal} {Phys. Rev. E}\ }\textbf {\bibinfo {volume} {82}},\ \bibinfo {pages} {066302} (\bibinfo {year} {2010})}\BibitemShut {NoStop}%
\bibitem [{\citenamefont {Cherubini}\ \emph {et~al.}(2011)\citenamefont {Cherubini}, \citenamefont {De~Palma}, \citenamefont {Robinet},\ and\ \citenamefont {Bottaro}}]{Cherubini11}%
  \BibitemOpen
  \bibfield  {author} {\bibinfo {author} {\bibfnamefont {S.}~\bibnamefont {Cherubini}}, \bibinfo {author} {\bibfnamefont {P.}~\bibnamefont {De~Palma}}, \bibinfo {author} {\bibfnamefont {J.-C.}\ \bibnamefont {Robinet}},\ and\ \bibinfo {author} {\bibfnamefont {A.}~\bibnamefont {Bottaro}},\ }\bibfield  {title} {\bibinfo {title} {The minimal seed of turbulent transition in the boundary layer},\ }\href@noop {} {\bibfield  {journal} {\bibinfo  {journal} {J. Fluid Mech.}\ }\textbf {\bibinfo {volume} {689}},\ \bibinfo {pages} {221–253} (\bibinfo {year} {2011})}\BibitemShut {NoStop}%
\bibitem [{\citenamefont {Duguet}\ \emph {et~al.}(2010)\citenamefont {Duguet}, \citenamefont {Brandt},\ and\ \citenamefont {Larsson}}]{Duguet10}%
  \BibitemOpen
  \bibfield  {author} {\bibinfo {author} {\bibfnamefont {Y.}~\bibnamefont {Duguet}}, \bibinfo {author} {\bibfnamefont {L.}~\bibnamefont {Brandt}},\ and\ \bibinfo {author} {\bibfnamefont {B.}~\bibnamefont {Larsson}},\ }\bibfield  {title} {\bibinfo {title} {Towards minimal perturbations in transitional plane {C}ouette flow},\ }\href@noop {} {\bibfield  {journal} {\bibinfo  {journal} {Phys. Rev. E}\ }\textbf {\bibinfo {volume} {82}},\ \bibinfo {pages} {026316} (\bibinfo {year} {2010})}\BibitemShut {NoStop}%
\bibitem [{\citenamefont {Monokrousos}\ \emph {et~al.}(2011)\citenamefont {Monokrousos}, \citenamefont {Bottaro}, \citenamefont {Brandt}, \citenamefont {Di~Vita},\ and\ \citenamefont {Henningson}}]{Monokrousos11}%
  \BibitemOpen
  \bibfield  {author} {\bibinfo {author} {\bibfnamefont {A.}~\bibnamefont {Monokrousos}}, \bibinfo {author} {\bibfnamefont {A.}~\bibnamefont {Bottaro}}, \bibinfo {author} {\bibfnamefont {L.}~\bibnamefont {Brandt}}, \bibinfo {author} {\bibfnamefont {A.}~\bibnamefont {Di~Vita}},\ and\ \bibinfo {author} {\bibfnamefont {D.}~\bibnamefont {Henningson}},\ }\bibfield  {title} {\bibinfo {title} {Nonequilibrium thermodynamics and the optimal path to turbulence in shear flows},\ }\href@noop {} {\bibfield  {journal} {\bibinfo  {journal} {Phys. Rev. Lett.}\ }\textbf {\bibinfo {volume} {106}},\ \bibinfo {pages} {134502} (\bibinfo {year} {2011})}\BibitemShut {NoStop}%
\bibitem [{\citenamefont {Rabin}\ \emph {et~al.}(2012)\citenamefont {Rabin}, \citenamefont {Caulfield},\ and\ \citenamefont {Kerswell}}]{Rabin12}%
  \BibitemOpen
  \bibfield  {author} {\bibinfo {author} {\bibfnamefont {S.}~\bibnamefont {Rabin}}, \bibinfo {author} {\bibfnamefont {C.}~\bibnamefont {Caulfield}},\ and\ \bibinfo {author} {\bibfnamefont {R.}~\bibnamefont {Kerswell}},\ }\bibfield  {title} {\bibinfo {title} {Triggering turbulence efficiently in plane {C}ouette flow},\ }\href@noop {} {\bibfield  {journal} {\bibinfo  {journal} {J. Fluid Mech.}\ }\textbf {\bibinfo {volume} {712}},\ \bibinfo {pages} {244–272} (\bibinfo {year} {2012})}\BibitemShut {NoStop}%
\bibitem [{\citenamefont {Cherubini}\ and\ \citenamefont {De~Palma}(2013)}]{Cherubini13}%
  \BibitemOpen
  \bibfield  {author} {\bibinfo {author} {\bibfnamefont {S.}~\bibnamefont {Cherubini}}\ and\ \bibinfo {author} {\bibfnamefont {P.}~\bibnamefont {De~Palma}},\ }\bibfield  {title} {\bibinfo {title} {Nonlinear optimal perturbations in a {C}ouette flow: bursting and transition},\ }\href@noop {} {\bibfield  {journal} {\bibinfo  {journal} {J. Fluid Mech.}\ }\textbf {\bibinfo {volume} {716}},\ \bibinfo {pages} {251–279} (\bibinfo {year} {2013})}\BibitemShut {NoStop}%
\bibitem [{\citenamefont {Farano}\ \emph {et~al.}(2016)\citenamefont {Farano}, \citenamefont {Cherubini}, \citenamefont {Robinet},\ and\ \citenamefont {De~Palma}}]{Farano16}%
  \BibitemOpen
  \bibfield  {author} {\bibinfo {author} {\bibfnamefont {M.}~\bibnamefont {Farano}}, \bibinfo {author} {\bibfnamefont {S.}~\bibnamefont {Cherubini}}, \bibinfo {author} {\bibfnamefont {J.-C.}\ \bibnamefont {Robinet}},\ and\ \bibinfo {author} {\bibfnamefont {P.}~\bibnamefont {De~Palma}},\ }\bibfield  {title} {\bibinfo {title} {Subcritical transition scenarios via linear and nonlinear localized optimal perturbations in plane {P}oiseuille flow},\ }\href@noop {} {\bibfield  {journal} {\bibinfo  {journal} {Fluid Dyn. Res.}\ }\textbf {\bibinfo {volume} {48}},\ \bibinfo {pages} {061409} (\bibinfo {year} {2016})}\BibitemShut {NoStop}%
\bibitem [{\citenamefont {Schmid}(2007)}]{Schmid07}%
  \BibitemOpen
  \bibfield  {author} {\bibinfo {author} {\bibfnamefont {P.}~\bibnamefont {Schmid}},\ }\bibfield  {title} {\bibinfo {title} {Nonmodal stability theory},\ }\href@noop {} {\bibfield  {journal} {\bibinfo  {journal} {Annu. Rev. Fluid Mech.}\ }\textbf {\bibinfo {volume} {39}},\ \bibinfo {pages} {129} (\bibinfo {year} {2007})}\BibitemShut {NoStop}%
\bibitem [{\citenamefont {Sipp}(2012)}]{Sipp12}%
  \BibitemOpen
  \bibfield  {author} {\bibinfo {author} {\bibfnamefont {D.}~\bibnamefont {Sipp}},\ }\bibfield  {title} {\bibinfo {title} {Open-loop control of cavity oscillations with harmonic forcings},\ }\href@noop {} {\bibfield  {journal} {\bibinfo  {journal} {J. Fluid Mech.}\ }\textbf {\bibinfo {volume} {708}},\ \bibinfo {pages} {439} (\bibinfo {year} {2012})}\BibitemShut {NoStop}%
\bibitem [{\citenamefont {Boujo}\ and\ \citenamefont {Gallaire}(2015)}]{Boujo15}%
  \BibitemOpen
  \bibfield  {author} {\bibinfo {author} {\bibfnamefont {E.}~\bibnamefont {Boujo}}\ and\ \bibinfo {author} {\bibfnamefont {F.}~\bibnamefont {Gallaire}},\ }\bibfield  {title} {\bibinfo {title} {Sensitivity and open-loop control of stochastic response in a noise amplifier flow: the backward-facing step},\ }\href@noop {} {\bibfield  {journal} {\bibinfo  {journal} {J. Fluid Mech.}\ }\textbf {\bibinfo {volume} {762}},\ \bibinfo {pages} {361–392} (\bibinfo {year} {2015})}\BibitemShut {NoStop}%
\bibitem [{\citenamefont {Garnaud}\ \emph {et~al.}(2013)\citenamefont {Garnaud}, \citenamefont {Lesshafft}, \citenamefont {Schmid},\ and\ \citenamefont {Huerre}}]{Garnaud13B}%
  \BibitemOpen
  \bibfield  {author} {\bibinfo {author} {\bibfnamefont {X.}~\bibnamefont {Garnaud}}, \bibinfo {author} {\bibfnamefont {L.}~\bibnamefont {Lesshafft}}, \bibinfo {author} {\bibfnamefont {P.}~\bibnamefont {Schmid}},\ and\ \bibinfo {author} {\bibfnamefont {P.}~\bibnamefont {Huerre}},\ }\bibfield  {title} {\bibinfo {title} {The preferred mode of incompressible jets: linear frequency response analysis},\ }\href@noop {} {\bibfield  {journal} {\bibinfo  {journal} {J. Fluid Mech.}\ }\textbf {\bibinfo {volume} {716}},\ \bibinfo {pages} {189–202} (\bibinfo {year} {2013})}\BibitemShut {NoStop}%
\bibitem [{\citenamefont {Huerre}\ and\ \citenamefont {Rossi}(1998)}]{Huerre98}%
  \BibitemOpen
  \bibfield  {author} {\bibinfo {author} {\bibfnamefont {P.}~\bibnamefont {Huerre}}\ and\ \bibinfo {author} {\bibfnamefont {M.}~\bibnamefont {Rossi}},\ }in\ \href@noop {} {\emph {\bibinfo {booktitle} {Hydrodynamics and Nonlinear Instabilities}}},\ \bibinfo {series and number} {Collection Alea-Saclay: Monographs and Texts in Statistical Physics}\ (\bibinfo  {publisher} {Cambridge University Press},\ \bibinfo {year} {1998})\ p.\ \bibinfo {pages} {81–294}\BibitemShut {NoStop}%
\bibitem [{\citenamefont {Beneddine}\ \emph {et~al.}(2016)\citenamefont {Beneddine}, \citenamefont {Sipp}, \citenamefont {Arnault}, \citenamefont {Dandois},\ and\ \citenamefont {Lesshafft}}]{Beneddine16}%
  \BibitemOpen
  \bibfield  {author} {\bibinfo {author} {\bibfnamefont {S.}~\bibnamefont {Beneddine}}, \bibinfo {author} {\bibfnamefont {D.}~\bibnamefont {Sipp}}, \bibinfo {author} {\bibfnamefont {A.}~\bibnamefont {Arnault}}, \bibinfo {author} {\bibfnamefont {J.}~\bibnamefont {Dandois}},\ and\ \bibinfo {author} {\bibfnamefont {L.}~\bibnamefont {Lesshafft}},\ }\bibfield  {title} {\bibinfo {title} {Conditions for validity of mean flow stability analysis},\ }\href@noop {} {\bibfield  {journal} {\bibinfo  {journal} {J. Fluid Mech.}\ }\textbf {\bibinfo {volume} {798}},\ \bibinfo {pages} {485–504} (\bibinfo {year} {2016})}\BibitemShut {NoStop}%
\bibitem [{\citenamefont {Jeun}\ \emph {et~al.}(2016)\citenamefont {Jeun}, \citenamefont {Nichols},\ and\ \citenamefont {Jovanovi\'c}}]{Jeun16}%
  \BibitemOpen
  \bibfield  {author} {\bibinfo {author} {\bibfnamefont {J.}~\bibnamefont {Jeun}}, \bibinfo {author} {\bibfnamefont {J.}~\bibnamefont {Nichols}},\ and\ \bibinfo {author} {\bibfnamefont {M.}~\bibnamefont {Jovanovi\'c}},\ }\bibfield  {title} {\bibinfo {title} {Input-output analysis of high-speed axisymmetric isothermal jet noise},\ }\href@noop {} {\bibfield  {journal} {\bibinfo  {journal} {Phys. Fluids}\ }\textbf {\bibinfo {volume} {28}},\ \bibinfo {pages} {047101} (\bibinfo {year} {2016})}\BibitemShut {NoStop}%
\bibitem [{\citenamefont {Semeraro}\ \emph {et~al.}(2016)\citenamefont {Semeraro}, \citenamefont {Lesshafft}, \citenamefont {Jaunet},\ and\ \citenamefont {Jordan}}]{Semeraro16}%
  \BibitemOpen
  \bibfield  {author} {\bibinfo {author} {\bibfnamefont {O.}~\bibnamefont {Semeraro}}, \bibinfo {author} {\bibfnamefont {L.}~\bibnamefont {Lesshafft}}, \bibinfo {author} {\bibfnamefont {V.}~\bibnamefont {Jaunet}},\ and\ \bibinfo {author} {\bibfnamefont {P.}~\bibnamefont {Jordan}},\ }\bibfield  {title} {\bibinfo {title} {Modeling of coherent structures in a turbulent jet as global linear instability wavepackets: Theory and experiment},\ }\href@noop {} {\bibfield  {journal} {\bibinfo  {journal} {Intl. J. Heat Fluid Flow}\ }\textbf {\bibinfo {volume} {62}},\ \bibinfo {pages} {24} (\bibinfo {year} {2016})}\BibitemShut {NoStop}%
\bibitem [{\citenamefont {Schmidt}\ \emph {et~al.}(2018)\citenamefont {Schmidt}, \citenamefont {Towne}, \citenamefont {Rigas}, \citenamefont {Colonius},\ and\ \citenamefont {Br\'es}}]{Schmidt18}%
  \BibitemOpen
  \bibfield  {author} {\bibinfo {author} {\bibfnamefont {O.}~\bibnamefont {Schmidt}}, \bibinfo {author} {\bibfnamefont {A.}~\bibnamefont {Towne}}, \bibinfo {author} {\bibfnamefont {G.}~\bibnamefont {Rigas}}, \bibinfo {author} {\bibfnamefont {T.}~\bibnamefont {Colonius}},\ and\ \bibinfo {author} {\bibfnamefont {G.}~\bibnamefont {Br\'es}},\ }\bibfield  {title} {\bibinfo {title} {Spectral analysis of jet turbulence},\ }\href@noop {} {\bibfield  {journal} {\bibinfo  {journal} {J. Fluid Mech.}\ }\textbf {\bibinfo {volume} {855}},\ \bibinfo {pages} {953–982} (\bibinfo {year} {2018})}\BibitemShut {NoStop}%
\bibitem [{\citenamefont {Mantič-Lugo}\ and\ \citenamefont {Gallaire}(2016{\natexlab{a}})}]{Lugo16}%
  \BibitemOpen
  \bibfield  {author} {\bibinfo {author} {\bibfnamefont {V.}~\bibnamefont {Mantič-Lugo}}\ and\ \bibinfo {author} {\bibfnamefont {F.}~\bibnamefont {Gallaire}},\ }\bibfield  {title} {\bibinfo {title} {Self-consistent model for the saturation mechanism of the response to harmonic forcing in the backward-facing step flow},\ }\href@noop {} {\bibfield  {journal} {\bibinfo  {journal} {J. Fluid Mech.}\ }\textbf {\bibinfo {volume} {793}},\ \bibinfo {pages} {777–97} (\bibinfo {year} {2016}{\natexlab{a}})}\BibitemShut {NoStop}%
\bibitem [{\citenamefont {Rigas}\ \emph {et~al.}(2021)\citenamefont {Rigas}, \citenamefont {Sipp},\ and\ \citenamefont {Colonius}}]{Rigas21}%
  \BibitemOpen
  \bibfield  {author} {\bibinfo {author} {\bibfnamefont {G.}~\bibnamefont {Rigas}}, \bibinfo {author} {\bibfnamefont {D.}~\bibnamefont {Sipp}},\ and\ \bibinfo {author} {\bibfnamefont {T.}~\bibnamefont {Colonius}},\ }\bibfield  {title} {\bibinfo {title} {Nonlinear input/output analysis: application to boundary layer transition},\ }\href@noop {} {\bibfield  {journal} {\bibinfo  {journal} {J. Fluid Mech.}\ }\textbf {\bibinfo {volume} {911}},\ \bibinfo {pages} {A15} (\bibinfo {year} {2021})}\BibitemShut {NoStop}%
\bibitem [{\citenamefont {Pier}\ and\ \citenamefont {Huerre}(2001)}]{Pier01b}%
  \BibitemOpen
  \bibfield  {author} {\bibinfo {author} {\bibfnamefont {B.}~\bibnamefont {Pier}}\ and\ \bibinfo {author} {\bibfnamefont {P.}~\bibnamefont {Huerre}},\ }\bibfield  {title} {\bibinfo {title} {Nonlinear self-sustained structures and fronts in spatially developing wake flows},\ }\href@noop {} {\bibfield  {journal} {\bibinfo  {journal} {J. Fluid Mech.}\ }\textbf {\bibinfo {volume} {435}},\ \bibinfo {pages} {145–174} (\bibinfo {year} {2001})}\BibitemShut {NoStop}%
\bibitem [{\citenamefont {Pier}(2003)}]{Pier03}%
  \BibitemOpen
  \bibfield  {author} {\bibinfo {author} {\bibfnamefont {B.}~\bibnamefont {Pier}},\ }\bibfield  {title} {\bibinfo {title} {Open-loop control of absolutely unstable domains},\ }\href@noop {} {\bibfield  {journal} {\bibinfo  {journal} {Proc. R. Soc. London, Ser. A}\ }\textbf {\bibinfo {volume} {459}},\ \bibinfo {pages} {1105} (\bibinfo {year} {2003})}\BibitemShut {NoStop}%
\bibitem [{\citenamefont {Farrell}\ and\ \citenamefont {Ioannou}(1993)}]{Farrell93}%
  \BibitemOpen
  \bibfield  {author} {\bibinfo {author} {\bibfnamefont {B.}~\bibnamefont {Farrell}}\ and\ \bibinfo {author} {\bibfnamefont {P.}~\bibnamefont {Ioannou}},\ }\bibfield  {title} {\bibinfo {title} {Stochastic forcing of the linearized {N}avier–{S}tokes equations},\ }\href@noop {} {\bibfield  {journal} {\bibinfo  {journal} {Phys. Fluids A}\ }\textbf {\bibinfo {volume} {5}},\ \bibinfo {pages} {2600} (\bibinfo {year} {1993})}\BibitemShut {NoStop}%
\bibitem [{\citenamefont {Farrell}\ and\ \citenamefont {Ioannou}(1994)}]{Farrell94}%
  \BibitemOpen
  \bibfield  {author} {\bibinfo {author} {\bibfnamefont {B.}~\bibnamefont {Farrell}}\ and\ \bibinfo {author} {\bibfnamefont {P.}~\bibnamefont {Ioannou}},\ }\bibfield  {title} {\bibinfo {title} {Variance maintained by stochastic forcing of non-normal dynamical systems associated with linearly stable shear flows},\ }\href@noop {} {\bibfield  {journal} {\bibinfo  {journal} {Phys. Rev. Lett.}\ }\textbf {\bibinfo {volume} {72}},\ \bibinfo {pages} {1188} (\bibinfo {year} {1994})}\BibitemShut {NoStop}%
\bibitem [{\citenamefont {Farrell}\ and\ \citenamefont {Ioannou}(1996)}]{Farrell96}%
  \BibitemOpen
  \bibfield  {author} {\bibinfo {author} {\bibfnamefont {B.}~\bibnamefont {Farrell}}\ and\ \bibinfo {author} {\bibfnamefont {P.}~\bibnamefont {Ioannou}},\ }\bibfield  {title} {\bibinfo {title} {Generalized stability theory. part i: Autonomous operators},\ }\href@noop {} {\bibfield  {journal} {\bibinfo  {journal} {J. Atmos. Sci.}\ }\textbf {\bibinfo {volume} {53}},\ \bibinfo {pages} {2025–2040} (\bibinfo {year} {1996})}\BibitemShut {NoStop}%
\bibitem [{\citenamefont {Fontane}\ \emph {et~al.}(2008)\citenamefont {Fontane}, \citenamefont {Brancher},\ and\ \citenamefont {Fabre}}]{Fontane08}%
  \BibitemOpen
  \bibfield  {author} {\bibinfo {author} {\bibfnamefont {J.}~\bibnamefont {Fontane}}, \bibinfo {author} {\bibfnamefont {P.}~\bibnamefont {Brancher}},\ and\ \bibinfo {author} {\bibfnamefont {D.}~\bibnamefont {Fabre}},\ }\bibfield  {title} {\bibinfo {title} {Stochastic forcing of the {L}amb–{O}seen vortex},\ }\href@noop {} {\bibfield  {journal} {\bibinfo  {journal} {J. Fluid Mech.}\ }\textbf {\bibinfo {volume} {613}},\ \bibinfo {pages} {233–254} (\bibinfo {year} {2008})}\BibitemShut {NoStop}%
\bibitem [{\citenamefont {Dergham}\ \emph {et~al.}(2013)\citenamefont {Dergham}, \citenamefont {Sipp},\ and\ \citenamefont {Robinet}}]{Dergham2013}%
  \BibitemOpen
  \bibfield  {author} {\bibinfo {author} {\bibfnamefont {G.}~\bibnamefont {Dergham}}, \bibinfo {author} {\bibfnamefont {D.}~\bibnamefont {Sipp}},\ and\ \bibinfo {author} {\bibfnamefont {J.-C.}\ \bibnamefont {Robinet}},\ }\bibfield  {title} {\bibinfo {title} {Stochastic dynamics and model reduction of amplifier flows: the backward-facing step flow},\ }\href@noop {} {\bibfield  {journal} {\bibinfo  {journal} {J. Fluid Mech.}\ }\textbf {\bibinfo {volume} {719}},\ \bibinfo {pages} {406–430} (\bibinfo {year} {2013})}\BibitemShut {NoStop}%
\bibitem [{\citenamefont {Farrell}\ and\ \citenamefont {Ioannou}(2003)}]{Farrell03}%
  \BibitemOpen
  \bibfield  {author} {\bibinfo {author} {\bibfnamefont {B.}~\bibnamefont {Farrell}}\ and\ \bibinfo {author} {\bibfnamefont {P.}~\bibnamefont {Ioannou}},\ }\bibfield  {title} {\bibinfo {title} {Structural stability of turbulent jets},\ }\href@noop {} {\bibfield  {journal} {\bibinfo  {journal} {J. Atmos. Sci.}\ }\textbf {\bibinfo {volume} {60}},\ \bibinfo {pages} {2101} (\bibinfo {year} {2003})}\BibitemShut {NoStop}%
\bibitem [{\citenamefont {Farrell}\ and\ \citenamefont {Ioannou}(2012)}]{Farrell12}%
  \BibitemOpen
  \bibfield  {author} {\bibinfo {author} {\bibfnamefont {B.}~\bibnamefont {Farrell}}\ and\ \bibinfo {author} {\bibfnamefont {P.}~\bibnamefont {Ioannou}},\ }\bibfield  {title} {\bibinfo {title} {Dynamics of streamwise rolls and streaks in turbulent wall-bounded shear flow},\ }\href@noop {} {\bibfield  {journal} {\bibinfo  {journal} {J. Fluid Mech.}\ }\textbf {\bibinfo {volume} {708}},\ \bibinfo {pages} {149–196} (\bibinfo {year} {2012})}\BibitemShut {NoStop}%
\bibitem [{\citenamefont {Marston}\ \emph {et~al.}(2016)\citenamefont {Marston}, \citenamefont {Chini},\ and\ \citenamefont {Tobias}}]{Marston16}%
  \BibitemOpen
  \bibfield  {author} {\bibinfo {author} {\bibfnamefont {J.}~\bibnamefont {Marston}}, \bibinfo {author} {\bibfnamefont {G.}~\bibnamefont {Chini}},\ and\ \bibinfo {author} {\bibfnamefont {S.}~\bibnamefont {Tobias}},\ }\bibfield  {title} {\bibinfo {title} {Generalized quasilinear approximation: Application to zonal jets},\ }\href@noop {} {\bibfield  {journal} {\bibinfo  {journal} {Phys. Rev. Lett.}\ }\textbf {\bibinfo {volume} {116}},\ \bibinfo {pages} {214501} (\bibinfo {year} {2016})}\BibitemShut {NoStop}%
\bibitem [{\citenamefont {Marston}\ and\ \citenamefont {Tobias}(2023)}]{Marston23}%
  \BibitemOpen
  \bibfield  {author} {\bibinfo {author} {\bibfnamefont {J.}~\bibnamefont {Marston}}\ and\ \bibinfo {author} {\bibfnamefont {S.}~\bibnamefont {Tobias}},\ }\bibfield  {title} {\bibinfo {title} {Recent developments in theories of inhomogeneous and anisotropic turbulence},\ }\href@noop {} {\bibfield  {journal} {\bibinfo  {journal} {Annu. Rev. Fluid Mech.}\ }\textbf {\bibinfo {volume} {55}},\ \bibinfo {pages} {351} (\bibinfo {year} {2023})}\BibitemShut {NoStop}%
\bibitem [{\citenamefont {Mantič-Lugo}\ and\ \citenamefont {Gallaire}(2016{\natexlab{b}})}]{Lugo16s}%
  \BibitemOpen
  \bibfield  {author} {\bibinfo {author} {\bibfnamefont {V.}~\bibnamefont {Mantič-Lugo}}\ and\ \bibinfo {author} {\bibfnamefont {F.}~\bibnamefont {Gallaire}},\ }\bibfield  {title} {\bibinfo {title} {Saturation of the response to stochastic forcing in two-dimensional backward-facing step flow: a self-consistent approximation},\ }\href@noop {} {\bibfield  {journal} {\bibinfo  {journal} {Phys. Fluids}\ }\textbf {\bibinfo {volume} {1}} (\bibinfo {year} {2016}{\natexlab{b}})}\BibitemShut {NoStop}%
\bibitem [{\citenamefont {Fauve}(1998)}]{Fauve98}%
  \BibitemOpen
  \bibfield  {author} {\bibinfo {author} {\bibfnamefont {S.}~\bibnamefont {Fauve}},\ }\bibinfo {title} {Pattern forming instabilities},\ in\ \href@noop {} {\emph {\bibinfo {booktitle} {Hydrodynamics and Nonlinear Instabilities}}}\ (\bibinfo  {publisher} {Cambridge University Press},\ \bibinfo {year} {1998})\ pp.\ \bibinfo {pages} {387--492}\BibitemShut {NoStop}%
\bibitem [{\citenamefont {Golubitsky}\ and\ \citenamefont {Stewart}(1985)}]{Golubitsky85}%
  \BibitemOpen
  \bibfield  {author} {\bibinfo {author} {\bibfnamefont {M.}~\bibnamefont {Golubitsky}}\ and\ \bibinfo {author} {\bibfnamefont {I.}~\bibnamefont {Stewart}},\ }\bibfield  {title} {\bibinfo {title} {{H}opf bifurcation in the presence of symmetry},\ }\href@noop {} {\bibfield  {journal} {\bibinfo  {journal} {Arch. Ration. Mech. and Anal.}\ }\textbf {\bibinfo {volume} {87}},\ \bibinfo {pages} {107} (\bibinfo {year} {1985})}\BibitemShut {NoStop}%
\bibitem [{\citenamefont {Cross}(1986)}]{Cross86}%
  \BibitemOpen
  \bibfield  {author} {\bibinfo {author} {\bibfnamefont {M.}~\bibnamefont {Cross}},\ }\bibfield  {title} {\bibinfo {title} {Traveling and standing waves in binary-fluid convection in finite geometries},\ }\href@noop {} {\bibfield  {journal} {\bibinfo  {journal} {Phys. Rev. Lett.}\ }\textbf {\bibinfo {volume} {57}},\ \bibinfo {pages} {2935} (\bibinfo {year} {1986})}\BibitemShut {NoStop}%
\bibitem [{\citenamefont {Crawford}\ \emph {et~al.}(1988)\citenamefont {Crawford}, \citenamefont {Golubitsky},\ and\ \citenamefont {Langford}}]{Crawford88}%
  \BibitemOpen
  \bibfield  {author} {\bibinfo {author} {\bibfnamefont {J.}~\bibnamefont {Crawford}}, \bibinfo {author} {\bibfnamefont {M.}~\bibnamefont {Golubitsky}},\ and\ \bibinfo {author} {\bibfnamefont {W.}~\bibnamefont {Langford}},\ }\bibfield  {title} {\bibinfo {title} {Modulated rotating waves in {O}(2) mode interactions},\ }\href@noop {} {\bibfield  {journal} {\bibinfo  {journal} {Dyn. Stab. Syst.}\ }\textbf {\bibinfo {volume} {3}},\ \bibinfo {pages} {159} (\bibinfo {year} {1988})}\BibitemShut {NoStop}%
\bibitem [{\citenamefont {Chossat}\ and\ \citenamefont {Iooss}(1994)}]{Chossat94}%
  \BibitemOpen
  \bibfield  {author} {\bibinfo {author} {\bibfnamefont {P.}~\bibnamefont {Chossat}}\ and\ \bibinfo {author} {\bibfnamefont {G.}~\bibnamefont {Iooss}},\ }\href@noop {} {\emph {\bibinfo {title} {The {C}ouette-{T}aylor Problem}}},\ {A}pplied {M}athematical {S}ciences\ (\bibinfo  {publisher} {Springer-Verlag},\ \bibinfo {year} {1994})\BibitemShut {NoStop}%
\bibitem [{\citenamefont {Chiffaudel}\ and\ \citenamefont {Fauve}(1987)}]{Chiffaudel87}%
  \BibitemOpen
  \bibfield  {author} {\bibinfo {author} {\bibfnamefont {A.}~\bibnamefont {Chiffaudel}}\ and\ \bibinfo {author} {\bibfnamefont {S.}~\bibnamefont {Fauve}},\ }\bibfield  {title} {\bibinfo {title} {Strong resonance in forced oscillatory convection},\ }\href@noop {} {\bibfield  {journal} {\bibinfo  {journal} {Phys. Rev. A}\ }\textbf {\bibinfo {volume} {35}},\ \bibinfo {pages} {4004} (\bibinfo {year} {1987})}\BibitemShut {NoStop}%
\bibitem [{\citenamefont {Sipp}\ and\ \citenamefont {Lebedev}(2007)}]{Sipp07}%
  \BibitemOpen
  \bibfield  {author} {\bibinfo {author} {\bibfnamefont {D.}~\bibnamefont {Sipp}}\ and\ \bibinfo {author} {\bibfnamefont {A.}~\bibnamefont {Lebedev}},\ }\bibfield  {title} {\bibinfo {title} {Global stability of base and mean flows : a general approach and its applications to cylinder and open cavity flows},\ }\href@noop {} {\bibfield  {journal} {\bibinfo  {journal} {J. Fluid Mech.}\ }\textbf {\bibinfo {volume} {593}},\ \bibinfo {pages} {333} (\bibinfo {year} {2007})}\BibitemShut {NoStop}%
\bibitem [{\citenamefont {Meliga}\ \emph {et~al.}(2009)\citenamefont {Meliga}, \citenamefont {Chomaz},\ and\ \citenamefont {Sipp}}]{Meliga09}%
  \BibitemOpen
  \bibfield  {author} {\bibinfo {author} {\bibfnamefont {P.}~\bibnamefont {Meliga}}, \bibinfo {author} {\bibfnamefont {J.-M.}\ \bibnamefont {Chomaz}},\ and\ \bibinfo {author} {\bibfnamefont {D.}~\bibnamefont {Sipp}},\ }\bibfield  {title} {\bibinfo {title} {Global mode interaction and pattern selection in the wake of a disk: a weakly nonlinear expansion},\ }\href@noop {} {\bibfield  {journal} {\bibinfo  {journal} {J. Fluid Mech.}\ }\textbf {\bibinfo {volume} {633}},\ \bibinfo {pages} {159} (\bibinfo {year} {2009})}\BibitemShut {NoStop}%
\bibitem [{\citenamefont {Meliga}\ \emph {et~al.}(2012)\citenamefont {Meliga}, \citenamefont {Gallaire},\ and\ \citenamefont {Chomaz}}]{Meliga12}%
  \BibitemOpen
  \bibfield  {author} {\bibinfo {author} {\bibfnamefont {P.}~\bibnamefont {Meliga}}, \bibinfo {author} {\bibfnamefont {F.}~\bibnamefont {Gallaire}},\ and\ \bibinfo {author} {\bibfnamefont {J.-M.}\ \bibnamefont {Chomaz}},\ }\bibfield  {title} {\bibinfo {title} {A weakly nonlinear mechanism for mode selection in swirling jets},\ }\href@noop {} {\bibfield  {journal} {\bibinfo  {journal} {J. Fluid Mech.}\ }\textbf {\bibinfo {volume} {699}},\ \bibinfo {pages} {216–262} (\bibinfo {year} {2012})}\BibitemShut {NoStop}%
\bibitem [{\citenamefont {Guckenheimer}\ and\ \citenamefont {Holmes}(1983)}]{Guckenheimer83}%
  \BibitemOpen
  \bibfield  {author} {\bibinfo {author} {\bibfnamefont {J.}~\bibnamefont {Guckenheimer}}\ and\ \bibinfo {author} {\bibfnamefont {P.}~\bibnamefont {Holmes}},\ }\href@noop {} {\emph {\bibinfo {title} {Nonlinear oscillations, dynamical systems, and bifurcations of vector fields}}}\ (\bibinfo  {publisher} {Springer, New York, NY},\ \bibinfo {year} {1983})\BibitemShut {NoStop}%
\bibitem [{\citenamefont {Haragus}\ and\ \citenamefont {Iooss}(2011)}]{Haragus11}%
  \BibitemOpen
  \bibfield  {author} {\bibinfo {author} {\bibfnamefont {M.}~\bibnamefont {Haragus}}\ and\ \bibinfo {author} {\bibfnamefont {G.}~\bibnamefont {Iooss}},\ }\href@noop {} {\emph {\bibinfo {title} {Local bifurcations, center manifolds, and normal forms in infinite-dimensional dynamical systems}}}\ (\bibinfo  {publisher} {Springer-Verlag London},\ \bibinfo {year} {2011})\BibitemShut {NoStop}%
\bibitem [{\citenamefont {Cox}\ and\ \citenamefont {Roberts}(1991)}]{Cox91}%
  \BibitemOpen
  \bibfield  {author} {\bibinfo {author} {\bibfnamefont {S.}~\bibnamefont {Cox}}\ and\ \bibinfo {author} {\bibfnamefont {A.}~\bibnamefont {Roberts}},\ }\bibfield  {title} {\bibinfo {title} {Centre manifolds of forced dynamical systems},\ }\href@noop {} {\bibfield  {journal} {\bibinfo  {journal} {J. Austral. Math. Soc. Ser. B, Appl. Math}\ }\textbf {\bibinfo {volume} {32}},\ \bibinfo {pages} {401–436} (\bibinfo {year} {1991})}\BibitemShut {NoStop}%
\bibitem [{\citenamefont {Carini}\ \emph {et~al.}(2015)\citenamefont {Carini}, \citenamefont {Auteri},\ and\ \citenamefont {Giannetti}}]{Carini15}%
  \BibitemOpen
  \bibfield  {author} {\bibinfo {author} {\bibfnamefont {M.}~\bibnamefont {Carini}}, \bibinfo {author} {\bibfnamefont {F.}~\bibnamefont {Auteri}},\ and\ \bibinfo {author} {\bibfnamefont {F.}~\bibnamefont {Giannetti}},\ }\bibfield  {title} {\bibinfo {title} {Centre-manifold reduction of bifurcating flows},\ }\href@noop {} {\bibfield  {journal} {\bibinfo  {journal} {J. Fluid Mech.}\ }\textbf {\bibinfo {volume} {767}},\ \bibinfo {pages} {109–145} (\bibinfo {year} {2015})}\BibitemShut {NoStop}%
\bibitem [{\citenamefont {Negi}(2024)}]{Negi24}%
  \BibitemOpen
  \bibfield  {author} {\bibinfo {author} {\bibfnamefont {P.}~\bibnamefont {Negi}},\ }\href@noop {} {\bibinfo {title} {Asymptotic center--manifold for the {N}avier--{S}tokes}} (\bibinfo {year} {2024}),\ \Eprint {https://arxiv.org/abs/2411.03727} {arXiv:2411.03727 [physics.flu-dyn]} \BibitemShut {NoStop}%
\bibitem [{\citenamefont {Knobloch}\ and\ \citenamefont {Guckenheimer}(1983)}]{Knobloch83}%
  \BibitemOpen
  \bibfield  {author} {\bibinfo {author} {\bibfnamefont {E.}~\bibnamefont {Knobloch}}\ and\ \bibinfo {author} {\bibfnamefont {J.}~\bibnamefont {Guckenheimer}},\ }\bibfield  {title} {\bibinfo {title} {Convective transitions induced by a varying aspect ratio},\ }\href@noop {} {\bibfield  {journal} {\bibinfo  {journal} {Phys. Rev. A}\ }\textbf {\bibinfo {volume} {27}},\ \bibinfo {pages} {408} (\bibinfo {year} {1983})}\BibitemShut {NoStop}%
\bibitem [{\citenamefont {Haller}\ and\ \citenamefont {Ponsioen}(2016)}]{Haller16}%
  \BibitemOpen
  \bibfield  {author} {\bibinfo {author} {\bibfnamefont {G.}~\bibnamefont {Haller}}\ and\ \bibinfo {author} {\bibfnamefont {S.}~\bibnamefont {Ponsioen}},\ }\bibfield  {title} {\bibinfo {title} {Nonlinear normal modes and spectral submanifolds: existence, uniqueness and use in model reduction},\ }\href@noop {} {\bibfield  {journal} {\bibinfo  {journal} {Nonlinear Dyn.}\ }\textbf {\bibinfo {volume} {86}},\ \bibinfo {pages} {1493} (\bibinfo {year} {2016})}\BibitemShut {NoStop}%
\bibitem [{\citenamefont {Li}\ \emph {et~al.}(2022)\citenamefont {Li}, \citenamefont {Jain},\ and\ \citenamefont {Haller}}]{Li22}%
  \BibitemOpen
  \bibfield  {author} {\bibinfo {author} {\bibfnamefont {M.}~\bibnamefont {Li}}, \bibinfo {author} {\bibfnamefont {S.}~\bibnamefont {Jain}},\ and\ \bibinfo {author} {\bibfnamefont {G.}~\bibnamefont {Haller}},\ }\bibfield  {title} {\bibinfo {title} {Nonlinear analysis of forced mechanical systems with internal resonance using spectral submanifolds, part i: Periodic response and forced response curve},\ }\href@noop {} {\bibfield  {journal} {\bibinfo  {journal} {Nonlinear Dyn.}\ }\textbf {\bibinfo {volume} {110}},\ \bibinfo {pages} {1005} (\bibinfo {year} {2022})}\BibitemShut {NoStop}%
\bibitem [{\citenamefont {Buza}\ \emph {et~al.}(2021)\citenamefont {Buza}, \citenamefont {Jain},\ and\ \citenamefont {Haller}}]{Buza21}%
  \BibitemOpen
  \bibfield  {author} {\bibinfo {author} {\bibfnamefont {G.}~\bibnamefont {Buza}}, \bibinfo {author} {\bibfnamefont {S.}~\bibnamefont {Jain}},\ and\ \bibinfo {author} {\bibfnamefont {G.}~\bibnamefont {Haller}},\ }\bibfield  {title} {\bibinfo {title} {Using spectral submanifolds for optimal mode selection in nonlinear model reduction},\ }\href@noop {} {\bibfield  {journal} {\bibinfo  {journal} {Proc. R. Soc. A.}\ }\textbf {\bibinfo {volume} {477}},\ \bibinfo {pages} {20200725} (\bibinfo {year} {2021})}\BibitemShut {NoStop}%
\bibitem [{\citenamefont {Touz\'e}\ \emph {et~al.}(2021)\citenamefont {Touz\'e}, \citenamefont {Vizzaccaro},\ and\ \citenamefont {Thomas}}]{Touze21}%
  \BibitemOpen
  \bibfield  {author} {\bibinfo {author} {\bibfnamefont {C.}~\bibnamefont {Touz\'e}}, \bibinfo {author} {\bibfnamefont {A.}~\bibnamefont {Vizzaccaro}},\ and\ \bibinfo {author} {\bibfnamefont {O.}~\bibnamefont {Thomas}},\ }\bibfield  {title} {\bibinfo {title} {Model order reduction methods for geometrically nonlinear structures: a review of nonlinear techniques},\ }\href@noop {} {\bibfield  {journal} {\bibinfo  {journal} {Nonlinear Dyn.}\ }\textbf {\bibinfo {volume} {105}},\ \bibinfo {pages} {1141} (\bibinfo {year} {2021})}\BibitemShut {NoStop}%
\bibitem [{\citenamefont {Kasz\'as}\ \emph {et~al.}(2022)\citenamefont {Kasz\'as}, \citenamefont {Cenedese},\ and\ \citenamefont {Haller}}]{Kaszas22}%
  \BibitemOpen
  \bibfield  {author} {\bibinfo {author} {\bibfnamefont {B.}~\bibnamefont {Kasz\'as}}, \bibinfo {author} {\bibfnamefont {M.}~\bibnamefont {Cenedese}},\ and\ \bibinfo {author} {\bibfnamefont {G.}~\bibnamefont {Haller}},\ }\bibfield  {title} {\bibinfo {title} {Dynamics-based machine learning of transitions in {C}ouette flow},\ }\href@noop {} {\bibfield  {journal} {\bibinfo  {journal} {Phys. Rev. Fluids}\ }\textbf {\bibinfo {volume} {7}},\ \bibinfo {pages} {L082402} (\bibinfo {year} {2022})}\BibitemShut {NoStop}%
\bibitem [{\citenamefont {Gallaire}\ \emph {et~al.}(2016)\citenamefont {Gallaire}, \citenamefont {Boujo}, \citenamefont {Mantič-Lugo}, \citenamefont {Arratia}, \citenamefont {Thiria},\ and\ \citenamefont {Meliga}}]{Gallaire16}%
  \BibitemOpen
  \bibfield  {author} {\bibinfo {author} {\bibfnamefont {F.}~\bibnamefont {Gallaire}}, \bibinfo {author} {\bibfnamefont {E.}~\bibnamefont {Boujo}}, \bibinfo {author} {\bibfnamefont {V.}~\bibnamefont {Mantič-Lugo}}, \bibinfo {author} {\bibfnamefont {C.}~\bibnamefont {Arratia}}, \bibinfo {author} {\bibfnamefont {B.}~\bibnamefont {Thiria}},\ and\ \bibinfo {author} {\bibfnamefont {P.}~\bibnamefont {Meliga}},\ }\bibfield  {title} {\bibinfo {title} {Pushing amplitude equations far from threshold: application to the supercritical {H}opf bifurcation in the cylinder wake},\ }\href@noop {} {\bibfield  {journal} {\bibinfo  {journal} {Fluid Dyn. Res.}\ }\textbf {\bibinfo {volume} {48}},\ \bibinfo {pages} {061401} (\bibinfo {year} {2016})}\BibitemShut {NoStop}%
\bibitem [{\citenamefont {Pham}\ and\ \citenamefont {Suslov}(2018)}]{Pham18}%
  \BibitemOpen
  \bibfield  {author} {\bibinfo {author} {\bibfnamefont {K.}~\bibnamefont {Pham}}\ and\ \bibinfo {author} {\bibfnamefont {S.}~\bibnamefont {Suslov}},\ }\bibfield  {title} {\bibinfo {title} {On the definition of {L}andau constants in amplitude equations away from a critical point},\ }\href@noop {} {\bibfield  {journal} {\bibinfo  {journal} {R. Soc. Open Sci.}\ }\textbf {\bibinfo {volume} {5}},\ \bibinfo {pages} {180746} (\bibinfo {year} {2018})}\BibitemShut {NoStop}%
\bibitem [{\citenamefont {Noack}\ and\ \citenamefont {Eckelmann}(1994)}]{Noack94}%
  \BibitemOpen
  \bibfield  {author} {\bibinfo {author} {\bibfnamefont {B.}~\bibnamefont {Noack}}\ and\ \bibinfo {author} {\bibfnamefont {H.}~\bibnamefont {Eckelmann}},\ }\bibfield  {title} {\bibinfo {title} {{A low‐dimensional {G}alerkin method for the three‐dimensional flow around a circular cylinder}},\ }\href@noop {} {\bibfield  {journal} {\bibinfo  {journal} {Phys. Fluids}\ }\textbf {\bibinfo {volume} {6}},\ \bibinfo {pages} {124} (\bibinfo {year} {1994})}\BibitemShut {NoStop}%
\bibitem [{\citenamefont {Noack}\ \emph {et~al.}(2003)\citenamefont {Noack}, \citenamefont {Afanasiev}, \citenamefont {Morzyński}, \citenamefont {Tadmor},\ and\ \citenamefont {Thiele}}]{Noack03}%
  \BibitemOpen
  \bibfield  {author} {\bibinfo {author} {\bibfnamefont {B.}~\bibnamefont {Noack}}, \bibinfo {author} {\bibfnamefont {K.}~\bibnamefont {Afanasiev}}, \bibinfo {author} {\bibfnamefont {M.}~\bibnamefont {Morzyński}}, \bibinfo {author} {\bibfnamefont {G.}~\bibnamefont {Tadmor}},\ and\ \bibinfo {author} {\bibfnamefont {F.}~\bibnamefont {Thiele}},\ }\bibfield  {title} {\bibinfo {title} {A hierarchy of low-dimensional models for the transient and post-transient cylinder wake},\ }\href@noop {} {\bibfield  {journal} {\bibinfo  {journal} {J. Fluid Mech.}\ }\textbf {\bibinfo {volume} {497}},\ \bibinfo {pages} {335–363} (\bibinfo {year} {2003})}\BibitemShut {NoStop}%
\bibitem [{\citenamefont {Noack}\ \emph {et~al.}(2005)\citenamefont {Noack}, \citenamefont {Papas},\ and\ \citenamefont {Monkewitz}}]{Noack05}%
  \BibitemOpen
  \bibfield  {author} {\bibinfo {author} {\bibfnamefont {B.}~\bibnamefont {Noack}}, \bibinfo {author} {\bibfnamefont {P.}~\bibnamefont {Papas}},\ and\ \bibinfo {author} {\bibfnamefont {P.}~\bibnamefont {Monkewitz}},\ }\bibfield  {title} {\bibinfo {title} {The need for a pressure-term representation in empirical {G}alerkin models of incompressible shear flows},\ }\href@noop {} {\bibfield  {journal} {\bibinfo  {journal} {J. Fluid Mech.}\ }\textbf {\bibinfo {volume} {523}},\ \bibinfo {pages} {339–365} (\bibinfo {year} {2005})}\BibitemShut {NoStop}%
\bibitem [{\citenamefont {Le~Diz{\`e}s}\ \emph {et~al.}(1993)\citenamefont {Le~Diz{\`e}s}, \citenamefont {Huerre}, \citenamefont {Chomaz},\ and\ \citenamefont {Monkewitz}}]{LeDizes93}%
  \BibitemOpen
  \bibfield  {author} {\bibinfo {author} {\bibfnamefont {S.}~\bibnamefont {Le~Diz{\`e}s}}, \bibinfo {author} {\bibfnamefont {P.}~\bibnamefont {Huerre}}, \bibinfo {author} {\bibfnamefont {J.-M.}\ \bibnamefont {Chomaz}},\ and\ \bibinfo {author} {\bibfnamefont {P.~A.}\ \bibnamefont {Monkewitz}},\ }\bibfield  {title} {\bibinfo {title} {Nonlinear stability analysis of slowly-diverging flows: Limitations of the weakly nonlinear approach},\ }in\ \href@noop {} {\emph {\bibinfo {booktitle} {Bluff-Body Wakes, Dynamics and Instabilities}}},\ \bibinfo {editor} {edited by\ \bibinfo {editor} {\bibfnamefont {H.}~\bibnamefont {Eckelmann}}, \bibinfo {editor} {\bibfnamefont {J.~M.~R.}\ \bibnamefont {Graham}}, \bibinfo {editor} {\bibfnamefont {P.}~\bibnamefont {Huerre}},\ and\ \bibinfo {editor} {\bibfnamefont {P.}~\bibnamefont {Monkewitz}}}\ (\bibinfo  {publisher} {Springer Berlin Heidelberg},\ \bibinfo {address} {Berlin, Heidelberg},\ \bibinfo {year} {1993})\ pp.\ \bibinfo {pages} {147--152}\BibitemShut {NoStop}%
\bibitem [{\citenamefont {Le~Diz{\`e}s}(1994)}]{LeDizes94}%
  \BibitemOpen
  \bibfield  {author} {\bibinfo {author} {\bibfnamefont {S.}~\bibnamefont {Le~Diz{\`e}s}},\ }\emph {\bibinfo {title} {Modes globaux dans les écoulements faiblement inhomogènes}},\ \href@noop {} {Ph.D. thesis},\ \bibinfo  {school} {Palaiseau, Ecole Polytechnique} (\bibinfo {year} {1994})\BibitemShut {NoStop}%
\bibitem [{\citenamefont {Schmid}\ and\ \citenamefont {Brandt}(2014)}]{Schmid14}%
  \BibitemOpen
  \bibfield  {author} {\bibinfo {author} {\bibfnamefont {P.}~\bibnamefont {Schmid}}\ and\ \bibinfo {author} {\bibfnamefont {L.}~\bibnamefont {Brandt}},\ }\bibfield  {title} {\bibinfo {title} {{Analysis of Fluid Systems: Stability, Receptivity, Sensitivity}},\ }\href@noop {} {\bibfield  {journal} {\bibinfo  {journal} {Appl. Mech. Rev}\ }\textbf {\bibinfo {volume} {66}},\ \bibinfo {pages} {024803} (\bibinfo {year} {2014})}\BibitemShut {NoStop}%
\bibitem [{\citenamefont {Ducimeti\`{e}re}\ \emph {et~al.}(2022)\citenamefont {Ducimeti\`{e}re}, \citenamefont {Boujo},\ and\ \citenamefont {Gallaire}}]{Ducimetiere22}%
  \BibitemOpen
  \bibfield  {author} {\bibinfo {author} {\bibfnamefont {Y.-M.}\ \bibnamefont {Ducimeti\`{e}re}}, \bibinfo {author} {\bibfnamefont {E.}~\bibnamefont {Boujo}},\ and\ \bibinfo {author} {\bibfnamefont {F.}~\bibnamefont {Gallaire}},\ }\bibfield  {title} {\bibinfo {title} {Weak nonlinearity for strong non-normality},\ }\href@noop {} {\bibfield  {journal} {\bibinfo  {journal} {J. Fluid Mech.}\ }\textbf {\bibinfo {volume} {947}},\ \bibinfo {pages} {A43} (\bibinfo {year} {2022})}\BibitemShut {NoStop}%
\bibitem [{\citenamefont {Symon}\ \emph {et~al.}(2018)\citenamefont {Symon}, \citenamefont {Rosenberg}, \citenamefont {Dawson},\ and\ \citenamefont {McKeon}}]{Symon18}%
  \BibitemOpen
  \bibfield  {author} {\bibinfo {author} {\bibfnamefont {S.}~\bibnamefont {Symon}}, \bibinfo {author} {\bibfnamefont {K.}~\bibnamefont {Rosenberg}}, \bibinfo {author} {\bibfnamefont {S.}~\bibnamefont {Dawson}},\ and\ \bibinfo {author} {\bibfnamefont {B.}~\bibnamefont {McKeon}},\ }\bibfield  {title} {\bibinfo {title} {Non-normality and classification of amplification mechanisms in stability and resolvent analysis},\ }\href@noop {} {\bibfield  {journal} {\bibinfo  {journal} {Phys. Rev. Fluids}\ }\textbf {\bibinfo {volume} {3}},\ \bibinfo {pages} {053902} (\bibinfo {year} {2018})}\BibitemShut {NoStop}%
\bibitem [{\citenamefont {Scott}(2013)}]{Scott13}%
  \BibitemOpen
  \bibfield  {author} {\bibinfo {author} {\bibfnamefont {M.}~\bibnamefont {Scott}},\ }\href@noop {} {\emph {\bibinfo {title} {Applied Stochastic Processes in Science and Engineering}}}\ (\bibinfo  {publisher} {University of Waterloo},\ \bibinfo {year} {2013})\BibitemShut {NoStop}%
\bibitem [{\citenamefont {Rolland}(2024)}]{Rolland24}%
  \BibitemOpen
  \bibfield  {author} {\bibinfo {author} {\bibfnamefont {J.}~\bibnamefont {Rolland}},\ }\bibfield  {title} {\bibinfo {title} {Rare, noise-induced, bypass transition in plane {C}ouette flow can bypass instantons},\ }\href@noop {} {\bibfield  {journal} {\bibinfo  {journal} {Phys. Rev. E}\ }\textbf {\bibinfo {volume} {110}},\ \bibinfo {pages} {065106} (\bibinfo {year} {2024})}\BibitemShut {NoStop}%
\bibitem [{\citenamefont {Sapsis}\ and\ \citenamefont {Lermusiaux}(2009)}]{Sapsis09}%
  \BibitemOpen
  \bibfield  {author} {\bibinfo {author} {\bibfnamefont {T.}~\bibnamefont {Sapsis}}\ and\ \bibinfo {author} {\bibfnamefont {P.}~\bibnamefont {Lermusiaux}},\ }\bibfield  {title} {\bibinfo {title} {Dynamically orthogonal field equations for continuous stochastic dynamical systems},\ }\href@noop {} {\bibfield  {journal} {\bibinfo  {journal} {Physica D}\ }\textbf {\bibinfo {volume} {238}},\ \bibinfo {pages} {2347} (\bibinfo {year} {2009})}\BibitemShut {NoStop}%
\bibitem [{\citenamefont {Ducimetière}\ \emph {et~al.}(2022)\citenamefont {Ducimetière}, \citenamefont {Boujo},\ and\ \citenamefont {Gallaire}}]{Ducimetiere22b}%
  \BibitemOpen
  \bibfield  {author} {\bibinfo {author} {\bibfnamefont {Y.-M.}\ \bibnamefont {Ducimetière}}, \bibinfo {author} {\bibfnamefont {E.}~\bibnamefont {Boujo}},\ and\ \bibinfo {author} {\bibfnamefont {F.}~\bibnamefont {Gallaire}},\ }\bibfield  {title} {\bibinfo {title} {Weakly nonlinear evolution of stochastically driven non-normal systems},\ }\href@noop {} {\bibfield  {journal} {\bibinfo  {journal} {J. Fluid Mech.}\ }\textbf {\bibinfo {volume} {951}},\ \bibinfo {pages} {R3} (\bibinfo {year} {2022})}\BibitemShut {NoStop}%
\bibitem [{\citenamefont {Bergstr{\"o}m}(2004)}]{Bergstrom04}%
  \BibitemOpen
  \bibfield  {author} {\bibinfo {author} {\bibfnamefont {L.~B.}\ \bibnamefont {Bergstr{\"o}m}},\ }\bibfield  {title} {\bibinfo {title} {{Nonmodal growth of three-dimensional disturbances on plane {C}ouette–{P}oiseuille flows}},\ }\href@noop {} {\bibfield  {journal} {\bibinfo  {journal} {Phys. Fluids}\ }\textbf {\bibinfo {volume} {17}},\ \bibinfo {pages} {014105} (\bibinfo {year} {2004})}\BibitemShut {NoStop}%
\bibitem [{\citenamefont {Shuai}\ \emph {et~al.}(2023)\citenamefont {Shuai}, \citenamefont {Liu},\ and\ \citenamefont {Gayme}}]{Shuai22}%
  \BibitemOpen
  \bibfield  {author} {\bibinfo {author} {\bibfnamefont {Y.}~\bibnamefont {Shuai}}, \bibinfo {author} {\bibfnamefont {C.}~\bibnamefont {Liu}},\ and\ \bibinfo {author} {\bibfnamefont {D.~F.}\ \bibnamefont {Gayme}},\ }\bibfield  {title} {\bibinfo {title} {Structured input–output analysis of oblique laminar–turbulent patterns in plane {C}ouette–{P}oiseuille flow},\ }\href@noop {} {\bibfield  {journal} {\bibinfo  {journal} {Intl. J. Heat Fluid Flow}\ }\textbf {\bibinfo {volume} {103}},\ \bibinfo {pages} {109207} (\bibinfo {year} {2023})}\BibitemShut {NoStop}%
\bibitem [{\citenamefont {Klotz}\ \emph {et~al.}(2017)\citenamefont {Klotz}, \citenamefont {Lemoult}, \citenamefont {Frontczak}, \citenamefont {Tuckerman},\ and\ \citenamefont {Wesfreid}}]{Klotz17}%
  \BibitemOpen
  \bibfield  {author} {\bibinfo {author} {\bibfnamefont {L.}~\bibnamefont {Klotz}}, \bibinfo {author} {\bibfnamefont {G.}~\bibnamefont {Lemoult}}, \bibinfo {author} {\bibfnamefont {I.}~\bibnamefont {Frontczak}}, \bibinfo {author} {\bibfnamefont {L.~S.}\ \bibnamefont {Tuckerman}},\ and\ \bibinfo {author} {\bibfnamefont {J.~E.}\ \bibnamefont {Wesfreid}},\ }\bibfield  {title} {\bibinfo {title} {{C}ouette-{P}oiseuille flow experiment with zero mean advection velocity: Subcritical transition to turbulence},\ }\href@noop {} {\bibfield  {journal} {\bibinfo  {journal} {Phys. Rev. Fluids}\ }\textbf {\bibinfo {volume} {2}},\ \bibinfo {pages} {043904} (\bibinfo {year} {2017})}\BibitemShut {NoStop}%
\bibitem [{\citenamefont {Klotz}\ \emph {et~al.}(2021)\citenamefont {Klotz}, \citenamefont {Pavlenko},\ and\ \citenamefont {Wesfreid}}]{Klotz21}%
  \BibitemOpen
  \bibfield  {author} {\bibinfo {author} {\bibfnamefont {L.}~\bibnamefont {Klotz}}, \bibinfo {author} {\bibfnamefont {A.}~\bibnamefont {Pavlenko}},\ and\ \bibinfo {author} {\bibfnamefont {J.}~\bibnamefont {Wesfreid}},\ }\bibfield  {title} {\bibinfo {title} {Experimental measurements in plane {C}ouette–{P}oiseuille flow: dynamics of the large- and small-scale flow},\ }\href@noop {} {\bibfield  {journal} {\bibinfo  {journal} {J. Fluid Mech.}\ }\textbf {\bibinfo {volume} {912}},\ \bibinfo {pages} {A24} (\bibinfo {year} {2021})}\BibitemShut {NoStop}%
\bibitem [{\citenamefont {Liu}\ \emph {et~al.}(2021)\citenamefont {Liu}, \citenamefont {Semin}, \citenamefont {Klotz}, \citenamefont {Godoy-Diana}, \citenamefont {Wesfreid},\ and\ \citenamefont {Mullin}}]{Liup21}%
  \BibitemOpen
  \bibfield  {author} {\bibinfo {author} {\bibfnamefont {T.}~\bibnamefont {Liu}}, \bibinfo {author} {\bibfnamefont {B.}~\bibnamefont {Semin}}, \bibinfo {author} {\bibfnamefont {L.}~\bibnamefont {Klotz}}, \bibinfo {author} {\bibfnamefont {R.}~\bibnamefont {Godoy-Diana}}, \bibinfo {author} {\bibfnamefont {J.}~\bibnamefont {Wesfreid}},\ and\ \bibinfo {author} {\bibfnamefont {T.}~\bibnamefont {Mullin}},\ }\bibfield  {title} {\bibinfo {title} {Decay of streaks and rolls in plane {C}ouette–{P}oiseuille flow},\ }\href@noop {} {\bibfield  {journal} {\bibinfo  {journal} {J. Fluid Mech.}\ }\textbf {\bibinfo {volume} {915}},\ \bibinfo {pages} {A65} (\bibinfo {year} {2021})}\BibitemShut {NoStop}%
\bibitem [{\citenamefont {Pralits}\ \emph {et~al.}(2015)\citenamefont {Pralits}, \citenamefont {Bottaro},\ and\ \citenamefont {Cherubini}}]{Pralits15}%
  \BibitemOpen
  \bibfield  {author} {\bibinfo {author} {\bibfnamefont {J.}~\bibnamefont {Pralits}}, \bibinfo {author} {\bibfnamefont {A.}~\bibnamefont {Bottaro}},\ and\ \bibinfo {author} {\bibfnamefont {S.}~\bibnamefont {Cherubini}},\ }\bibfield  {title} {\bibinfo {title} {Weakly nonlinear optimal perturbations},\ }\href@noop {} {\bibfield  {journal} {\bibinfo  {journal} {J. Fluid Mech.}\ }\textbf {\bibinfo {volume} {785}},\ \bibinfo {pages} {135–151} (\bibinfo {year} {2015})}\BibitemShut {NoStop}%
\bibitem [{\citenamefont {Nolan}\ and\ \citenamefont {Farrell}(1999)}]{Nolan99}%
  \BibitemOpen
  \bibfield  {author} {\bibinfo {author} {\bibfnamefont {D.}~\bibnamefont {Nolan}}\ and\ \bibinfo {author} {\bibfnamefont {B.}~\bibnamefont {Farrell}},\ }\bibfield  {title} {\bibinfo {title} {The intensification of two-dimensional swirling flows by stochastic asymmetric forcing},\ }\href@noop {} {\bibfield  {journal} {\bibinfo  {journal} {J. Atmos. Sci.}\ }\textbf {\bibinfo {volume} {56}},\ \bibinfo {pages} {3937 } (\bibinfo {year} {1999})}\BibitemShut {NoStop}%
\bibitem [{\citenamefont {Denoix}\ \emph {et~al.}(1994)\citenamefont {Denoix}, \citenamefont {Sommeria},\ and\ \citenamefont {Thess}}]{Denoix94}%
  \BibitemOpen
  \bibfield  {author} {\bibinfo {author} {\bibfnamefont {M.-A.}\ \bibnamefont {Denoix}}, \bibinfo {author} {\bibfnamefont {J.}~\bibnamefont {Sommeria}},\ and\ \bibinfo {author} {\bibfnamefont {A.}~\bibnamefont {Thess}},\ }\bibinfo {title} {Two-dimensional turbulence: The prediction of coherent structures by statistical mechanics},\ in\ \href@noop {} {\emph {\bibinfo {booktitle} {Progress in Turbulence Research}}}\ (\bibinfo {year} {1994})\ pp.\ \bibinfo {pages} {88--107}\BibitemShut {NoStop}%
\bibitem [{\citenamefont {Van~Heijst}\ and\ \citenamefont {Kloosterziel}(1989)}]{VanHeijst89}%
  \BibitemOpen
  \bibfield  {author} {\bibinfo {author} {\bibfnamefont {G.}~\bibnamefont {Van~Heijst}}\ and\ \bibinfo {author} {\bibfnamefont {R.}~\bibnamefont {Kloosterziel}},\ }\bibfield  {title} {\bibinfo {title} {Tripolar vortices in a rotating fluid},\ }\href@noop {} {\bibfield  {journal} {\bibinfo  {journal} {Nature}\ }\textbf {\bibinfo {volume} {338}},\ \bibinfo {pages} {569–571} (\bibinfo {year} {1989})}\BibitemShut {NoStop}%
\bibitem [{\citenamefont {Van~Heijst}\ \emph {et~al.}(1991)\citenamefont {Van~Heijst}, \citenamefont {Kloosterziel},\ and\ \citenamefont {Williams}}]{VanHeijst91}%
  \BibitemOpen
  \bibfield  {author} {\bibinfo {author} {\bibfnamefont {G.}~\bibnamefont {Van~Heijst}}, \bibinfo {author} {\bibfnamefont {R.}~\bibnamefont {Kloosterziel}},\ and\ \bibinfo {author} {\bibfnamefont {C.}~\bibnamefont {Williams}},\ }\bibfield  {title} {\bibinfo {title} {Formation of a tripolar vortex in a rotating fluid},\ }\href@noop {} {\bibfield  {journal} {\bibinfo  {journal} {Phys. Fluids A}\ }\textbf {\bibinfo {volume} {3}},\ \bibinfo {pages} {2033} (\bibinfo {year} {1991})}\BibitemShut {NoStop}%
\bibitem [{\citenamefont {Kloosterziel}\ and\ \citenamefont {van Heijst}(1991)}]{Kloosterziel91}%
  \BibitemOpen
  \bibfield  {author} {\bibinfo {author} {\bibfnamefont {R.}~\bibnamefont {Kloosterziel}}\ and\ \bibinfo {author} {\bibfnamefont {G.}~\bibnamefont {van Heijst}},\ }\bibfield  {title} {\bibinfo {title} {An experimental study of unstable barotropic vortices in a rotating fluid},\ }\href@noop {} {\bibfield  {journal} {\bibinfo  {journal} {J. Fluid Mech.}\ }\textbf {\bibinfo {volume} {223}},\ \bibinfo {pages} {1–24} (\bibinfo {year} {1991})}\BibitemShut {NoStop}%
\bibitem [{\citenamefont {Knobloch}(2015)}]{Knobloch15}%
  \BibitemOpen
  \bibfield  {author} {\bibinfo {author} {\bibfnamefont {E.}~\bibnamefont {Knobloch}},\ }\bibfield  {title} {\bibinfo {title} {Spatial localization in dissipative systems},\ }\href@noop {} {\bibfield  {journal} {\bibinfo  {journal} {Annu. Rev. Condens. Matter Phys}\ }\textbf {\bibinfo {volume} {6}},\ \bibinfo {pages} {325} (\bibinfo {year} {2015})}\BibitemShut {NoStop}%
\bibitem [{\citenamefont {Ducimeti\`ere}\ \emph {et~al.}(2024)\citenamefont {Ducimeti\`ere}, \citenamefont {Boujo},\ and\ \citenamefont {Gallaire}}]{Ducimetiere24}%
  \BibitemOpen
  \bibfield  {author} {\bibinfo {author} {\bibfnamefont {Y.-M.}\ \bibnamefont {Ducimeti\`ere}}, \bibinfo {author} {\bibfnamefont {E.}~\bibnamefont {Boujo}},\ and\ \bibinfo {author} {\bibfnamefont {F.}~\bibnamefont {Gallaire}},\ }\bibfield  {title} {\bibinfo {title} {Noise-induced transitions past the onset of a steady symmetry-breaking bifurcation: The case of the sudden expansion},\ }\href@noop {} {\bibfield  {journal} {\bibinfo  {journal} {Phys. Rev. Fluids}\ }\textbf {\bibinfo {volume} {9}},\ \bibinfo {pages} {053905} (\bibinfo {year} {2024})}\BibitemShut {NoStop}%
\bibitem [{\citenamefont {McMullen}\ and\ \citenamefont {Gallis}(2024)}]{McMullen24}%
  \BibitemOpen
  \bibfield  {author} {\bibinfo {author} {\bibfnamefont {R.}~\bibnamefont {McMullen}}\ and\ \bibinfo {author} {\bibfnamefont {M.}~\bibnamefont {Gallis}},\ }\bibfield  {title} {\bibinfo {title} {Hydrodynamic fluctuations near a {H}opf bifurcation: Stochastic onset of vortex shedding behind a circular cylinder},\ }\href@noop {} {\bibfield  {journal} {\bibinfo  {journal} {Phys. Rev. E}\ }\textbf {\bibinfo {volume} {110}},\ \bibinfo {pages} {045104} (\bibinfo {year} {2024})}\BibitemShut {NoStop}%
\bibitem [{\citenamefont {Ha}\ \emph {et~al.}(2021)\citenamefont {Ha}, \citenamefont {Chomaz},\ and\ \citenamefont {Ortiz}}]{Ha21}%
  \BibitemOpen
  \bibfield  {author} {\bibinfo {author} {\bibfnamefont {K.}~\bibnamefont {Ha}}, \bibinfo {author} {\bibfnamefont {J.-M.}\ \bibnamefont {Chomaz}},\ and\ \bibinfo {author} {\bibfnamefont {S.}~\bibnamefont {Ortiz}},\ }\bibfield  {title} {\bibinfo {title} {Transient growth, edge states, and repeller in rotating solid and fluid},\ }\href@noop {} {\bibfield  {journal} {\bibinfo  {journal} {Phys. Rev. E}\ }\textbf {\bibinfo {volume} {103}},\ \bibinfo {pages} {033102} (\bibinfo {year} {2021})}\BibitemShut {NoStop}%
\bibitem [{\citenamefont {Asllani}\ \emph {et~al.}(2018)\citenamefont {Asllani}, \citenamefont {Lambiotte},\ and\ \citenamefont {Carletti}}]{Asllani18}%
  \BibitemOpen
  \bibfield  {author} {\bibinfo {author} {\bibfnamefont {M.}~\bibnamefont {Asllani}}, \bibinfo {author} {\bibfnamefont {R.}~\bibnamefont {Lambiotte}},\ and\ \bibinfo {author} {\bibfnamefont {T.}~\bibnamefont {Carletti}},\ }\bibfield  {title} {\bibinfo {title} {Structure and dynamical behavior of non-normal networks},\ }\href@noop {} {\bibfield  {journal} {\bibinfo  {journal} {Sci. Adv.}\ }\textbf {\bibinfo {volume} {4}},\ \bibinfo {pages} {eaau9403} (\bibinfo {year} {2018})}\BibitemShut {NoStop}%
\bibitem [{\citenamefont {Asllani}\ and\ \citenamefont {Carletti}(2018)}]{Asllani18b}%
  \BibitemOpen
  \bibfield  {author} {\bibinfo {author} {\bibfnamefont {M.}~\bibnamefont {Asllani}}\ and\ \bibinfo {author} {\bibfnamefont {T.}~\bibnamefont {Carletti}},\ }\bibfield  {title} {\bibinfo {title} {Topological resilience in non-normal networked systems},\ }\href@noop {} {\bibfield  {journal} {\bibinfo  {journal} {Phys. Rev. E}\ }\textbf {\bibinfo {volume} {97}},\ \bibinfo {pages} {042302} (\bibinfo {year} {2018})}\BibitemShut {NoStop}%
\bibitem [{\citenamefont {Nicolaou}\ \emph {et~al.}(2020)\citenamefont {Nicolaou}, \citenamefont {Nishikawa}, \citenamefont {Nicholson}, \citenamefont {Green},\ and\ \citenamefont {Motter}}]{Zachary20}%
  \BibitemOpen
  \bibfield  {author} {\bibinfo {author} {\bibfnamefont {Z.~G.}\ \bibnamefont {Nicolaou}}, \bibinfo {author} {\bibfnamefont {T.}~\bibnamefont {Nishikawa}}, \bibinfo {author} {\bibfnamefont {S.~B.}\ \bibnamefont {Nicholson}}, \bibinfo {author} {\bibfnamefont {J.~R.}\ \bibnamefont {Green}},\ and\ \bibinfo {author} {\bibfnamefont {A.~E.}\ \bibnamefont {Motter}},\ }\bibfield  {title} {\bibinfo {title} {Non-normality and non-monotonic dynamics in complex reaction networks},\ }\href@noop {} {\bibfield  {journal} {\bibinfo  {journal} {Phys. Rev. Res.}\ }\textbf {\bibinfo {volume} {2}},\ \bibinfo {pages} {043059} (\bibinfo {year} {2020})}\BibitemShut {NoStop}%
\bibitem [{\citenamefont {Sornette}\ \emph {et~al.}(2023)\citenamefont {Sornette}, \citenamefont {Lera}, \citenamefont {Lin},\ and\ \citenamefont {Wu}}]{Sornette23}%
  \BibitemOpen
  \bibfield  {author} {\bibinfo {author} {\bibfnamefont {D.}~\bibnamefont {Sornette}}, \bibinfo {author} {\bibfnamefont {C.}~\bibnamefont {Lera}}, \bibinfo {author} {\bibfnamefont {J.}~\bibnamefont {Lin}},\ and\ \bibinfo {author} {\bibfnamefont {K.}~\bibnamefont {Wu}},\ }\bibfield  {title} {\bibinfo {title} {Non-normal interactions create socio-economic bubbles},\ }\href@noop {} {\bibfield  {journal} {\bibinfo  {journal} {Commun. Phys.}\ }\textbf {\bibinfo {volume} {6}},\ \bibinfo {pages} {261} (\bibinfo {year} {2023})}\BibitemShut {NoStop}%
\bibitem [{\citenamefont {Weinreich}(1977)}]{Weinreich77}%
  \BibitemOpen
  \bibfield  {author} {\bibinfo {author} {\bibfnamefont {G.}~\bibnamefont {Weinreich}},\ }\bibfield  {title} {\bibinfo {title} {{Coupled piano strings}},\ }\href@noop {} {\bibfield  {journal} {\bibinfo  {journal} {J. Acoust. Soc. Am.}\ }\textbf {\bibinfo {volume} {62}},\ \bibinfo {pages} {1474} (\bibinfo {year} {1977})}\BibitemShut {NoStop}%
\bibitem [{\citenamefont {Politzer}(2015)}]{Politzer15}%
  \BibitemOpen
  \bibfield  {author} {\bibinfo {author} {\bibfnamefont {D.}~\bibnamefont {Politzer}},\ }\bibfield  {title} {\bibinfo {title} {{The plucked string: An example of non-normal dynamics}},\ }\href@noop {} {\bibfield  {journal} {\bibinfo  {journal} {Am. J. Phys.}\ }\textbf {\bibinfo {volume} {83}},\ \bibinfo {pages} {395} (\bibinfo {year} {2015})}\BibitemShut {NoStop}%
\bibitem [{\citenamefont {Biancalani}\ \emph {et~al.}(2017)\citenamefont {Biancalani}, \citenamefont {Jafarpour},\ and\ \citenamefont {Goldenfeld}}]{Biancalani17}%
  \BibitemOpen
  \bibfield  {author} {\bibinfo {author} {\bibfnamefont {T.}~\bibnamefont {Biancalani}}, \bibinfo {author} {\bibfnamefont {F.}~\bibnamefont {Jafarpour}},\ and\ \bibinfo {author} {\bibfnamefont {N.}~\bibnamefont {Goldenfeld}},\ }\bibfield  {title} {\bibinfo {title} {Giant amplification of noise in fluctuation-induced pattern formation},\ }\href@noop {} {\bibfield  {journal} {\bibinfo  {journal} {Phys. Rev. Lett.}\ }\textbf {\bibinfo {volume} {118}},\ \bibinfo {pages} {018101} (\bibinfo {year} {2017})}\BibitemShut {NoStop}%
\bibitem [{\citenamefont {Klika}(2017)}]{Vaclav17}%
  \BibitemOpen
  \bibfield  {author} {\bibinfo {author} {\bibfnamefont {V.}~\bibnamefont {Klika}},\ }\bibfield  {title} {\bibinfo {title} {{Significance of non-normality-induced patterns: Transient growth versus asymptotic stability}},\ }\href@noop {} {\bibfield  {journal} {\bibinfo  {journal} {Chaos}\ }\textbf {\bibinfo {volume} {27}},\ \bibinfo {pages} {073120} (\bibinfo {year} {2017})}\BibitemShut {NoStop}%
\bibitem [{\citenamefont {Jaramillo}\ \emph {et~al.}(2021)\citenamefont {Jaramillo}, \citenamefont {Macedo},\ and\ \citenamefont {Sheikh}}]{Jaramillo21}%
  \BibitemOpen
  \bibfield  {author} {\bibinfo {author} {\bibfnamefont {J.}~\bibnamefont {Jaramillo}}, \bibinfo {author} {\bibfnamefont {R.}~\bibnamefont {Macedo}},\ and\ \bibinfo {author} {\bibfnamefont {L.}~\bibnamefont {Sheikh}},\ }\bibfield  {title} {\bibinfo {title} {Pseudospectrum and black hole quasinormal mode instability},\ }\href@noop {} {\bibfield  {journal} {\bibinfo  {journal} {Phys. Rev. X}\ }\textbf {\bibinfo {volume} {11}},\ \bibinfo {pages} {031003} (\bibinfo {year} {2021})}\BibitemShut {NoStop}%
\bibitem [{\citenamefont {Neubert}\ and\ \citenamefont {Caswell}(1997)}]{Neubert97}%
  \BibitemOpen
  \bibfield  {author} {\bibinfo {author} {\bibfnamefont {M.}~\bibnamefont {Neubert}}\ and\ \bibinfo {author} {\bibfnamefont {H.}~\bibnamefont {Caswell}},\ }\bibfield  {title} {\bibinfo {title} {Alternatives to resilience for measuring the responses of ecological systems to perturbations},\ }\href@noop {} {\bibfield  {journal} {\bibinfo  {journal} {Ecology}\ }\textbf {\bibinfo {volume} {78}},\ \bibinfo {pages} {653} (\bibinfo {year} {1997})}\BibitemShut {NoStop}%
\end{thebibliography}%

\end{document}